\pdfoutput=1

\documentclass[ twoside,openright,titlepage,numbers=noenddot,headinclude,
                footinclude=true,cleardoublepage=empty,abstractoff, 
                BCOR=5mm,paper=a4,fontsize=13.5pt,
                american,%
                ]{scrreprt}

\PassOptionsToPackage{eulerchapternumbers,listings,pdfspacing,
				 subfig,beramono,eulermath,parts}{classicthesis}	

\usepackage{ifthen}
\newboolean{enable-backrefs} 
\setboolean{enable-backrefs}{false} 

\newcommand{\myTitle}{Drag-free Spacecraft Technologies: criticalities in the initialization of geodesic motion\xspace}

\newcommand{\myName}{Carlo Zanoni\xspace}

\newcommand{\mySupervisor}{prof. Daniele Bortoluzzi (tutor)\xspace}
\newcommand{\myFaculty}{PhD school in \textit{Engineering of Civil and Mechanical Structural Systems} \xspace}

\newcommand{\myUni}{University of Trento\xspace}

\newcommand{\myYear}{AA 2013/14\xspace}
\newcommand{\myTime}{24\textsuperscript{th} April 2015}
\newcommand{\myVersion}{version 4.1\xspace}

\newcounter{dummy} 
\providecommand{\mLyX}{L\kern-.1667em\lower.25em\hbox{Y}\kern-.125emX\@}

\PassOptionsToPackage{utf8}{inputenc}	
 \usepackage{inputenc}				

 \usepackage[english, italian]{babel}

\PassOptionsToPackage{square,numbers}{natbib}
 \usepackage{natbib}				

\PassOptionsToPackage{fleqn}{amsmath}		
 \usepackage{amsmath}

\usepackage{siunitx}

\PassOptionsToPackage{T1}{fontenc} 
	\usepackage{fontenc}     
\usepackage{textcomp} 
\usepackage{scrhack} 
\usepackage{xspace} 
\usepackage{mparhack} 
\usepackage{fixltx2e} 
\PassOptionsToPackage{smaller}{acronym}
	\usepackage{acronym} 
\usepackage{textgreek}

\usepackage{tabularx} 
\usepackage{caption}
\usepackage{subfig}  

\usepackage{listings} 
\PassOptionsToPackage{pdftex,hyperfootnotes=false,pdfpagelabels}{hyperref}
	\usepackage{hyperref}  
\PassOptionsToPackage{pdftex}{graphicx}
	\usepackage{graphicx,wrapfig} 

\newcommand{\backrefnotcitedstring}{\relax}
\newcommand{\backrefcitedsinglestring}[1]{(Cited on page~#1.)}
\newcommand{\backrefcitedmultistring}[1]{(Cited on pages~#1.)}
\ifthenelse{\boolean{enable-backrefs}}%
{%
		\PassOptionsToPackage{hyperpageref}{backref}
		\usepackage{backref} 
		   \renewcommand*{\backref}[1]{}  
		   \renewcommand*{\backrefalt}[4]{
		      \ifcase #1 %
		         \backrefnotcitedstring%
		      \or%
		         \backrefcitedsinglestring{#2}%
		      \else%
		         \backrefcitedmultistring{#2}%
		      \fi}%
}{\relax}    

\hypersetup{%
    colorlinks=true, linktocpage=true, pdfstartpage=1, pdfstartview=FitV,%
    breaklinks=true, pdfpagemode=UseNone, pageanchor=true, pdfpagemode=UseOutlines,%
    plainpages=false, bookmarksnumbered, bookmarksopen=true, bookmarksopenlevel=1,%
    hypertexnames=true, pdfhighlight=/O,
    urlcolor=webbrown, linkcolor=RoyalBlue, citecolor=webgreen, 
    pdftitle={\myTitle},%
    pdfauthor={\textcopyright\ \myName, \myUni, \myFaculty},%
    pdfsubject={},%
    pdfkeywords={},%
    pdfcreator={pdfLaTeX},%
    pdfproducer={LaTeX with hyperref and classicthesis}%
}   

\makeatletter
\@ifpackageloaded{babel}%
    {%
       \addto\extrasamerican{%
				}%
       \addto\extrasngerman{%
				}%
    }{\relax}
\makeatother

\usepackage{classicthesis} 

\begin{document}

\frenchspacing
\raggedbottom
\selectlanguage{american} 
\pagenumbering{roman}
\pagestyle{plain}
\begin{titlepage}
	\begin{addmargin}[0cm]{-1cm}
    \begin{center}
\sffamily
		
		\begin{flushright}
		{\scriptsize		
		University of Trento \\
		University of Brescia \\
		University of Bergamo \\
		University of Padova \\
		University of Trieste \\
		University of Udine \\
		University IUAV of Venezia \\
		}

        \large  

        \hfill

        \vfill

        \myName

        \begingroup
            \color{Black}\spacedallcaps{\myTitle} \\ \bigskip
        \endgroup

        \vfill

				\normalsize
				
				\mySupervisor \\ \bigskip
				
        \large 
        \myYear

		\end{flushright}
    \end{center}  
  \end{addmargin}       
\end{titlepage}   
\begin{titlepage}
			\sffamily
		
		\begin{flushleft}

		\textbf{UNIVERSITY OF TRENTO} \\
		Doctoral School in Engineering of Civil and Mechanical Structural Systems \\
		Cycle XXVII \\

        \hfill

        \vfill

        \myTime

        \vfill


				\bigskip

		\end{flushleft} 
\end{titlepage}   
\thispagestyle{empty}

\hfill

\vfill

\small
\noindent\myName: \textit{\myTitle,} 
\textcopyright\ \myTime

%
%
%
%
%

\cleardoublepage
\thispagestyle{empty}
\refstepcounter{dummy}
\pdfbookmark[1]{Dedication}{Dedication}

\vspace*{3cm}

\begin{center}
	Creatures of my dreams raise up and dance with me! \\
	Now and forever, I'm your king! \\ \medskip
    --- M83. ''Outro.'' Hurry up! We're dreaming.    
\end{center}

\medskip

\begin{center}
    Sing yourself with fife and drum, \\
		sing yourself to overcome \\
		the thought that someone’s lost \\
		and someone else has won. \\ \smallskip
   --- U2. ''Soon.''
\end{center}
\cleardoublepage
\pdfbookmark[1]{Abstract}{Abstract}
\begingroup
\let\clearpage\relax
\let\cleardoublepage\relax
\let\cleardoublepage\relax

\chapter*{Abstract}

Present and future space missions rely on systems of increasingly demanding performance for being successful. Drag-free technology is one of the technologies that is fundamental for LISA-Pathfinder, a European Space Agency mission whose launch is planned for the end of September 2015.

A purely drag-free object is defined by the absence of all external forces other than gravity. This is not a natural condition and therefore a shield has to be used in order to eliminate the effect of undesired interactions. In space, this is achieved by properly designing the spacecraft that surrounds the object, usually called test mass (TM). Once the TM is subjected to gravity alone its motion is used as a reference for the spacecraft orbit. The satellite orbit is controlled by measuring the relative TM-to-spacecraft position and feeding back the command to the propulsion system that counteracts any non gravitational force acting on the spacecraft. Ideally, the TM should be free from all forces and the hosting spacecraft should follow a pure geodesic orbit. However, the purity of the orbit depends on the spacecraft’s capability of protecting the TM from disturbances, which indeed has limitations.

According to a NASA study, such a concept is capable of decreasing operation and fuel costs, increasing navigation accuracy. At the same time, a drag-free motion is required in many missions of fundamental physics.

eLISA is an ESA concept mission aimed at opening a new window to the universe, black holes, and massive binary systems by means of gravitational waves. This mission will be extremely challenging and needs to be demonstrated in flight. LISA-Pathfinder is in charge of proving this concept by demonstrating the possibility of reducing the non-gravitational disturbance below a certain demanding threshold. The success of this mission relies on recent technologies in the field of propulsion, interferometry, and space mechanisms.

In this frame, the system holding the TM during launch and releasing it in free-fall before the science phase represents a single point of failure for the whole mission. This thesis describes the phenomena, operations, issues, tests, activities, and simulations linked to the release following a system engineering approach. Great emphasis is given to the adhesion (or cold welding) that interferes with the release. Experimental studies have been carried out to investigate this phenomenon in conditions representative of the LISA-Pathfinder flight environment.

The last part of the thesis is dedicated to the preliminary design of the housing of the TM in the frame for a low-cost mission conceived at Stanford (USA). Analysis and results are through out presented and discussed.

The goal of this thesis is a summary of the activities aimed at a successful LISA-Pathfinder mission. The ambition is to increase the maturity of the technology needed in drag-free projects and therefore provide a starting point for future fascinating and challenging missions of this kind.

\vfill

\pdfbookmark[1]{Sommario}{Sommario}
\chapter*{Sommario}

\selectlanguage{italian}

Il successo di missioni spaziali sempre più ambiziose richiede tecnologie sempre più avanzate. Una di queste tecnologie è la cosiddetta tecnologia drag-free, la cui traduzione letterale in italiano sarebbe \textit{libera da resistenza areodinamica} (riferito ad un orbita o traiettoria), che può essere generalizzata come \textit{libera da effetti non gravitazionali}. Tale tecnologia è la base su cui si fonda la missione LISA-Pathfinder dell'Agenzia Spaziale Europea (ESA), il cui lancio è previsto per la fine di settembre 2015.

Un oggetto drag-free è caratterizzato dal fatto che la gravità è l'unica forza agente su di esso. Perchè ciò sia possibile, uno schermo opportuno deve essere costruito attorno all'oggetto che si desidera essere drag-free. Un satellite drag-free è progettato per eseguire questa funzione schermando una massa di riferimento (TM, dall'inglese test mass) dalle forze che non siano gravità e minimizzando i disturbi che esso stesso genera. Una volta che la TM si comporta come un oggetto drag-free, può essere utilizzata a sua volta come riferimento per la traiettoria del satellite. Infatti, la posizione della TM all'interno del satellite è misurata ed analizzata attraverso un algoritmo di controllo con cui si comandano i propulsori facendo sì che il satellite segua la TM. Idealmente, la TM dovrebbe risultare in volo libero, che è un altro modo per definire il moto drag-free, così come dovrebbe esserlo il satellite, che ne segue la traiettoria. Come è facile intuire, invece, non esiste lo schermo perfetto e quindi delle forze di disturbo non-gravitazionali, seppur mitigate, continuano ad agire perturbando il moto ideale.

Secondo uno studio della NASA, questo schema di funzionamento può portare significative economie sia nell'uso del carburante che nei costi operativi di correzione dell'orbita del satellite. Allo stesso tempo, la tecnologia drag-free è la base necessaria per diverse missioni scientifiche finalizzate allo studio della gravità e alla parte di fisica fondamentale in cui la gravità ricopre un ruolo primario.

eLISA è una proposta di missione dell'ESA con l'ambizione di aprire un intero nuovo campo della fisica basato sulle onde gravitazionali per studiare fenomeni quali buchi neri e sistemi binari. Lo studio delle onde gravitazionali è già stato approvato per uno slot per una missione a budget elevato. Il concetto di eLISA è laborioso e complesso e richiede un'intera missione spaziale perché ne sia dimostrata la fattibilità. Lo scopo primario di LISA-Pathfinder è quindi dimostrare che l'idea alla base di eLISA è realizzabile provando che è possibile mitigare i disturbi non gravitazionali al di sotto di una certa soglia. Il successo di questa missione dipende da sviluppi tecnologici anche nell'ambito della propulsione, dell'interferometria e dei meccanismi spaziali.

Nell'ambito di LISA-Pathfinder, il meccanismo incaricato di vincolare le TM durante il lancio e rilasciarle in volo libero una volta in orbita è particolarmente critico. Il suo fallimento determinerebbe il fallimento di tutta la missione. Questa tesi descrive i fenomeni, le operazioni, i problemi, i test, le attività e le simulazioni realizzate nell'ambito del rilascio. Particolare attenzione è dedicata al fenomeno adesivo che si oppone alla separazione della TM dal meccanismo di rilascio. Una sostanziale attività sperimentale è stata portata avanti per caratterizzare questo fenomeno nelle condizioni di interesse per LISA-Pathfinder. 

L'ultima parte della tesi è dedicata alla progettazione dell'alloggiamento della TM nell'ambito di una missione low-cost ideata a Stanford (USA). Le analisi e i risultati sono presentati e discussi.

Nel complesso, questa tesi si prefigge di riassumere un insieme di attività di qualifica della fase di rilascio, finalizzate alla buona riuscita di LISA-Pathfinder. Parallelamente, ha l'ambizione di ampliare la maturità tecnologica necessaria a realizzare un satellite drag-free, fungendo quindi da punto di partenza per affascinanti ed ambiziosi progetti futuri con problematiche simili, tra cui, appunto, eLISA.

\endgroup			

\vfill
\cleardoublepage
\pdfbookmark[1]{Publications}{publications}
\chapter*{Publications}
Some paragraphs, figures and ideas have appeared previously in the following publications:

\bigskip

{\small C. Zanoni and D. Bortoluzzi. \textit{Experimental-analytical qualification of a piezo-electric mechanism for a critical space application}. IEEE/ASME Transactions on Mechatronics. 20: 427-437. 2015.

D. Bortoluzzi, C. Zanoni, and S. Vitale. \textit{Improvements in the measurement of metallic adhesion dynamics}. Mechanical Systems and Signal Processing. 52-53: 600-613. 2014.

C. Zanoni, D. Bortoluzzi, J.W. Conklin, I. K\"{o}ker, C.G. Marirrodriga, P.M. Nellen, and S. Vitale. \textit{Testing the injection of the LISA-Pathfinder Test Mass into geodesic conditions}. In Proceedings of the 15th ESMATS conference. 2013.

C. Zanoni, A. Alfauwaz, A. Aljadaan, et al. \textit{The design of a drag-free CubeSat and the housing for its gravitational reference sensor}. In 2nd IAA Conference On University Satellite Missions And Cubesat Workshop. 2013.

D. Bortoluzzi, S. Vitale, and C. Zanoni. \textit{Test mass release testing - release test summary}. Technical report. ESA. 2013.

D. Bortoluzzi, J.W. Conklin, and C. Zanoni. \textit{Prediction of the LISA-Pathfinder release mechanism in-flight performance}. Advances in Space Research. 51:1145–1156. 2013.

D. Bortoluzzi, M. Benedetti, C. Zanoni, J.W. Conklin, and S. Vitale. \textit{Investigation of dynamic failure of metallic adhesion: a space-technology related case of study}. In Proceedings of the SEM International Conference \& Exposition on Experimental and Applied Mechanics. 2013.

C. Zanoni, D. Bortoluzzi, and J.W. Conklin. \textit{Simulation of a critical task of the LISA release mechanism: the injection of the Test Mass into geodesic}. In The 9th LISA Symposium. 2012.

D. Bortoluzzi, M. Benedetti, C. Zanoni, and J.W. Conklin. \textit{Test mass release testing extended preload test campaign}. Technical report. ESA. 2012.

M. Benedetti, D. Bortoluzzi, C. Zanoni, and J.W. Conklin. \textit{Measurement of metallic adhesion force-to-elongation profile under high separation-rate conditions}. In Proceedings of the SEM International Conference \& Exposition on Experimental and Applied Mechanics. 2012.

D. Bortoluzzi, M. Benedetti, L. Baglivo, C. Zanoni, and S. Vitale. \textit{Test-mass release testing summary of criticalities of the test-mass release}. Technical report. ESA. 2010.}
\pagestyle{scrheadings}
\cleardoublepage
\refstepcounter{dummy}
\pdfbookmark[1]{\contentsname}{tableofcontents}
\setcounter{tocdepth}{2} 
\setcounter{secnumdepth}{3} 
\manualmark
\markboth{\spacedlowsmallcaps{\contentsname}}{\spacedlowsmallcaps{\contentsname}}
\tableofcontents 
\automark[section]{chapter}
\renewcommand{\chaptermark}[1]{\markboth{\spacedlowsmallcaps{#1}}{\spacedlowsmallcaps{#1}}}
\renewcommand{\sectionmark}[1]{\markright{\thesection\enspace\spacedlowsmallcaps{#1}}}
\clearpage

\begingroup 
    \let\clearpage\relax
    \let\cleardoublepage\relax
    \let\cleardoublepage\relax
    \refstepcounter{dummy}
    \pdfbookmark[1]{\listfigurename}{lof}
    \listoffigures

    \vspace*{8ex}

    \refstepcounter{dummy}
    \pdfbookmark[1]{\listtablename}{lot}
    \listoftables
        
    \vspace*{8ex}
    

       
    \refstepcounter{dummy}
    \pdfbookmark[1]{Acronyms}{acronyms}
    \markboth{\spacedlowsmallcaps{Acronyms}}{\spacedlowsmallcaps{Acronyms}}
    \chapter*{Acronyms}
    \begin{acronym}[UML]
				\acro{CMV}{Caging and Vent Mechanism}
				\acro{DFACS}{Drag Free and Attitude Control System}
				\acro{EH}{Electrode Housing}
        \acro{GPRM}{Grabbing, Positioning and Release Mechanism}
				\acro{ISS}{Inertial Sensor Subsystems}
        \acro{LISA}{Lisa Interferometer Space Antenna}
				\acro{RT}{Release Tip}
				\acro{RTmu}{Release Tip mock-up}
        \acro{TM}{Test Mass}
        \acro{TMmu}{Test Mass mock-up}
				\acro{TMMF}{Transferred Momentum Measurement Facility}
    \end{acronym}                     
\endgroup

\cleardoublepage
\pagenumbering{arabic}
\cleardoublepage
\chapter{Introduction}\label{ch:Introduction}

\emergencystretch=.3em

A purely free-falling object is defined by the absence of all external forces other than gravity. Such a motion is also called geodesic or drag-free, as a reference to the main source of non-gravitational interactions in Low Earth orbits.

In drag-free space missions, a free-floating object, usually called test mass (TM), is enclosed into a satellite. The spacecraft acts as a shield and protects the object from disturbances, due to forces other than gravity. These disturbances are the result of external interactions (air drag, solar pressure, Earth magnetic field) and spacecraft-generated effects (gravitational attraction, mechanical, electrical and thermal noise). The core of a drag-free satellite is the Gravitational Reference Sensor (GRS). This device is constituted by a housing and a set of sensors that measures the TM position. The TM trajectory is used as a reference for the orbit through a control system that commands the actuation of the thrusters and counteracts the non-gravitational forces applied on the spacecraft. Unfortunately, as seen later, a small fraction of forces, other than gravity, will always act also on the TM.

The drag-free trajectory is of great interest for both fundamental physics experiments and spacecraft navigation technologies. That is why drag-free missions have been conceived since the early 60s, \cite{Pugh59,Lange}. However, since DISCOS flight on TRIAD in 1972, \cite{TRIAD}, only few of them have been actually flown, because of the technical challenges. One of the main critical factors in obtaining drag-free motion is the transient between the launch phase and normal operations. As a matter of facts, in most of these missions the need arises to constrain the TM during launch. Such a constraint has then to be removed leaving the TM in a proper state for operation.

The phase between the constrained and free-floating motion is called \textit{release}. The importance of this phase has been recognized since the early times of drag-free technology, \cite{Dan1976}. A good performance of the release depends on the impulse transferred to the test mass in this phase, which has to be as close to zero as possible. One of the main obstacles in obtaining this result is the adhesion phenomenon that makes difficult the removal of mechanical constraints without applying a force on the test mass.

An example of the criticality of drag-free projects is the LISA-Pathfinder mission, whose launch is planned for the end of September 2015.

The content of this thesis is the summary of activities aimed at assessing the release phase for drag-free missions. Particular emphasis is given to the LISA-Pathfinder case, that constitutes the state of the art, up to now.

The next sections of the Introduction will highlight the motivation for drag-free technology for both scientific and technological reasons. 

A first overview of such motivations is given in Tab. \ref{tab:applications}, presented in \cite{DFCubeSat}. Such a table summarizes all the applications and scientific fields of drag-free technology and provides an estimate of the required level of purity of the gravitational action (i.e. the amount of other interactions) and the frequency range of interest. In some cases the measurement requires the presence of more than a single drag-free object, such that a relative quantity is sensed.

\begin{table}[!h]
\footnotesize
\renewcommand{\arraystretch}{1}
\caption{Applications of Drag-free Technology}
\label{tab:applications}
\centering
\begin{tabular}{p{2cm}|p{2.5cm}|p{2cm}|p{1.7cm}|p{2.3cm}}
\hline
\textbf{Category} & \textbf{Application} & \textbf{Drag-free Performance [$m/s^2 Hz^{1/2}$]} & \textbf{Frequency Range [Hz]}& \textbf{Metrology Required [m]} \\
\hline
Navigation & Autonomous orbit maintenance  & $< 10^{-10}$ & near-zero & N/A\\ \cline{2-5}
 & Future autonomous orbit maintenance  & $< 10^{-10}$ & near-zero & < 10, absolute\\ \cline{2-5}
 & Future precision real-time on-board navigation  & $< 10^{-10}$ & near-zero & < 3, absolute\\ \cline{1-5}
Earth Science & Aeronomy  & $< 10^{-10}$ & $10^{-2}-1$ & < 1, absolute\\ \cline{2-5}
 & Geodesy  & $< 10^{-10}$ & $10^{-2}-1$ & < $10^{-6}$, differential\\ \cline{2-5}
 & Future Earth Geodesy  & $< 10^{-12}$ & $10^{-2}-1$ & < $10^{-9}$, differential\\ \cline{1-5}
Fundamental Physics & Equivalence principle tests  & $< 10^{-10}$ & $10^{-2}-1$ & < $10^{-10}$, differential\\ \cline{2-5}
 & Tests of general relativity & $< 10^{-10}$ & near-zero & < 1, absolute\\ \cline{1-5}
Astrophysics & Gravitational waves & $< 10^{-14}$ & $10^{-4}-1$ & $10^{-11}$, differential\\
\hline
\end{tabular}
\end{table}

In the last section of this Introduction a summary of the thesis content is finally provided.

\section{Scientific Motivation}

Space is a privileged environment for fundamental physics experiments \cite{Worden2013}. It allows -for instance- long distances, long interaction times, large velocity changes and absence of seismic noise. Among all, the noise free environment with sometime harsh, but stable thermal, electric and magnetic conditions is extremely precious for most experiments \cite{LasersClocks}. It is well known how the human space age began under the push of military needs. Both the first rockets and the moon race were in truth weapons or strongly linked to weapons and to matters of geopolitical power. 

One of the first fully-scientific mission was Gravity Probe-A (1976), which measured the gravitational redshift with an accuracy still to be repeated. The interest in space-based experiments has greatly increased since that project. Today, a large number of scientific space mission are proposed each year. The European Space Agency itself has the goal of promoting ''for exclusively \textit{peaceful} purposes, cooperation among European States in space research and technology and their space applications, with a view to their being used for scientific purposes''\footnote{Article II, Purpose, Convention of establishment of a European Space Agency, SP-1271(E), 2003}. All this proves the increasing importance of the space environment for science.

Unfortunately, space missions have also peculiar disadvantages:
\begin{itemize}
	\item long time for preparation and development;
	\item no access to the payload during operation;
	\item no access to the payload for a post-experiment analysis.
\end{itemize}
At the same time, the complex and long development as well as the search for an ideal 100\% reliability have a negative impact on the costs that often become huge. 

In the frame of space-based experiments, drag-free technology is proven to be fundamental in a wide range of scientific fields \cite{inorbit}. Examples of them are:
\begin{itemize}
	\item aeronomy;
	\item study of the Earth's geoid;
	\item study of gravity;
	\item detection and analysis of gravitational waves.	
\end{itemize}
Indeed, these studies have very different essence, from short-term application to fundamental research. As example, the knowledge of the Earth's geoid would allow the prediction and study of catastrophic events due to scale change in water distribution and, at the same time, the detection and study of gravitational waves ''has the potential to transform much of physics and astronomy [...] and to reshape the science questions of the future'' \footnote{LISA science team: \textsl{LISA: Probing the Universe with Gravitational Waves}, 2009, ESA}.

Independently from the type of study, drag-free concept promises to open a whole new type of measurements promising for both a better knowledge of our planet and the understanding of the foundations of the universe.

\section{Technological Motivation}

Drag-free technology is not only a requirement for several physics space missions, but also a path to autonomous orbit control and maintenance and fuel and operational cost savings \cite{NASA2003}.

Such a NASA study analyzes the possibilities and the challenges of drag free technology. It highlights that missions whose drag is continuously compensated provide the following advantages:
\begin{enumerate}
	\item a 50\% fuel savings for satellites in LEO with altitude smaller than 350 km. The savings are computed with a comparison to a spacecraft whose orbit is corrected once every month;
	\item a 30\%-50\% reduction in navigation error; 
	\item substantial cost savings for a constellation of satellites in LEO.
\end{enumerate}
The 350 km figure depends on the capability of reducing the non gravitational noise generated on the TM by the spacecraft itself. Such a noise must be a negligible fraction of the other interactions. Up to now, this appears to be economically convenient in low orbits where drag is a substantial effect.
On top of the technological advantage, the use of drag-free makes the need of correcting satellite orbits\footnote{such a correction is required once every 2-4 weeks in LEO, due to the continuous decay of the orbit} less and less important. The operational costs would then be further reduced. 

\begin{center}
\begin{figure}[!ht]
\includegraphics[width=\columnwidth]{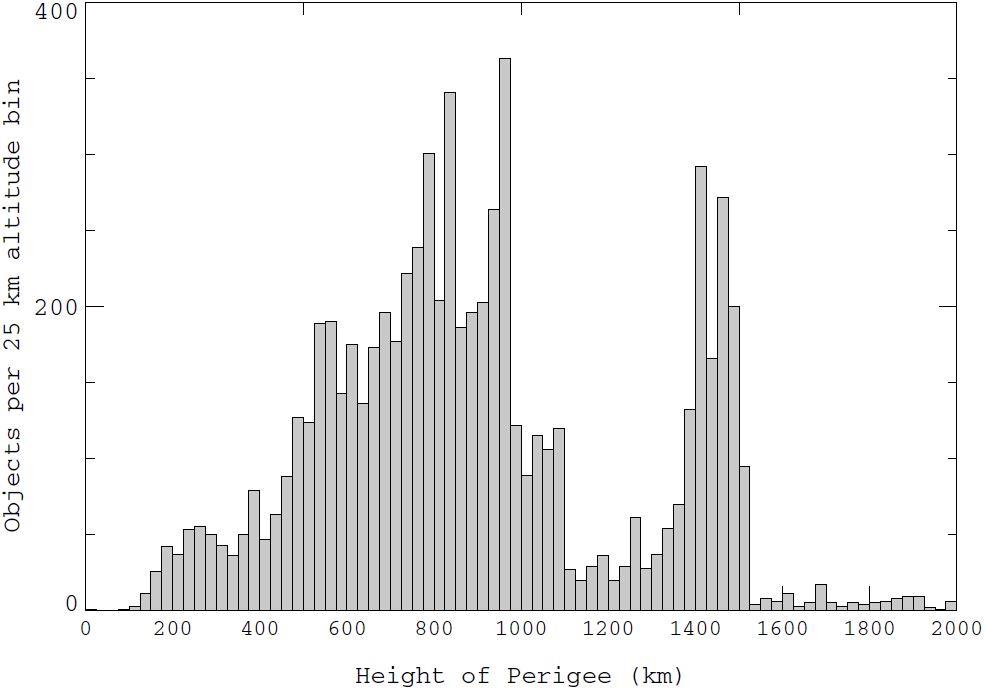}
\caption{Debris concentration as a function of the perigee of their orbit \cite{debris}.}
\label{fig:Debris}
\end{figure}
\end{center}

It is interesting to link this analysis with one of the main issues in modern space engineering: space debris. Fig.~\ref{fig:Debris} (from \cite{debris}) shows that the concentration of debris is limited below 400 km. This happens because atmospheric drag is able to naturally clean space in a reasonable time below that altitude. As a consequence, an increased economic profitability of orbits lower than 350 km would also have benefits for the risks of this missions, with additional savings, and for the cleanliness of space.

Lastly, the core of drag-free missions can be either a GRS or an accelerometer. In facts, the structure and functioning of a GRS is very similar to that one of precise accelerometers \cite{Accelerometer}, in which the TM is electrostatically constrained -or suspended- to the spacecraft and the measurement of the actuation needed on it is an estimate of the acceleration. The development of drag-free technology is then also a push for more accurate and precise accelerometers.

\subsection{Examples of space projects with key drag-free technologies}

As described above, the drag-free concept relies on a spacecraft following a TM, with no forces applied on the TM by the control system. The same concept can be inverted such that the TM is kept centered in its housing by the control loop while the spacecraft follows a certain trajectory. In such a case the force applied on the TM is the output, instead of the force applied on the spacecraft. The name of this concept is \textit{accelerometer mode} and has already been implemented on flown space missions. 

A few examples of projects relying on the drag-free or accelerometer concept are listed in the following. It is worth noting that the drag free mode can be implemented in slightly different ways (i.e. the force on the TM is not zero, but is only minimized according to a certain criterion) as is well summarized in \cite{Domenico}.
\begin{description}
	\item[TRIAD]. \textsl{Goal}: extend the time between ephemeris updates\footnote{Ephemeris provide the position of astronomical objects for navigation purposes. The more such a source is needed, the less the automatic navigation system is accurate.} to 10 days, \cite{TRIAD, DragJohn}. This means the residual acceleration has to be below $10^{-10} m s^{-2}$. TRIAD achieved an impressive $5\times 10^{-11} m s^{-2}$.
	\item[Gravity Probe-B, GP-B]. \textsl{Goal}: test of two unverified predictions of general relativity: the geodetic effect and frame-dragging. It was a combined effort of NASA and Stanford University and it was launched in 2004. The final results were shown in 2011, \cite{GPB}.
	\item[Satellite Test of the Equivalence Principle, STEP]. \textsl{Goal}: test of the weak equivalence principle, \cite{STEP}. This mission is in the preliminary design phase at Stanford, but the funding for a further completion is not secured yet.
	\item[GRACE]. \textsl{Goal}: measure Earth's gravity field anomalies. Gravity Recovery and Climate Experiment is a mission launched in 2002. It consists of 2 spacecraft in the same polar orbit. GRACE uses microwave ranging and an accelerometer in each satellite in order to accurately measure the relative change in speed and distance of the two spacecraft, \cite{GRACE}.
	\item[GOCE]. \textsl{Goal}: map the Earth's geoid with the highest accuracy and precision up to now, \cite{GOCE}. The Gravity Field and Steady-State Ocean Circulation Explorer main instrument was a highly sensitive gradiometer, consisting of an array of accelerometers (three pairs). The mission was launched in 2009, was extremely successful and ended its life in November 2013. The LEO had a 250 km altitude and, in order to minimize drag force, the spacecraft had an aerodynamic shape.
	\item[Microscope]. \textsl{Goal}: test of the weak equivalence principle with an accuracy of one hundred times better than the one obtained with experiments realized on Earth, \cite{MICROSCOPE, MicroscopeHudson2007}. Launch planned for 2016.

\end{description}

\subsection{Pushing new technologies: the hold-down and release case}

Successful drag-free missions require the further development of several novel technologies, that are already used in space projects, but are not able to guarantee the challenging performance needed for a drag-free satellite. Among them, reliable and long-life micro-N thrusters are indeed critical for an accurate control of the spacecraft orbit. Laser interferometry provides the mean for precise measurement of the TM position, especially if the relative distance between many TM is required. Finally the minimum impulse release of the TM must be guaranteed at the beginning of science or navigation operation, after LEOP\footnote{Launch and Early Orbit Phase}. This need represents well the transition from pyrotechnic to precise separation, that is happening in these days for safety and reliability reasons that go beyond the drag-free scope.

Space mechanisms are very often among the most critical systems in a spacecraft. They are so critical that the European Union lists the \textit{Space qualification of low shock Non-Explosive Actuators} among the urgent actions in the scope of its Horizon 2020 programme. This fact is of course true also in the drag-free missions case where the hold-down and release of the TM is a single-point of failure for the entire project.

In general, the systems of this kind can be classified in one of the following categories:
\begin{itemize}
	\item Pyrotechnics: the separation is produced by the firing of an explosive. Such systems are very energy efficient, quick and reliable, thanks to the long experience of use. However, they determine a shock on the spacecraft that may not be compliant with the payload. At the same time, the use of explosives always raises some concerns on the safety and testing on-ground.
	\item High Energy Paraffin actuators: that are based on the large expansion of paraffin wax from the solid to the liquid state. These systems can generate very high forces and can be repeatedly actuated. However, they are not reliable when a simultaneous action of more devices is needed.
	\item Burn wire mechanisms: the separate parts of the mechanism are kept together by a wire. The ignition of the release is triggered by the melting of this wire. These devices can be operated in a very quick time (order of magnitude: ms).
\end{itemize}

LISA-Pathfinder release mechanism has some analogies with the paraffin actuator in terms of performance. However, on top of that a very quick time of action is needed as well as a certain control of the position. For this reason the LISA-Pathfinder case represents also a \textit{first} in the field of hold-down and release devices and a mission critical function \cite{TaskGroup}.

\subsection{Object Injection in Geodesic Conditions}

The need to bring a test mass from the Earth environment to free-fall or suspended conditions requires the development of challenging technologies and deep studies. During launch, the forces applied to the TM are several orders of magnitude higher than during the science phase in space. At the same time, the TM and its housing electronics need to be properly protected from the harsh launch environment.

\begin{figure}[!ht]
\begin{center}
\includegraphics[width=0.92\columnwidth]{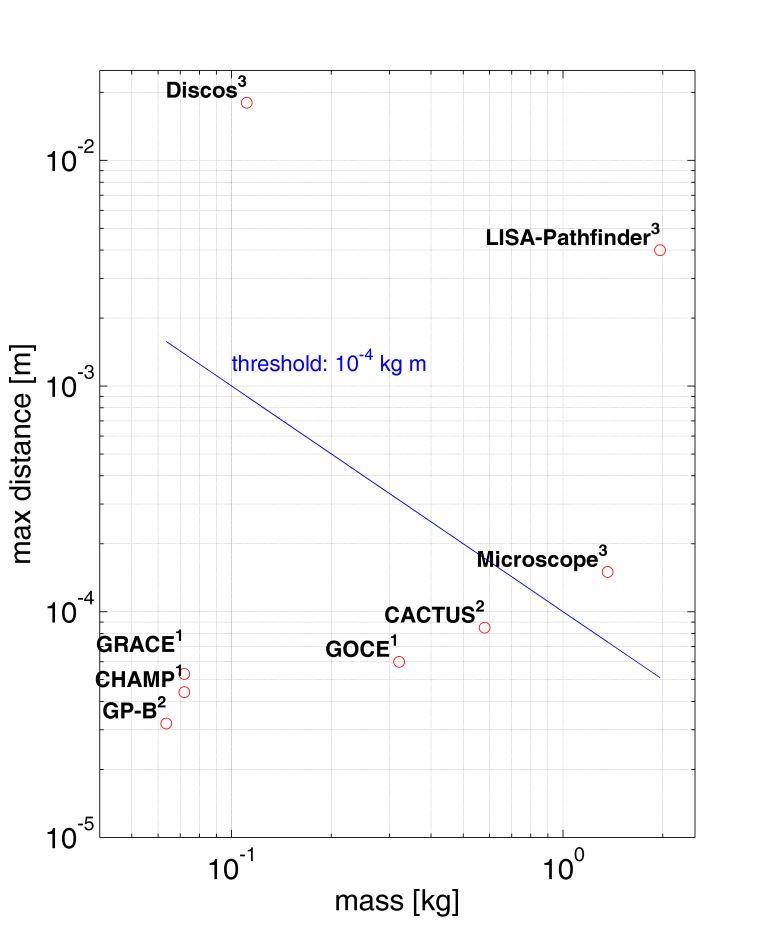}
\end{center}
\caption{Free mass vs. free-path in several space missions. $^1$ without caging, $^2$ without caging, but with gaseous or liquid damping, $^3$ with caging}
\label{fig:impact}
\end{figure}

\cite{ESAobjectinjection} highlights two different strategies for facing the launch issue: leave the TM free of moving inside the housing or design a caging system. In the first case, the main advantage is the simplicity. On the other hand, the safety of both TM and housing must be verified and guaranteed by careful design of the housing. In the second case, the development of a caging-dedicated system reduces the reliability of the project, but is more applicable in general as it leaves more freedom in the design of the housing.
A quantity that suggests which strategy is better in each case is the impact kinetic energy that the TM would have during launch \cite{Collins}:

\begin{equation}
	K_{impact} = \frac{1}{2}m v_{impact}^2 = \frac{1}{2}m (\sqrt{2ad})^2 = a m d
\end{equation}

where $m$ is the TM mass, $a$ the launcher mean acceleration while the TM moves, $d$ the potential free motion distance (in absence of any caging). Assuming the launcher performance is roughly the same, the main quantity is then the product between mass and free distance. Fig.~\ref{fig:impact} shows the value of mass and free space in a few missions, \cite{ESAobjectinjection}. The threshold value appears to be about a mass-distance product of $10^{-4}$ kg m.

A remarkable source of problems while injecting a body in free-fall or suspended conditions is the adhesion phenomenon that may arise at the mechanical contact. Such a phenomenon is enhanced by vacuum and the absence of gravity. However, if the TM is left free during launch, the loads and vibrations of that phase will break any adhesive contact that may have arisen in earlier test, integration or storage phases. A small joining force may be eventually present and an electrostatic system must be designed in order to provide a detaching force. On the other hand, the use of a caging system means that a constant high load is applied onto the TM during launch. Such a load may determine high adhesion. The caging system has to be designed accordingly.

Most of this thesis describes the LISA-Pathfinder example. In this mission, a caging system is needed and electrostatic actuation is very limited, compared to adhesion. The object injection strategy has then to deal with one of the most critical cases.

\section{Thesis Structure}

The main topic of this thesis is an example of the technical developments pushed by drag-free technology and represents a small step forward in space-mechanism tribology and release devices. The release mechanism conceived for LISA-Pathfinder is a \textit{first} in space history. As every separation device, it has a dual purpose: secure the payload (or part of it) against the launch loads and cleanly release it once in orbit.

The core of this work follows a system engineering approach, emphasizing all the different aspects of the release, and is organized in 7 chapters:

\begin{enumerate}
	\item The second chapter provides an introduction to eLISA, LISA-Pathfinder and the main systems involved in the release.
	\item The third chapter describes more in detail the release mechanism, presents an electro-mechanical model and the procedure followed for the identification of the model's parameters.
	\item The forth chapter overviews the adhesion phenomenon and some basic theory.
	\item In the fifth chapter a detailed description of the experimental facility and procedure is provided. Together with this description particular care emphasis is given to the presentation and preliminary discussion of the results.
	\item The sixth chapter presents the simulations that estimate the in-flight behavior of the release along with a worst-case figure relative to the release performance.
	\item The seventh chapter broadly summarizes the working principle and the performance of the systems and procedures involved in the release in a more general sense. That means the systems with an interaction with the TM after the mechanical release [more info can be requested to the author, this thesis version omits some data to safeguard Airbus Space property]. 
	\item The last part provides information on the preliminary design on the housing for the Test Mass in a Stanford-UFL\footnote{University of Florida}-NASA-KACST\footnote{King Abdulaziz City for Science and Technology} proposed Drag-Free CubeSat.
\end{enumerate}


\cleardoublepage
\chapter{The LISA Pathfinder Space Mission}\label{ch:LPF}

\emergencystretch=.3em
\hyphenation{gra-vi-ta-tio-nal}

This chapter overviews the eLISA and LISA-Pathfinder goals, main features and critical points as well as an introduction to the TM release issue, which is intimately linked to the success of both missions.

\section{\textnormal{\small{e}}LISA overview}

The evolving Laser Interferometer Space Antenna, eLISA, is an ESA Large class mission conceived to measure gravitational waves. In 2013, the science theme ''The Gravitational Universe'' \cite{whitepaper} has been selected by ESA for an L3 launch slot, but the history of this mission begins long before this milestone. Among all, it has been the object of several technological and scientific studies, a candidate for combined efforts of both ESA and NASA  \cite{LISAyellow} and for the just-European L1 slot (under the name of New Gravitational Observatory, \cite{NGO}). Unfortunately, budget constraints have often delayed the project, but this long process is not fully negative as it has allowed decades of intensive technological studies and preparation.

Almost the entirety of the space missions observe the Universe by means of electromagnetic radiation. However, most of the Universe is not visible electromagnetically and must be sensed in a different way. We can listen (\cite{nature}) to the Universe by means of gravitational waves that, thanks to their weak interaction with matter, can transport information for very long distances. Unfortunately, this weak interaction is also the reason why the detection of gravitational waves is so challenging.

The goals of eLISA are summarized in the following, \cite{whitepaper,Vit2014}:
\begin{enumerate}
	\item observe the merging of massive black holes and study their mass, spin and redshift. This will allow, for the first time, the exploration of the low-mass part of the black-hole population ($10^4-10^7\:M_{sun}$).
	\item analyze EMRI\footnote{Extreme Mass Ratio Inspiral, that is a compact start (neutron star...) pulled in a highly relativistic spiral orbit by a massive black hole} especially those happening in black holes with masses about the mass of the black hole that sits at the center of our galaxy.
	\item test General Relativity with un-precedent sensitivity, that could allow measuring even small deviations from Einstein's theory.
	\item observe compact binaries systems. Only a small fraction of the stars are companion-less, the majority is part of a binary or even multiple system. If the orbit separation is small, these systems evolve into compact systems.
	\item provide precious information on the nature of gravity and help the development of a physics theory capable of taking into account both particle and quantum elements and large scale effects. eLISA is a corner element of a larger picture of fundamental physics experiments that already includes high-energy particle accelerators, such as LHC \cite{LHC}, and cosmological probes, \cite{Worden2013}.
\end{enumerate}
These targets will be reached measuring the waves in the 0.1 mHz - 1 Hz range, which is not accessible on Earth due to phenomena such as seismic noise and tides and therefore justifies the need of a space mission.


\begin{center}
\begin{figure}[!ht]
\caption{Left: eLISA constellation. Right: spacecraft concept. \cite{whitepaper}}
\label{fig:eLISA}
\includegraphics[width=1.1\columnwidth]{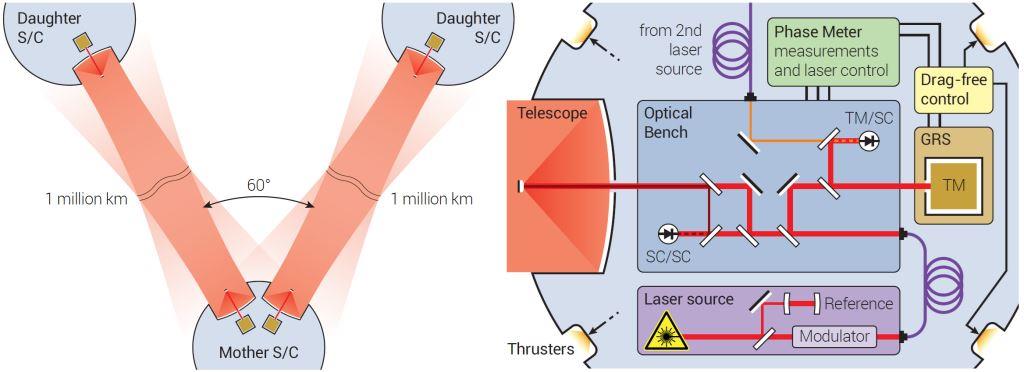}
\end{figure}
\end{center}

Thanks to the 20-year work on LISA, eLISA mission concept is already well defined. eLISA will be a 3 spacecraft mission shaped as a triangular constellation in heliocentric orbit. 

The squeezing of space-time due to the gravitational waves of interest for eLISA has an amplitude $h=\Delta L/L\approx10^{-20}$. The L quantity has then to be maximized in order to obtain a measurable $\Delta L$. The side length of this triangle will then be $1\times10^6$ km. In the current plan, the constellation will have one mother spacecraft and 2 daughters. The mother is linked with an interferometer arm with the daughters. The technical limits on the interferometer are the constraint on the triangle side length. The laser link will sense the relative positions of the free falling test masses, that are affected by gravitational waves.

The last concept provides 4 identical payloads, 2 mounted on the mother and 1 on each daughter. Each of these units consists of a Gravitational Reference Sensor (GRS) that hosts a 46 mm Au/Pt cubic TM, a UV discharge system, an electrostatic suspension controller and a caging and release mechanism (see next section for further details). Beside the GRS, each payload hosts a 2 W telescope, for spacecraft-to-spacecraft sensing, and an optical bench for the spacecraft-to-TM position. Indeed, the road towards a gravitational telescope is still long and challenging and the final concept may have differences from this overview.

\section{LISA Pathfinder}

Several of the technologies mentioned in the previous section are very critical for the mission success and are also hard to be tested on-ground. An entire space mission has then been conceived in order to facilitate technological development and demonstrate the eLISA concept. This mission is called LISA-Pathfinder and it is an ESA/NASA space probe to be launched in September 2015 \cite{CQG,CQGmission,AnzaLTP,Vit2014}.

\begin{center}
\begin{figure}[ht!]
\caption{LISA Pathfinder core, schematic view.}
\label{fig:LPFscheme}
\begin{center}
\includegraphics[width=1.05\columnwidth]{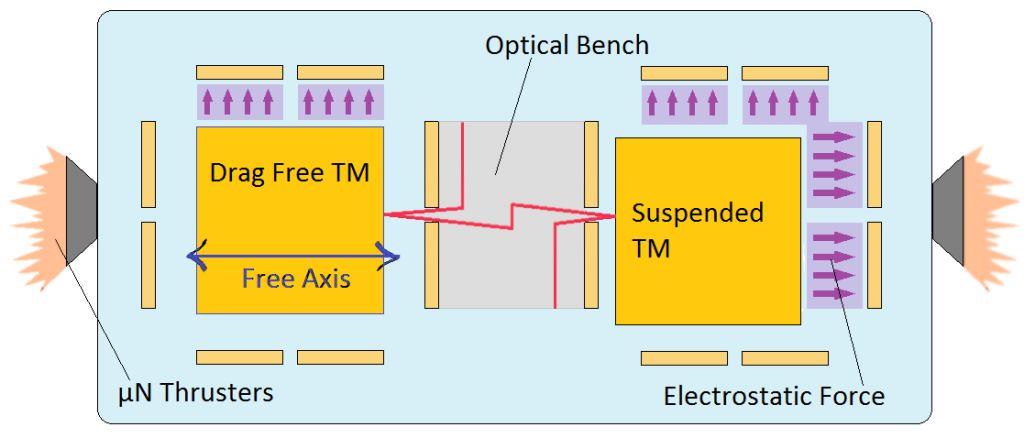}
\end{center}
\end{figure}
\end{center}

The basic idea is to reduce the arm length between a pair of TM from $10^6$ km to 35 cm. The two TM are then enclosed and shielded by the same spacecraft. In the main science experiment, one of the TM is drag-free along one linear DoF while the other DoF are controlled by the electrodes. The second TM is electrostatically actuated on all the DoF at low frequencies in order to follow the first TM and assess the residual acceleration. See Fig. \ref{fig:LPFscheme} for a visual explanation.

\begin{center}
\begin{figure}[ht!]
\caption{LISA Pathfinder in Thermal Vacuum Chamber, courtesy of Airbus Defence and Space.}
\label{fig:LPF}
\begin{center}
\includegraphics[width=\columnwidth]{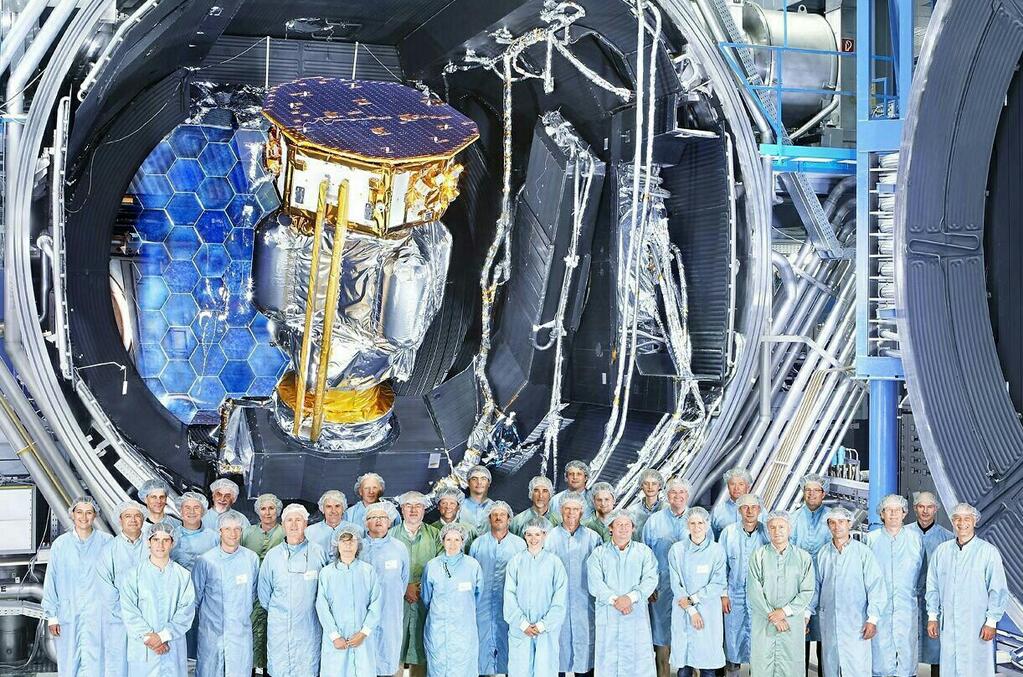}
\end{center}
\end{figure}
\end{center}

\begin{center}
\begin{figure}[ht!]
\caption{LISA Pathfinder chassis, courtesy of Airbus Defence and Space.}
\label{fig:LPFsc}
\begin{center}
\includegraphics[width=\columnwidth]{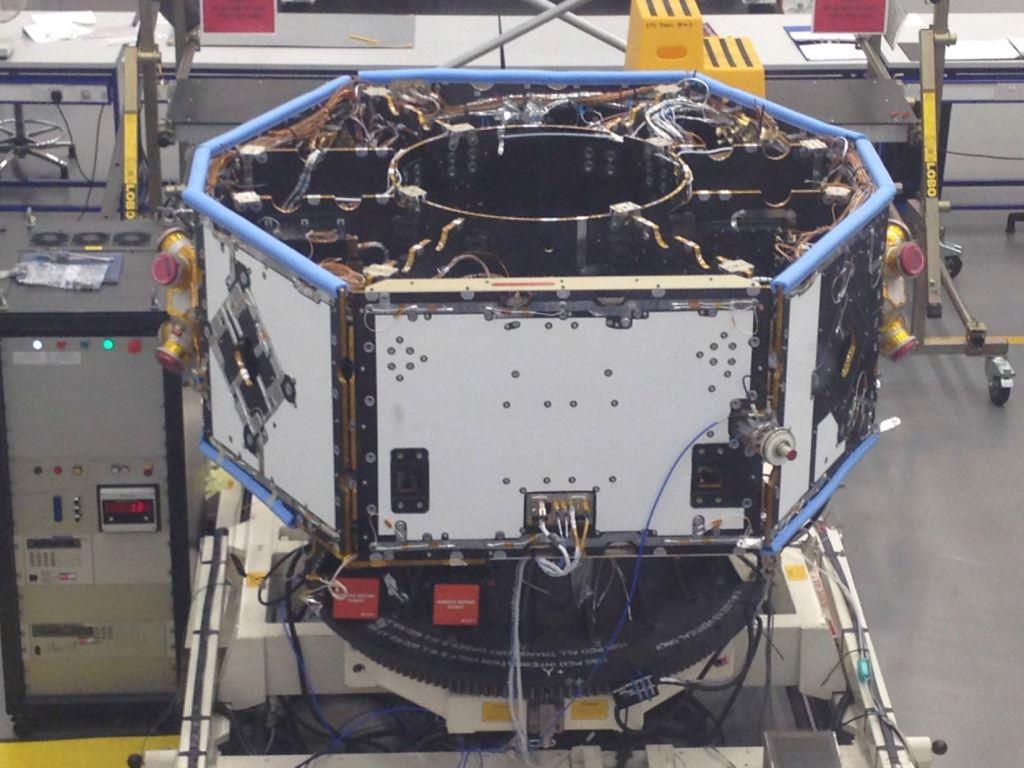}
\end{center}
\end{figure}
\end{center}

The goals of LISA Pathfinder (Fig.~\ref{fig:LPF}) are summarized in the following points:

\begin{enumerate}
	\item demonstrate the possibility of reducing the non-gravitational disturbance acceleration below:
		\begin{equation}
			\Delta a \leq 3 \times 10^{-14} \left[ 1 + \left(\frac{f}{3 mHz}\right)^2 \right] m s^{-2} Hz^{-1/2}
		\end{equation}
				over a frequency bandwidth, $f$, between 1 and 30 mHz \cite{CQGmission, CQGstatus}. This quantity is the relative acceleration between the two TM. It is worth noting that such an approach is conservative as the suspension noise of the non drag-free TM is not present in a full LISA configuration, \cite{freeflight}.
	\item demonstrate that the laser interferometer is capable of a resolution below:
			\begin{equation}
			\Delta x \leq 9 \times 10^{-12} \left[ 1 + \left(\frac{f}{3 mHz}\right)^2 \right] m Hz^{-1/2}
			\end{equation}
	\item assess a large set of critical technologies such as discharging system, micro-newtonian thrusters, drag-free control system and caging and release mechanism.
\end{enumerate}
Beside this objective, LISA-Pathfinder (Fig.~\ref{fig:LPFchassis}) will also be a space laboratory for testing the model of several non-gravitational forces in interplanetary space.

\begin{figure}[ht!]
\begin{center}
\caption{Sources of acceleration noise (design requirements), \cite{freeflight}.}
\label{fig:LPFreq}
\includegraphics[width=0.8\columnwidth]{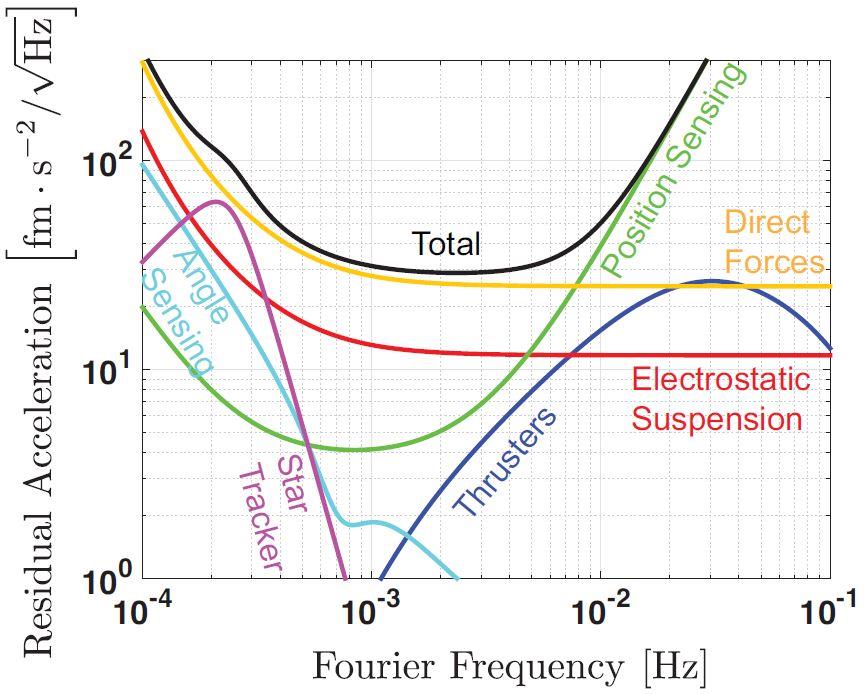}
\end{center}
\end{figure}

\begin{figure}[ht!]
\begin{center}
\caption{LISA Pathfinder chassis, courtesy of Airbus Defence and Space.}
\label{fig:LPFchassis}
\includegraphics[width=\columnwidth]{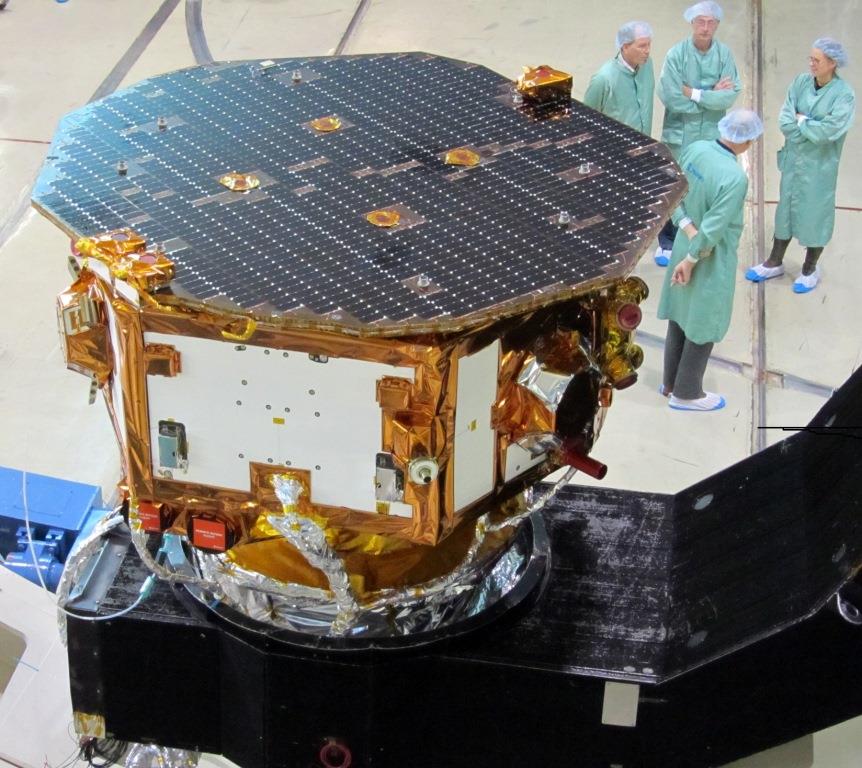}
\end{center}
\end{figure}


Traditional Space Science distinguishes clearly between payload and spacecraft systems, in LISA-Pathfinder this distinction is hardly applicable: spacecraft and payload act much more like a unique unit.

\begin{figure}[ht!]
\begin{center}
\caption{LISA Pathfinder Inertial Sensor Head. The optical windows for the laser light are substituted by plane flanges, courtesy of CGS Italy.}
\label{fig:ISH}
\includegraphics[width=\columnwidth]{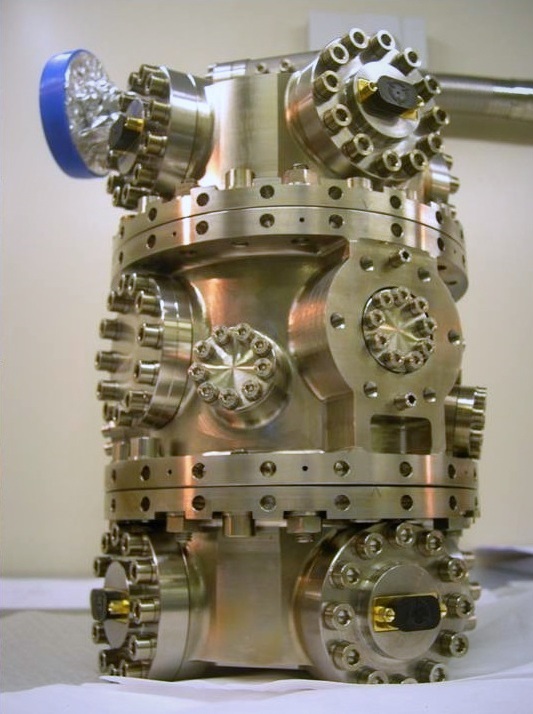}
\end{center}
\end{figure}

\begin{figure}[ht!]
\begin{center}
\caption{LISA Pathfinder Electrode Housing, courtesy of CGS Italy.}
\label{fig:EHpic}
\includegraphics[width=\columnwidth]{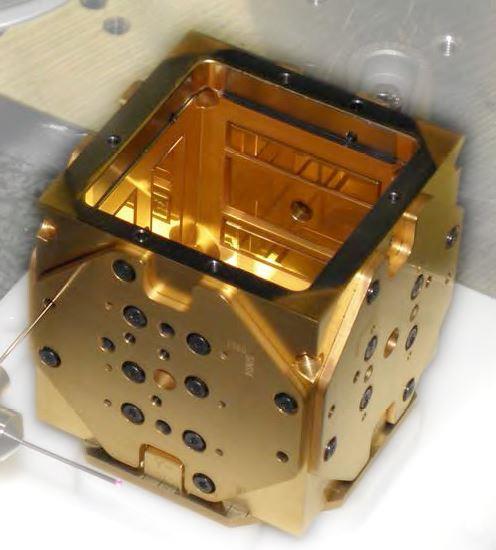}
\end{center}
\end{figure}

\begin{figure}[ht!]
\begin{center}
\caption{LISA Pathfinder Electrode Housings after thermal-vacuum testing, courtesy of CGS Italy.}
\label{fig:EHs}
\includegraphics[width=\columnwidth]{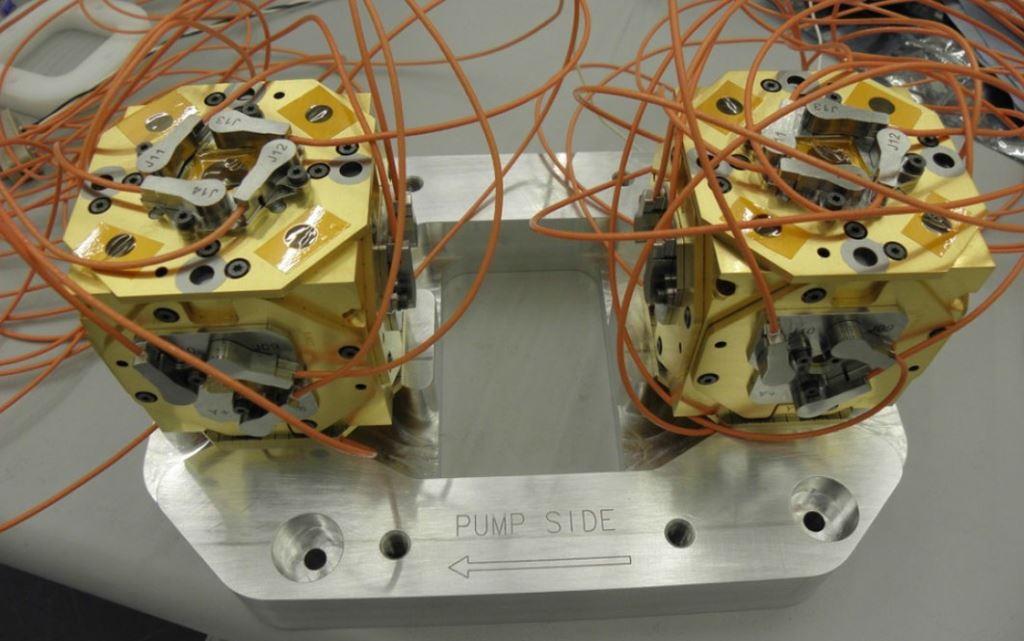}
\end{center}
\end{figure}

\begin{figure}[ht!]
\begin{center}
\caption{LISA Pathfinder Test Mass, courtesy of CGS Italy.}
\label{fig:TM}
\includegraphics[width=\columnwidth]{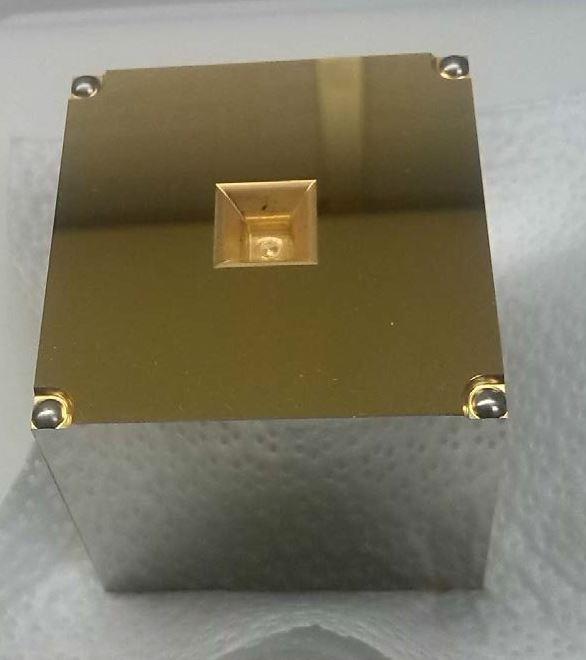}
\end{center}
\end{figure}

The most important subsystem is the Inertial Sensor Subsystem (ISS), Fig.~\ref{fig:ISH}. It includes the electrode housing (in charge of hosting the TM and sense/actuate its position, Fig.~\ref{fig:EHpic} and \ref{fig:EHs}), the front-end electronics, the vacuum system\footnote{although space is certainly a vacuum environment, the spacecraft is subjected to uncontrolled out-gassing.}, charge management (in charge of avoiding the accumulation of charge on the TM, due to cosmic rays and solar particles), the caging and release system and of course the TM (Fig. \ref{fig:TM}). Each TM is a 1.96 kg 30\% Pt-70 \% Au 46 mm side cube. It must have a very high density homogeneity ($<<1 \mu m$ pores) and the center of gravity must be closer than $2 \mu m$ to the geometrical center. The magnetic properties are also very challenging (magnetic susceptibility: $–(2.3 \pm 0.2 ) \times 10^{-5}$).

The sensing of the TM position is performed both with the housing electrodes (part of the ISS) and the interferometer (part of the OMS, see later). 

As far as the electrodes are concerned, a trade off has to be found between sensing and actuation authority and the disturbances due to uncontrolled electric fields inside the housing. These fields are due to localized time-varying changes in the charge density on the surface of both electrodes and TM. Indeed, such an effect applies disturbance forces and is particularly critical as every force on the TM reduces the purity of the drag-free motion, which is the main goal of the mission. Both the uncontrolled fields and the authority increase when the TM-to-Housing gap becomes small. A gap of about 4 mm has then been chosen. Indeed, such a choice has consequences also on other devices, as will be shown later.

Finally, the ISS is also in charge of minimizing the TM-to-spacecraft gravitational attraction by means of a set of compensation masses \cite{Gcompensation2005} and guarantee proper temperature stability.

\begin{figure}[ht!]
\begin{center}
\caption{LISA Pathfinder optical bench during integration, courtesy of University of Glasgow.}
\label{fig:opticalbench}
\includegraphics[width=\columnwidth]{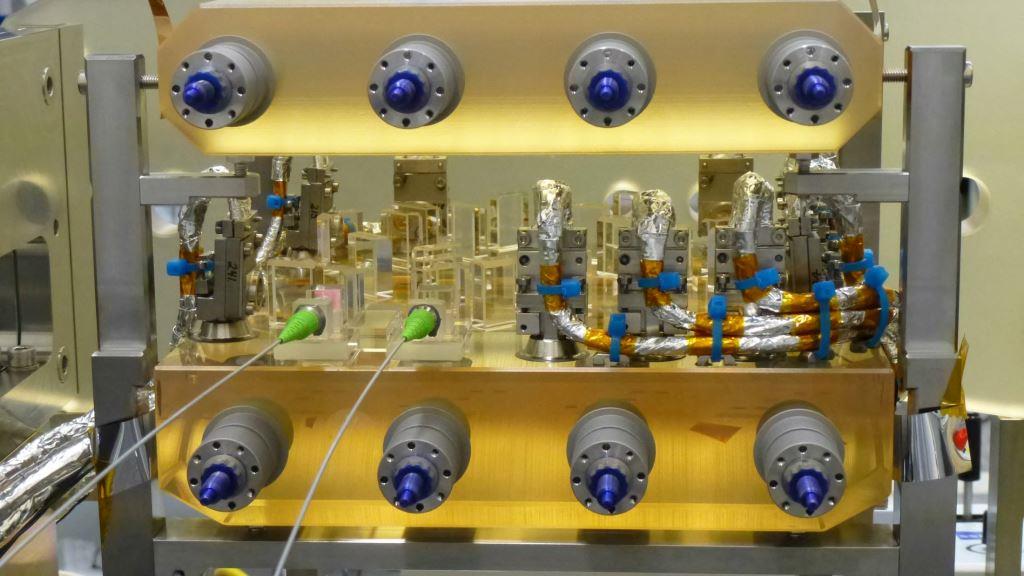}
\end{center}
\end{figure}

Another key subsystem is the Optical Metrology Subsystem (OMS), that provides a high resolution laser readout of the TM position.  Among the OMS devices we find the optical bench, the laser modulator, the phase-meter and the reference laser. Indeed this subsystem is the main sensor, for science purposes. Fig. \ref{fig:opticalbench} shows the optical bench during integration.

Finally, in order to maintain the spacecraft in a drag-free orbit compliant with the requirement, a micro-propulsion system is needed. The thrusters initially designed to this purpose were the Field Emission Electric Propulsors (FEEP). However, due to bad performance in terms of lifetime, a cold-gas system was eventually preferred. Such a system has also the advantage of being flight-proven, as it is the same mounted aboard the GAIA spacecraft.

It can be easily shown that all these sub-systems are intimately interlinked. The division is made for management purposes and for allowing the geographical distribution of tasks, as usually performed by ESA.

Table \ref{tab:contributions} shows the geographical distribution of the contributions to the LISA-Pathfinder technology.
\begin{table}[!ht]
\small
\renewcommand{\arraystretch}{1}
\caption{LISA Technology Package Contributions \cite{LPFOverview}}
\label{tab:contributions}
\centering
\begin{tabular}{p{3.5cm}|p{6cm}}
\hline
\textbf{Country} & \textbf{Responsibility} \\
\hline
France & Laser Modulator \\
\hline
Germany & Co-PI, Interferometer design, LTP architect, Reference laser unit \\
\hline
Italy & PI, Inertial sensor design, Inertial sensor subsystem, Test Mass, Electrode housing \\
\hline
The Netherlands & Inertial sensor check-out equipment \\
\hline
Spain & Data management unit, Data diagnostic system \\
\hline
Switzerland & Inertial sensor front-end electronics \\
\hline
UK & Phasemeter assembly, Optical bench interferometer, charge management system \\
\hline
ESA & Coordination, Caging \& release mechanism \\
\hline
\end{tabular}
\end{table}

\section{Caging and Releasing of the Test Mass}
\label{sec:Caging}

A very large portion of space missions rely on mechanisms for successfully completing their task. Space mechanisms are therefore easily identified as one of the most critical systems in a spacecraft \cite{RobertsSpaceTribology, Fortescue}. The environment in which they are operated is hugely varied from the high loads of launch to the extreme conditions of vacuum. At the same time their reliability must be maximized, for obvious reasons.

Mechanism have to be designed such that their performance are compliant with the requirement defined at system and subsystem level. On-ground environment is clearly different from space. The qualification and validation before flight are therefore always challenging.

Among the space mechanism, release devices are often the most critical as they have to fulfill a certain action in a certain instant, with certain performance and without a second chance. This is true for both the deployment of a launcher payload, solar arrays or for small subsystems. There is a large industrial interest in enhancing the performance of release devices by mitigating the shock.

At the same time, one of the most critical phenomena for space mechanism is tribology \cite{RobertsSpaceTribology,EidenTribology}, that is not of easy modeling and repeatability on-ground.

LISA-Pathfinder release device sums these criticalities as it has to both separate the TM from the spacecraft with a very limited relative velocity and also to face a tribological phenomenon that has never been studied in real dynamic conditions: adhesion. 
 
As mentioned, the Test Mass (TM) is a 1.96~kg gold coated Au/Pt cube and is hosted in an electrode housing for both capacitive sensing and actuation. The housing-to-TM gaps are about 4~mm in every direction. If compared with similar missions, the combination of large mass and large gaps is chosen to limit the acceleration produced by spurious forces and to decrease the force produced by surface charge patches located both on housing and TM. However, three critical drawbacks follow these design choice. First, the TM must be heavily constrained during launch, because its impact with the housing would produce serious damages \cite{ESAobjectinjection}. Second, not only the TM but also any other surrounding device must be either gold-based or gold-coated, in order to limit non-uniformity of surface work function and related uncontrolled stray force. Third, due to the large gaps to the electrodes the actuation authority is very weak ($5 \times 10^{-7}$ N), such that after the release the capacitive control system is able to stabilize the TM only if its initial state is limited by very demanding upper bounds as listed in Tab.~\ref{tab:req}.

\begin{table}[!ht]
\small
\renewcommand{\arraystretch}{1}
\caption{TM state requirement}
\label{tab:req}
\centering
\begin{tabular}{|c||c|}
\hline
\textbf{State} & \textbf{Value} \\
\hline
offset along x, y and z & $\pm$ 200~$\mu$m \\
\hline
linear velocity along x, y and z & $\pm$ 5~$\mu$m/s \\
\hline
angle around x, y and z & $\pm$ 2 mrad \\
\hline
angular rate around x, y and z & $\pm$ 100~$\mu$rad/s \\
\hline
\end{tabular}
\end{table}

A non compliant release of the TM hinders the initialization of the following science phase. As a consequence of the tight requirements on the TM centering, the major criticality is identified in the linear momentum rather than the angular. As a matter of facts, the linear momentum associated with the maximum velocity along z is converted in an angular velocity about 20 times smaller than the requirement.

As already experienced in space mechanisms \cite{ESASTM, RobertsSpaceTribology,JAXAUHV, EidenTribology}, adhesive interactions between the TM and a lock/release device are expected, enhanced by many boundary conditions such as vacuum, fretting and inhibition of surface oxydes formation. Moreover, the presence of gold coating on the TM surface and the exclusion of anti-adhesive coatings on any surface exposed to the TM limit any design reduction of adhesion in vacuum. The estimation of the adhesive interactions between the TM and a lock/release device makes it possible to rule out any static disengagement of the two bodies performed by the capacitive actuation, as they result orders of magnitude larger than the force authority. The most promising strategy is therefore to split the launch lock and in-flight release functions, that require forces of different orders of magnitude.

\subsection{Caging and Vent Mechanism: CVM}

The Caging and Vent Mechanism is in charge of:
\begin{itemize}
	\item hold the TM during launch.
	\item hand-over the TM to the Grabbing, Positioning and Release Mechanism (GPRM) as a first step of the release.
\end{itemize}

Such a system has to complain the following design constraints:
\begin{itemize}
	\item survive and sustain launcher vibrations
	\item no magnetic materials
	\item no liquid lubricants
	\item low-shock during hand-over ( < 200 g SRS\footnote{Shock Response Spectrum})
\end{itemize}

In the resulting design, the TM is held by a set of 8 fingers grasping the 8 TM corners.

\begin{figure}[ht!]
\caption{Caging and Vent Mechanism concept \cite{Esmats2013z}.}
\label{fig:CVM2}
\begin{center}
\includegraphics[width=1\columnwidth]{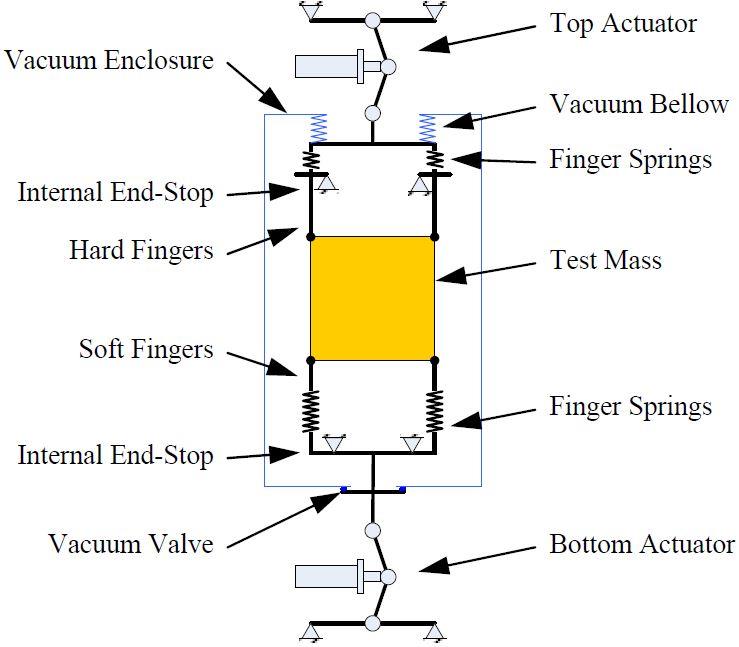}
\end{center}
\end{figure}

\begin{figure}[ht!]
\caption{Caging and Vent Mechanism (in drafted form) and Grabbing, Positioning and Release Mechanism (solid model)}
\label{fig:CVM1}
\begin{center}
\includegraphics[width=1.2\columnwidth]{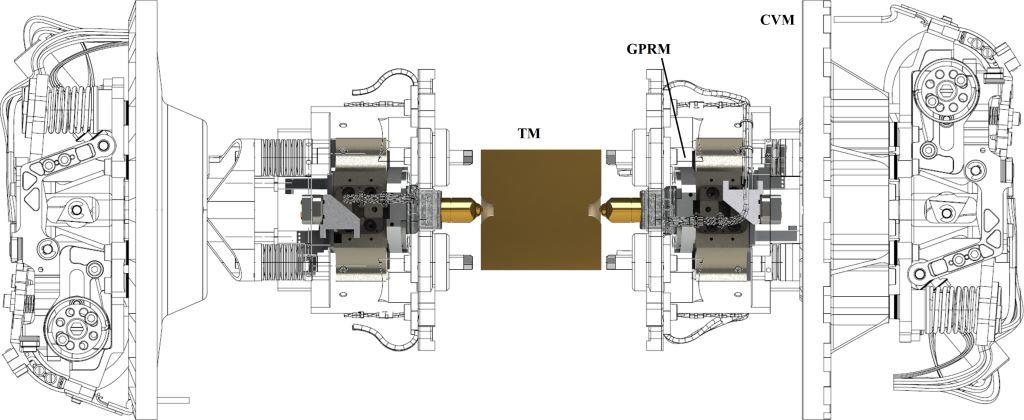}
\end{center}
\end{figure}

Eventually, the CVM (Fig. \ref{fig:CVM2} and \ref{fig:CVM1}) consists of two equal and opposed subsystems, which are assembled to firmly constrain the TM with a set of 8 fingers, engaging the cube on its corners with a force of about 300 N each \cite{Esmats2013z, CVMrep}. A preloaded toggle mechanism performs a single-shot spiral cam driven retraction of the fingers in order to break the strong adhesive interactions produced at the contacts by the high pressure. The strength of adhesion (on the order of N) and its asymmetry on the 8 contacts makes it possible to exclude that this retraction can release the TM with the requirements listed in Tab.~\ref{tab:req}. The mechanism chosen is not able to perform multiple strokes.

\subsection{Grabbing, Positioning and Release Mechanism: GPRM}

\begin{figure}[ht!]
\caption{Views and details of the GPRM, courtesy of RUAG Schweiz AG, RUAG Space and MAGNA International Europe.}
\label{fig:GPRM1}
\begin{center}
\includegraphics[width=1.05\columnwidth]{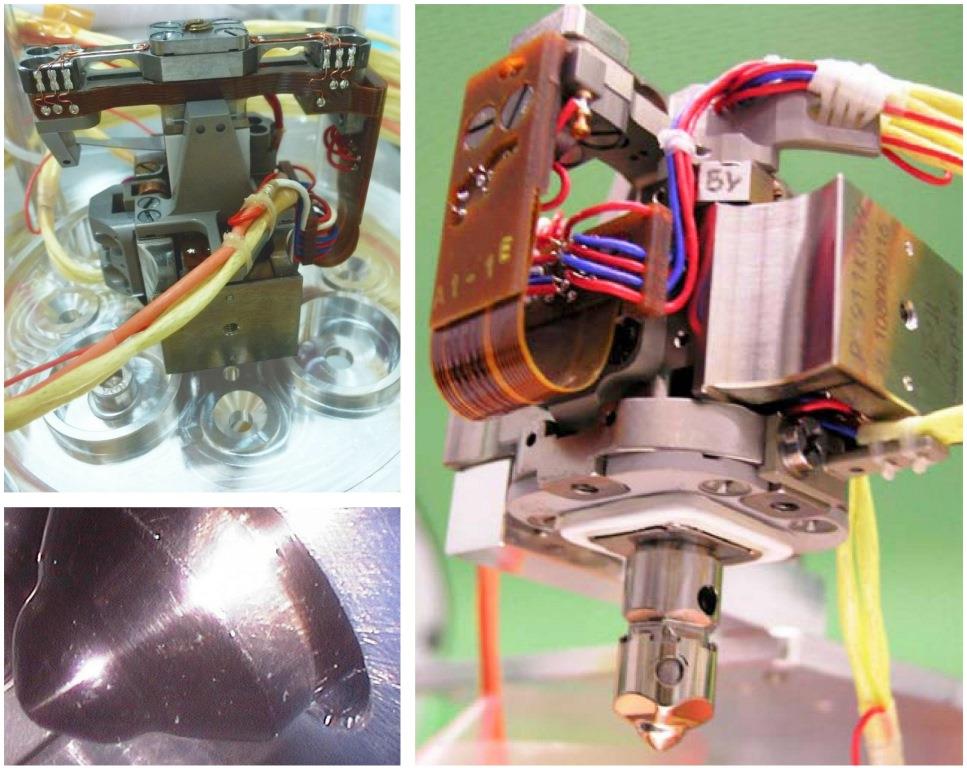}
\end{center}
\end{figure}

\begin{figure}[ht!]
\caption{Caging Mechanism during thermal test, upper view. The system in the center with blue and red wires is the GPRM, courtesy of CGS Italy.}
\label{fig:C;thermal}
\begin{center}
\includegraphics[width=1\columnwidth]{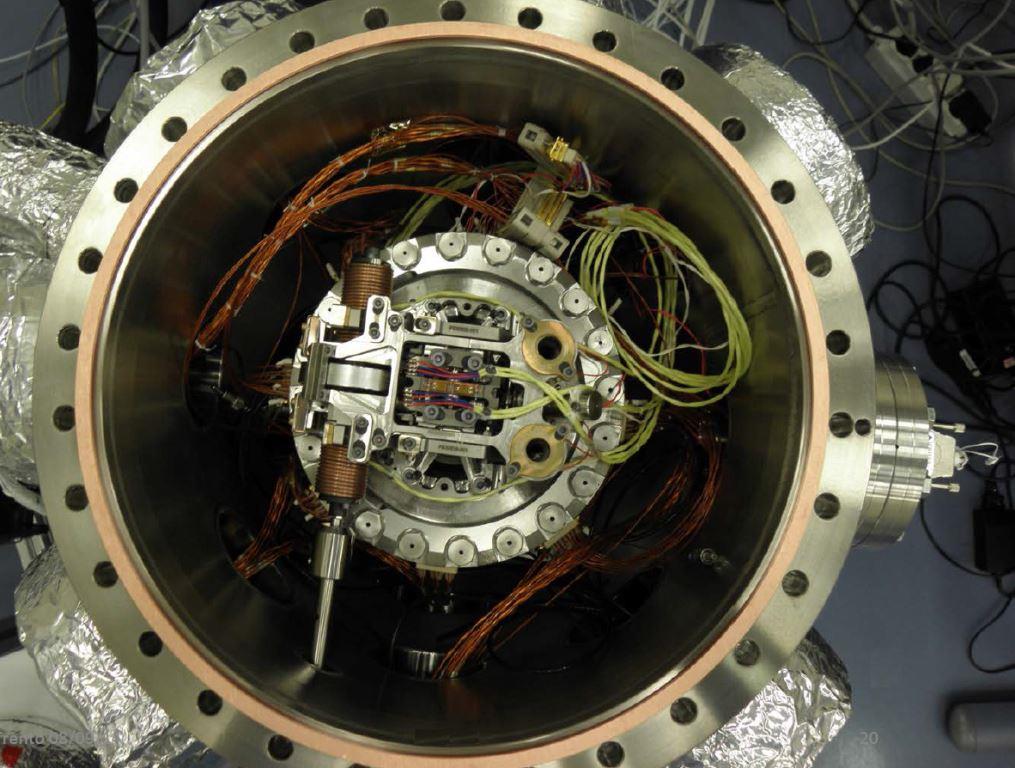}
\end{center}
\end{figure}

The Grabbing, Positioning and Release Mechanism \cite{GPRMdesign, esmats2007, ESMATS13i, ESMATS2009} is conceived for completing the CVM action and release the TM with limited contact load ($\approx$ 1 N). In order to meet the expected performance, the GPRM must: 
\begin{itemize}
	\item Grab the TM from all the positions/configurations possible inside the housing.
	\item Avoid damaging the TM.
	\item Allow multiple caging and release operations.
	\item Avoid magnetic materials.
\end{itemize}

These guidelines heavily constrain the geometry and the type of actuators allowed. The final GPRM is designed to grab and center the TM during in-orbit operations, applying the limited force required to handle it in absence of weight. It engages two cam-shaped wedges by means of opposite plungers, which are actuated by a long stroke (on the order of 10 mm) piezo-walk actuator. This actuator generates motion through succession of coordinated clamp/unclamp and expand/contract cycles of a set of smaller piezo stages. Conversely, the release phase is performed by two release tips, which are commanded by piezo-stack actuators. 

The Release Tip (RT) is the system included inside the plunger and is the last device mechanically constraining the TM. After the plunger has engaged the pyramidal recesses, starts the hand-over phase. In this phase the plunger is cyclically retracted of small amounts ($\approx\:\mu m$) while the RT is extracted. This coordinated movement allows a continuous contact between the TM and the GPRM and a load below 0.4 N.

The RT performs the real release, that must be compliant with the requirements of Tab.~\ref{tab:req}. Such a challenging request is due to the very weak actuation ($5 \times 10^{-7}$ N) which would not be able to guide the TM within the housing if the initial states exceeds the limits. The RT design is detailed later.

\begin{figure}[ht!]
\begin{center}
\caption{View of the GPRM during alignment tests at Magna.}
\end{center}
\label{fig:GPRM2}
\includegraphics[width=7cm]{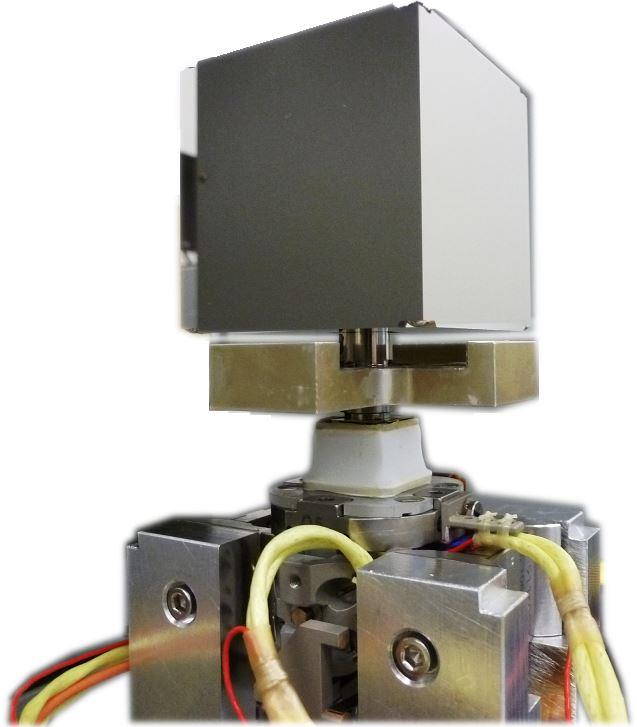}
\end{figure}

\begin{figure}[ht!]
\caption{Extraction and retraction of RT and plunger.}
\label{fig:procedure}
\begin{center}
\includegraphics[width=0.8\columnwidth]{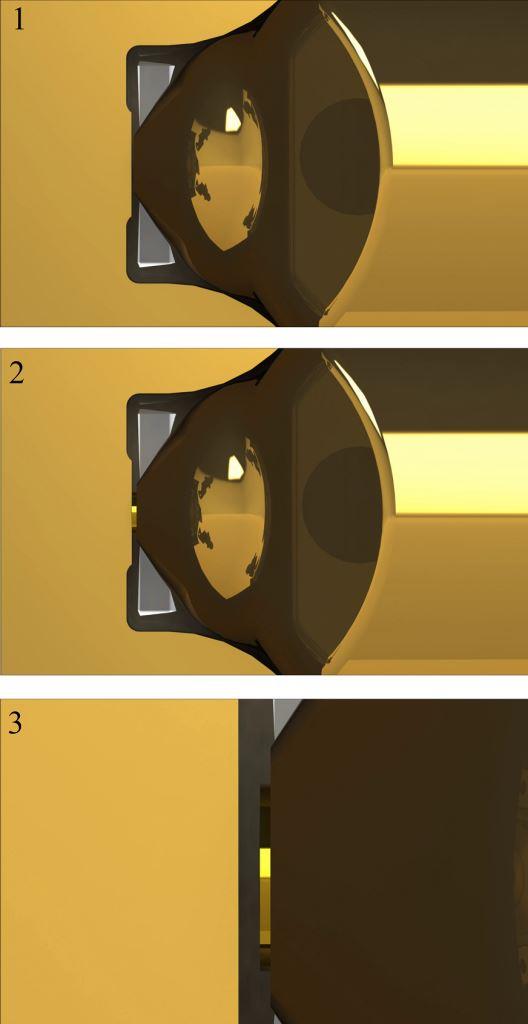}
\end{center}
\end{figure}

Summarizing, the release procedure consists of the following phases, Fig.~\ref{fig:procedure}:
\begin{enumerate}
	\item the TM is constrained by the CVM during LEOP\footnote{Launch and Early Orbit Operations}.
	\item the CVM hands the TM over to the GPRM.
	\item the GPRM plunger engages the TM pyramidal recesses.
	\item the plunger is slightly retracted while the RT touches the TM on the bottom of the recesses with a contact load below 0.4 N. The plunger looses contact with the TM by about 10 $\mu m$.
	\item the RT is quickly ($\approx 40 mm/s$) retracted leaving the TM freed of all the mechanical constraints.
\end{enumerate}

\subsection{Drag-Free and Attitude Control System: DFACS}
\label{DFACSintro}

The Drag Free and Attitude Control System (DFACS, \cite{Fichter2008}) is in charge of all the spacecraft actuation, that includes the two TMs indeed. It is therefore responsible of controlling TM attitude and position such that each TM is aligned and centered in the housing after the release. The DFACS is also responsible of all the control activities during the science phase.

A detailed description of the DFACS goes beyond the scope of this thesis. However, the weak electrostatic actuation is one of the reasons behind the challenges of the release and thus its basic principles along with the robust control concept are shown here. More on the DFACS and the actuation of the TM during the science phase is reported in \cite{Fichter2008,DFACSpaper}. Besides, useful insight in the drag-free control concept and in the flight-proven GOCE case is given in \cite{CanutoPap,CanutoConf}.

The mode in which the control system actuates the TM after the RT retraction is called \textit{Accelerometer Control Mode} and is aimed at:
\begin{itemize}
	\item providing 3 axis attitude control of the spacecraft.
	\item capture the 2 TM after the GPRM release.
	\item move the TM to their nominal position and attitude.
	\item compensate the disturbances.
\end{itemize}
As a consequence, in this phase, the spacecraft is not drag-free and the TM are forced to follow its trajectory.

\begin{center}
\begin{figure}
\caption{Electrodes surrounding the TM.}
\label{fig:EH}
\includegraphics[width=\columnwidth]{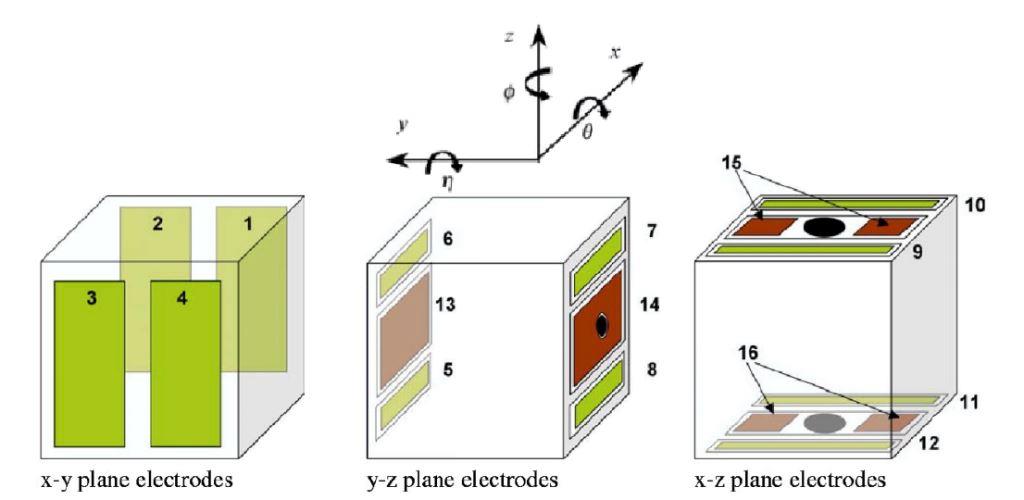}
\end{figure}
\end{center}

The actuation is provided by a set of 18 electrodes (Fig.~\ref{fig:EH}) at a roughly 4 mm distance from the TM. Once again this gap is a trade-off between the actuation authority and the disturbance rejection. 12 of these electrodes are dedicated to actuation. The other 6 provide a voltage bias and for sensing. The frequency of the bias voltage is much higher than the actuation. A detailed model of the electrostatic actuation is described in \cite{Nico, PaulThesis, TMactuation}.

In general the force between two bodies with a voltage difference is expressed as:
\begin{equation}
	F_{el} = \frac{1}{2}\frac{\partial C}{\partial q}\Delta V^2
	\label{Fel}
\end{equation}
where C is the capacitance and q is the axis (that is, if the force along x is desired, the partial derivative is computed with respect to x). The formula for the torques is equal. This same equation can be generalized to several bodies. The force along a certain axis q of a body is then:
\begin{equation}
	F_{el,N} = \sum_{j=1}^N \sum_{i=1}^N \frac{1}{2}\frac{\partial C_{i,j}}{\partial q}\Delta V_{i,j}^2
\end{equation}

Eq.~\ref{Fel} makes it possible to easily understand the weakness of the authority. Assuming a planar capacitor of area 5 $cm^2$, gap 3 $mm$ and excitation voltage about 50 $V$ the force between the 2 plates is $\approx 0.5\: \mu N$ because of the vacuum permittivity ($8.854\times10^{-12}$) at the numerator.

\subsubsection{The Sliding Mode Control}

The sliding mode controller is a non-linear control algorithm. Sliding mode approach guarantees robust control and, with respect to a classical PID it provides a reduced TM overshoot, a fast convergence and a better steady-state accuracy \cite{DFACS}. Besides, sliding mode control allows a more general design while PID has to be specified for a single case (the worst one). 

The following paragraphs provide a qualitative introduction to the topic, in order to put a frame around the mechanical release. For the details on this algorithm, the reader can refer to \cite{Slotine}.

Let's consider the system:
\begin{equation}
	\mathbf{x^{(n)}}=f(\mathbf{x})+b(\mathbf{x})u
\end{equation}
where $\mathbf{x}=[x,\dot{x},\dotsc,x^{n-1}]^T$ and b(x) is bounded $b_{min} < b(x) < b_{max}$.

Now, if $\mathbf{\widetilde{x}}$ is the tracking error, the function $s$ is then defined:
\begin{equation}
	s(\mathbf{\widetilde{x}},t)=\left( \lambda + \frac{d}{dt} \right)^{n-1}
\end{equation}
This becomes a sliding function once:
\begin{equation}
	s(\mathbf{\widetilde{x}},t)=0
	\label{slide}
\end{equation}
A Lyapunov function can be defined as:
\begin{equation}
	V = \frac{1}{2}s^2 >0 \:\: \forall s \neq 0
\end{equation}
whose derivative is:
\begin{equation}
	V = \frac{d}{dt} \left(\frac{1}{2} s^2 \right) \leq \:\: \forall s \neq 0
\end{equation}
which means that the system trajectories point towards $s = 0$.

\begin{center}
\begin{figure}
\caption{Sliding mode control applied to a free 2 kg mass.}
\label{fig:sliding}
\includegraphics[width=1.1\columnwidth]{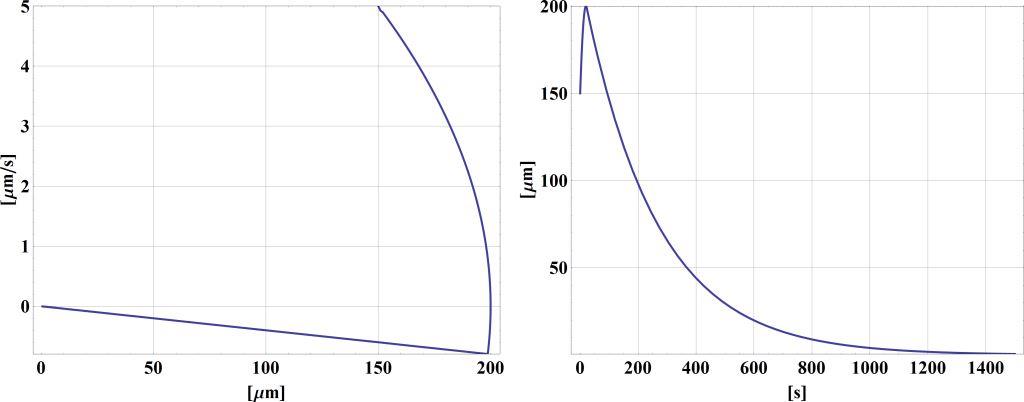}
\end{figure}
\end{center}

In the case of a free floating mass the sliding surface is:
\begin{equation}
	s = \lambda x_1 + x_2
\end{equation}
where $x_1$ is the position and $x_2$ the velocity.

The derivation of the control law in the DFACS is described in \cite{PaulThesis}. The basic idea is to constrain the TM trajectory on $s = 0$ with a discontinuous control such as:
\begin{equation}
	u_{disc}(s) = - k \text{sign}(s)
\end{equation}

Once on $s = 0$, the system evolution will be something like Fig.~\ref{fig:sliding}, where one can easily distinguish the reaching phase and the sliding one. Of course the actual system acts in a more complex way and in several degrees of freedom simultaneously. 

What is of interest, in the scope of this thesis, is that the TM motion will have an overshoot, that depends on the initial position and velocity, the actuation authority is non-linear and becomes weaker when the TM is far from the housing center and an unknown disturbance continuously acts on the TM in this phase. On top of the weak average authority of a single electrode at the mm range, there are several other aspects that limit the state the TM can have right after the mechanical release.

Finally, the LTP operation procedure after the release follows:
\begin{enumerate}
	\item t = 0+: RT retraction;
	\item 0 < t < 15 s: the Plungers move from the initial distance (equal to the Tip length, ~15 µm) to 500 µm;
	\item t = 15 s: start of observation and control;
	\item 15 s < t < 25 s: the Plungers remain at their position;
	\item t = 25 s: first estimation of the TM velocity with a less than 5 \% error;
	\item t > 25 s: if the velocities are assumed controllable the Plungers move in a safe position and the control keeps guiding the TM.
\end{enumerate}
Indeed, the observation of the TM velocity is not instantaneous. This adds extra challenges that, together with what has been described before, determine the requirement of Tab. \ref{tab:req}.

\begin{center}
\begin{figure}[ht!]
\caption{Timing of plunger retraction and DFACS initialization.}
\label{fig:fifteen}
\includegraphics[width=1\columnwidth]{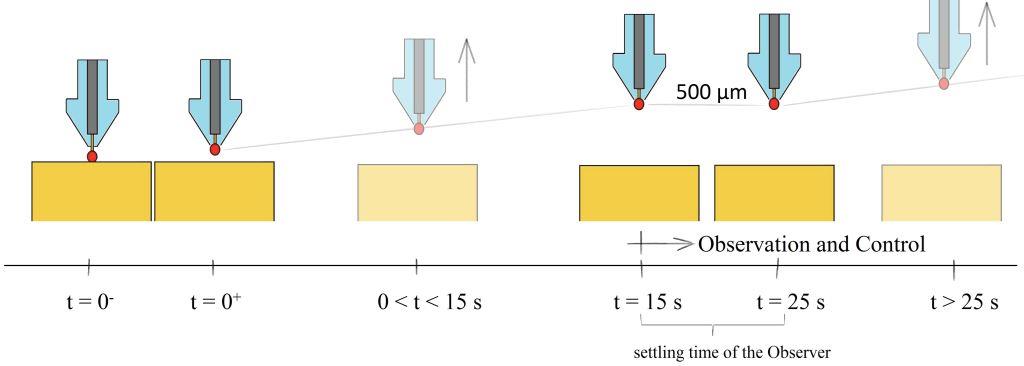}
\end{figure}
\end{center}



\chapter{Release Mechanism: model and identification}
\label{sec:RTmodel}

Piezo-electric devices are widely used for precise positioning both in on-ground and space applications \cite{TMpiezo,Bar_Cohen}. Several mathematical models are available in the literature to describe their dynamical behaviour. The focus here is on linear models \cite{IEEEpiezo}, in which relevant quantities like elastic, piezoelectric and dielectric coefficients are considered constant. \cite{Goldfarb} proposes a modelization for control purposes. Similarly to what is proposed in \cite{Adriaens}, the piezo stack electro-mechanical dynamics is here described by means of a lumped element model. Such a model is introduced in \cite{Antonello}, where the transfer function from commanded voltage to RT position is estimated by fitting the predicted dynamic behaviour of the mechanism to the measured one. In the following step, the physical constants defining the system dynamics (which are implicit in the transfer function coefficients) are found and the validated mathematical model is used to simulate the overall release phase by adding the TM dynamics and the effect of adhesion \cite{Bortoluzzi2013}. This last reference represents the first step of the simulation and prediction activity that sees here a substantial advance.

\begin{figure}[!ht]
\begin{center}
\includegraphics[width=6.5cm]{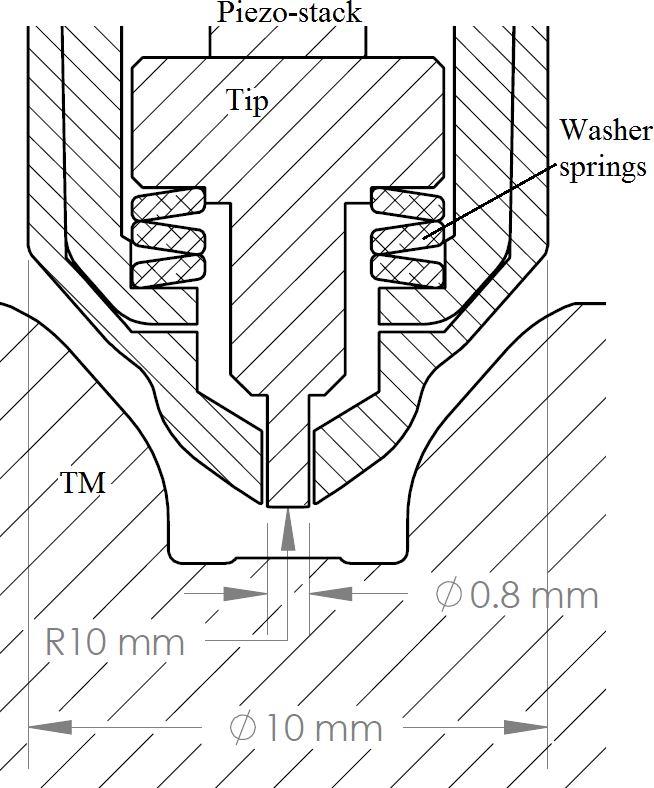}
\caption{Cut view of the GPRM plunger with the TM pyramidal recess.}
\label{fig:GPRMcut}
\end{center}
\end{figure}

\begin{figure}[!ht]
\begin{center}
\includegraphics[width=\columnwidth]{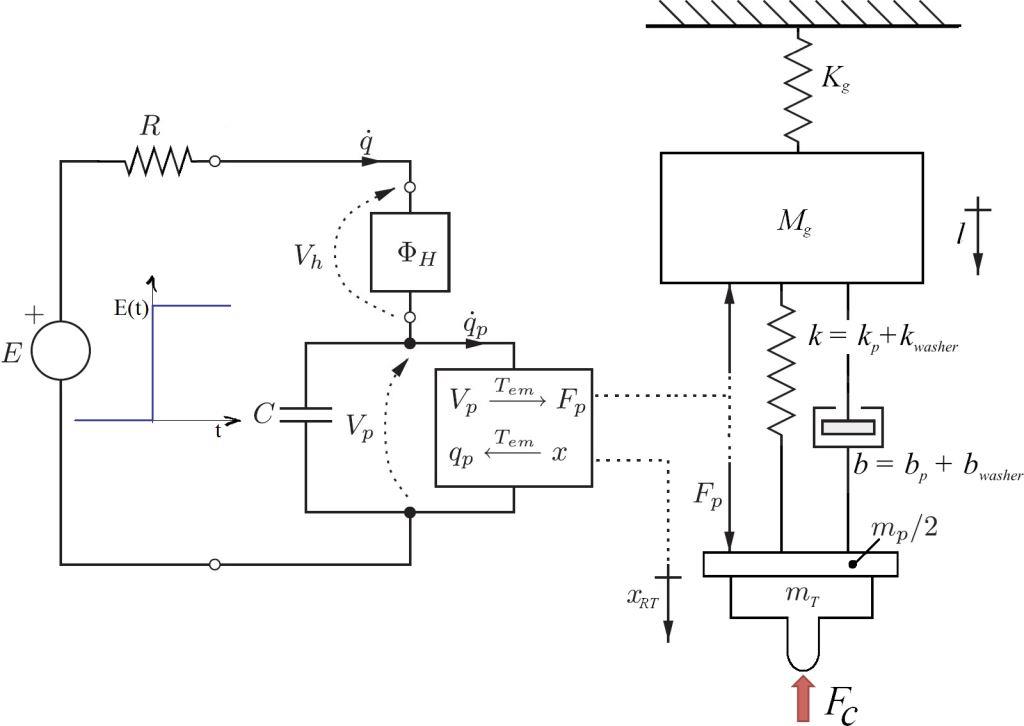}
\caption{Lumped-element diagram of the GPRM release mechanism.}
\label{fig:GPRMmodel}
\end{center}
\end{figure}

The GPRM release actuator is constituted by a $3 \times 3 \times 18$ mm piezo-stack commanding the small release tip, which has a spherical edge with 10 mm radius of curvature. The stack actuator is preloaded by a set of washer springs, as depicted in Fig.~\ref{fig:GPRMcut} and is commanded by a voltage applied to its ends. The voltage is increased during the pass-over phase from 0 V up to 120 V, at which the tip is fully extracted, and is short-circuited to 0 V through a resistor at the release, \cite{EICD}. Nominally, the voltage drops from 120 V to 0 V with an inverse-step function. In order to perform the release, the tip is extended and retracted, after few seconds, and no repetition of this action is foreseen. In the absence of a cyclic actuation, it is assumed sufficient to introduce a linear viscous element (labeled b in Fig.~\ref{fig:GPRMmodel}) to model the overall dissipation of the system.
The GPRM release mechanism electro-mechanical model is based on the lumped element scheme of Fig.~\ref{fig:GPRMmodel}.

$Fc$ is the contact force between RT and TM and can be either compressive, when the tip is initially holding the mass, or tractive, when the tip is retracted from the adhesive contact. A compressive $Fc$ force is set as initial condition of the solution and its behaviour turns tractive when, according to the dynamic response of the mechanism and of the TM, the bonds are elongated. As a consequence, back-action of adhesion is considered because in principle it may limit the quickness of the retraction of the RT and affect the impulse developed. The equations describing the piezo electro-mechanical dynamics are the following:
\begin{equation}
R\: C_a\: \dot{q}(t) + q(t)-T_{em}x_T(t)=C_a\:E(t)
\label{eqel}
\end{equation}
\begin{equation}
\begin{split}
M_g &\:\ddot{l}(t)+K_g\:\dot{l}_(t)-b\left(\dot{x}_T(t)-\dot{l}(t)\right) \\
& -\left(k+\frac{T_{em}^{2}}{C_a}\right) (x_T(t)-l(t)) + \frac{T_{em}}{C_a} q(t) = 0 
\end{split}
\label{eqMg}
\end{equation}
\begin{equation}
\begin{split}
m &\:\ddot{x}_{RT}(t)+b\left(\dot{x}_{RT}(t)-\dot{l}_(t)\right) \\
& +\left(k+\frac{T_{em}^{2}}{C_a}\right) (x_{RT}(t)-l(t)) - \frac{T_{em}}{C_a} q(t) + F_c= 0 
\end{split}
\label{eqpiez}
\end{equation}
where $q(t)$ is the charge accumulated on the piezo, $x_{RT}(t)$ is the release tip position, $R$ is the resistance of the electrical circuit, $C_{a}$ the capacitance of the piezo-stack, $T_{em}$ the electromechanical transducer (or the piezo effect), $m$ the mass of the release tip and half piezo-stack, $k$ the combined stiffness of the preloading washer spring and of the piezo , $b$ the combined damping, $F_c$ the contact force acting on the mechanism (that is 0 during identification) and $E(t)$ the voltage input. 

Differently from \cite{Antonello,Bortoluzzi2013}, the mass discretization here adopted considers the whole plunger ($M_g$) constrained to ground by a load cell (stiffness $K_g$) and the RT attached to the plunger by the preloaded piezo. Similarly to \cite{Bortoluzzi2013}, the physical parameters are estimated by means of dedicated tests performed on the mechanism, where the RT is retracted from the extended configuration in the absence of contact forces. The theoretical motion profile of the RT is fit to the measured motion profiles of the mechanism Flight Model (FM) \cite{RTtests}, and a covariance matrix is associated to each set of best fit parameters to describe their uncertainty. 

The uncertainty is expressed in terms of a covariance matrix. Such a matrix is associated to a best-fit set of parameters with the formula:
\begin{equation}
C_p=(J^T\Sigma^{-1}J)^{-1}
\end{equation}
where $J$ is a Jacobian matrix, defined as:
\begin{equation}
J_{i,j}=\frac{\partial x_i}{\partial p_j} 
\end{equation}
with $x_i$ the retraction at the $i^{th}$ time sample and $p_j$ the $j^{th}$ parameter of the model. The matrix $\Sigma$ is a diagonal matrix, built with the square of the residuals for each sample.

\begin{figure}[!ht]
\includegraphics[width=\columnwidth]{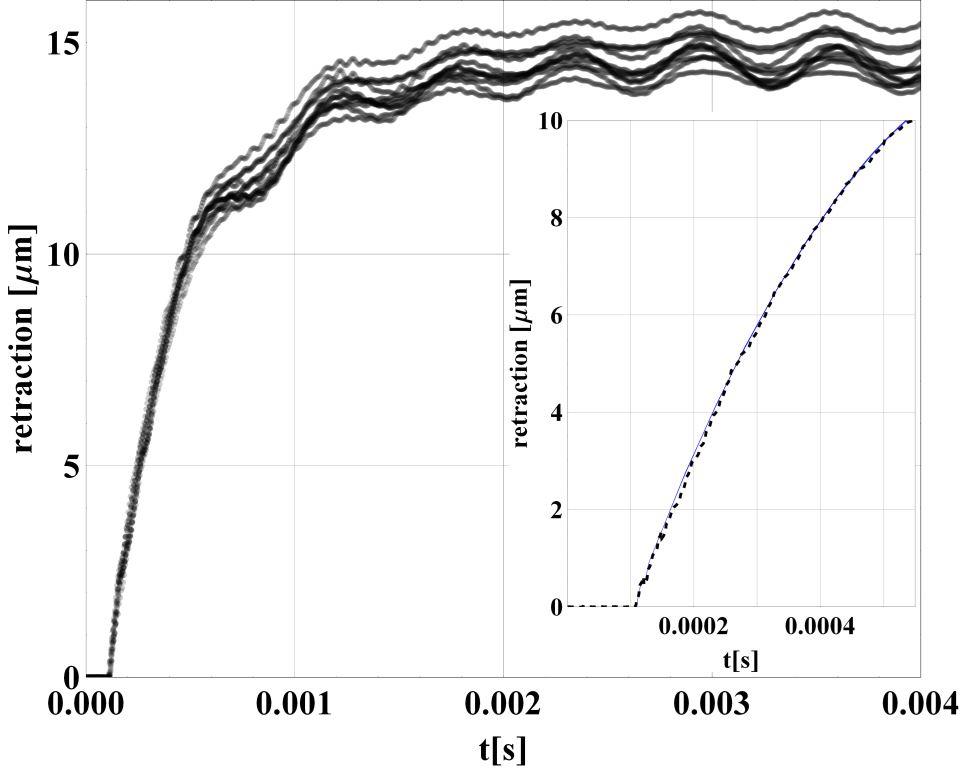}
\caption{Mesured retractions of the FM RT. The sub-frame shows a comparison with a best-fit.}
\label{fig:retractions}
\end{figure}

The measurement on 8 different mechanisms are available. They correspond to the 4 GPRM FM equipped with 2 piezo-stacks each (main and redundant). Since the in-flight release will be performed by a pair of such mechanisms, the parameter space can be restricted to the estimated physical quantities, still considering their uncertainty. Fig.~\ref{fig:retractions} shows the measured retractions and in the sub-frame an example of best fit, whose parameters are reported in Tab.~\ref{tab2}. The data are measured by a SIOS SPS 2000 vibrometer at 1 MHz sampling frequency.

If two mechanisms must be commanded synchronously, the need arises to estimate additional asymmetry of the retraction profiles produced by different voltage commands at the output channels of the control unit.

\begin{table}[!ht]
\small
\renewcommand{\arraystretch}{1}
\caption{Example of estimated model parameters}
\label{tab2}
\centering
\begin{tabular}{c|c|c|}
\hline
\textbf{Name} & \textbf{Best Fit} & \textbf{Units} \\
\hline
$R$ & 371.2  & $\Omega$ \\
\hline
$m$ & \num{9e-4}  & $kg$\\
\hline
$C_a$ & \num{2.3e-7} & $F$ \\
\hline
$k$ & \num{4.91e7} &  $N/m$\\
\hline
$K_g$ & \num{2.15e6} & $N/m$ \\
\hline
$b$ & 9.77   & $N\:s/m$\\
\hline
$M_g$ & $0.019$  & $kg$ \\
\hline
$T_{em}$ & 5.73  & $C/m$\\
\hline
\end{tabular}
\end{table}

The actual voltage time profile $E(t)$ commanded by the two channels of the FM control unit has been measured and differs from the nominal inverse-step assumed in \cite{Antonello,Bortoluzzi2013,TN3092}. A drop time on the order of microseconds is required to null the voltage and a limited repeatability characterizes the voltage time profile. For this reason $E(t)$ is expressed as:

 \begin{equation}
   E(t) = \left\{
     \begin{array}{lr}
       120 - em\:t  & : t < 120/em\\
       0 & : t \geq 120/em
     \end{array}
   \right.
\end{equation} 

where $em$ is the slope of the dropping voltage. The set of measured profiles is limited, therefore the uncertainty on $em$, whose distribution is assumed normal, is large. The mean value is estimated in $5 \times 10^{7}$ V/s with a standard deviation equal to $1.02 \times 10^{7}$ V/s.

\begin{figure}[!ht]
\includegraphics[width=\columnwidth]{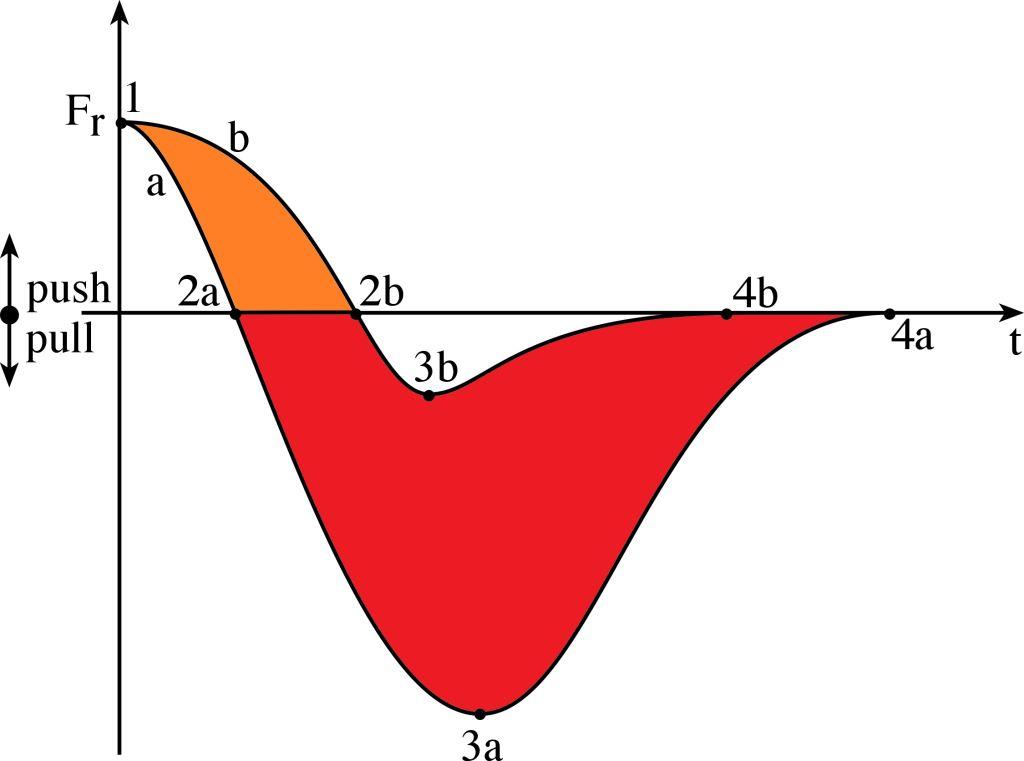}
\caption{Contact forces between TM and release tip during the release phase. The tip \textit{a} is assumed to be quicker and its adhesion with the TM stronger. The sum of the 2 areas is the momentum transferred to the TM.}
\label{fig3}
\end{figure}

The mathematical model of the release mechanism makes it possible to predict asymmetrical retractions of the RTs, in the limit of the uncertainty of the relevant parameters estimated.
The asymmetry of retraction of the two tips is critical because it can affect both the momentum transferred by pushing contact forces between RT and TM (i.e. without adhesion, light gray area in Fig.~\ref{fig3}) and that transferred by adhesion at the two contacts (dark gray area in Fig.~\ref{fig3}). This means that:
\begin{itemize}
	\item even with no adhesion, the presence of an initial contact force combined with the asymmetry of retraction can transfer momentum;
	\item even with equal adhesion strength and elongation at the two contacts, asymmetry of retraction can convert symmetric adhesion into momentum.
\end{itemize}
The latter point is enhanced by the low repeatability of adhesion, which adds further dispersion in the dynamic behaviour of the two opposed mechanisms, as will be clarified in the following.

\chapter{Adhesion Between Metals: A Quick Overview}
\label{sec:Adhesion}

Adhesion or cold-welding is a phenomenon that takes places between two metallic objects in contact. As a consequence the two objects remain attached even if a certain external force pulls them apart. 

Adhesion is a recognized issue in space applications \cite{ESASTM, ECSS, RobertsSpaceTribology}. It can be high enough to hinder the deployment of even large systems. A famous example is the high gain antenna in Galileo, that was only partially opened during operation around Jupiter. This antenna was shaped as an umbrella. Despite several attempts and different strategies, a few of the antenna ribs did not pop out of the holding cup during the deployment \cite{RobertsSpaceTribology, Galileo}. Galileo had to rely its success on a secondary antenna, with a much lower gain, making the result of the whole mission much harder than planned. This example highlights once again the criticality of adhesion and the need of proper studies and tests for each case.

In most applications, adhesion can be limited by a wise material or coating choice, by proper cleaning and lubrication and by minimizing the contact areas. But indeed, all this features have to be a trade off between several opposed needs.

Due to its poor repeatability, adhesion between the rough metallic surfaces of the RT and the TM is expected to be a relevant source of net force on the TM during the release. Unfortunately, in such a case, material, coating and lubrication design choices are constrained by the need of minimizing the non-gravitational force acting on the TM. Only the area of contact has been kept limited in light of adhesion and results in the design of a set of actuators, with gradual reduced area, operated in sequence (CVM, GPRM, RT).

In the following, a basic overview of the standard theories on adhesion is given . Although the real behaviour is different because of the use of real engineering surfaces, these theories can provide an initial understanding of the phenomenon, its main parameters and the main references. After this short overview, a detailed description of the experimental facility is provided. Such a facility was designed specifically for part of the qualification of the LISA-Pathfinder mechanism. The main results are also summarized. 

\section{Basic Theories: JKR and DMT}

The contact description according to Hertz theory \cite{Hertz} does not provide any explanation of the adhesion phenomenon. It is called unilateral problem, because only compressive stresses exist inside the contact patch. Hertz theory is the starting point for deriving the main adhesion models. The main results are here summarized.

In the case of a sphere-plane contact, the force-to-penetration law is:
\begin{equation}
	F = \frac{4}{3}E^*R^{1/2}\delta^{3/2}
\end{equation}
where R is the radius of the sphere, $\delta$:
\begin{equation}
	\delta = \frac{a^2}{R}
\end{equation}
the penetration, with $a$ radius of the contact patch, and $E^*$:
\begin{equation}
	\frac{1}{E^*} = \frac{1-\nu_1^2}{E_1} + \frac{1-\nu_2^2}{E_2}
\end{equation}
with $E_1$, $E_2$, $\nu_1$ and $\nu_2$ the elastic moduli and Poisson ratios of the materials of the sphere and of the plane.

For a contact between a flat punch and an elastic plane the force-to-penetration equation is:
\begin{equation}
	F = 2\:a\:E^{*}d
\end{equation}

\subsection{Johnson-Kendall-Roberts (JKR)}

The JKR model \cite{JKR1971,Maugis} adapts Hertz theory and describes adhesion as a balance between the elastic energy stored into the contact and the surface energy loss due to the surface union. Somehow JKR is similar to Griffith's theory of crack propagation. Of course tensile stresses due to adhesion can exist both inside and outside the Hertz contact area. JKR model considers only the effect of such stresses inside the contact.
  
First of all, let's consider an adhesionless contact given by a force $F_1$. The contact radius will be, according to Hertz:
\begin{equation}
	a = \left( \frac{3}{4}\frac{F_1R}{E^*} \right)^{1/3}
	\label{eqa}
\end{equation}
whose penetration is again:
\begin{equation}
	\delta = \frac{a^2}{R}
\end{equation}
and, the elastic energy associated is:
\begin{equation}
	U_{E,1} = \int F_1 d\delta = \frac{2}{5}\frac{F_1a^2}{R} = \frac{8}{15}\frac{a^5 E^*}{R^2}
\end{equation}
Starting from this load it is possible to imagine a reduction of the force applied until a value equal to $F$, while the contact area is kept constant. The penetration depth decreases by:
\begin{equation}
	\Delta \delta = \frac{F_1-F}{2\:a\:E^*}
\end{equation}
and the elastic energy by:
\begin{equation}
	\Delta U_E = \frac{F_1^2-F^2}{2\:a\:E^*}
\end{equation}
The total elastic energy becomes then:
\begin{equation}
	\Delta U_E = \frac{F^2}{4\:a\:E^*}+\frac{4}{45}\frac{a^5E^*}{R^2} = a E^* \left(\delta - \frac{a^2}{3R} \right)^2+\frac{4}{45}\frac{a^5E^*}{R^2}
\end{equation}

This last expression can be derived according to the definition of surface energy:
\begin{equation}
G = \left(\frac{\partial U_E}{\partial A}\right)_{\delta} = \frac{E^*}{2\pi\:a}\left(\delta - \frac{a^2}{R} \right)^2 = \frac{(F_1 - F)^2}{8\pi\:a^3E^*} = \gamma
\label{energyG}
\end{equation}
which, solved with respect to $F_1$ gives:
\begin{equation}
F_1 = F + 3\pi\gamma R + \sqrt{6\pi\gamma R F + (3\pi\gamma R)^2}
\label{eqF1}
\end{equation}
The relation between the radius of the contact area, a, and the force, F, comes from Eq.~\ref{eqa} and Eq.~\ref{eqF1}:
\begin{equation}
a^3 = \frac{3R}{4E^*}\left(F + 3\pi\gamma R + \sqrt{6\pi\gamma R F + (3\pi\gamma R)^2}\right)
\end{equation}
with the penetration given by:
\begin{equation}
\delta = \frac{a^2}{R}-\sqrt{\frac{2\pi a \gamma}{E^*}}
\end{equation}
applying Eq.~\ref{energyG}.

\begin{figure}[!ht]
\includegraphics[width=0.9\columnwidth]{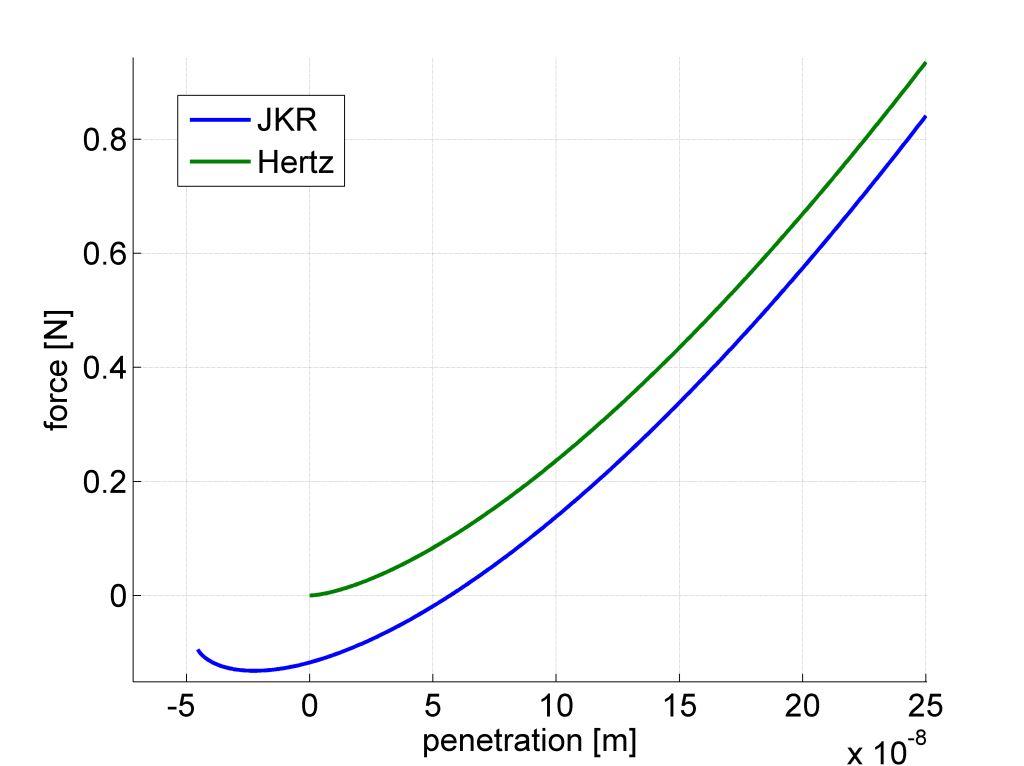}
\caption{Force-to-penetration law in the JKR model.}
\label{fig:JKR}
\end{figure}

Fig.~\ref{fig:JKR} shows the force-to-penetration law for a sphere on a plane according to the JKR theory ($R = 0.01 m$, $\gamma = 2.8 J/m^2$, $E^*=59 GPa$). It can be noted the 0.135 N pull force peak.

In certain conditions, the JKR model can be adapted for describing adhesion after a full plastic indentation, \cite{Maugis2}. After an indentation in the plastic regime, the elastic recovery produces an elastic contact of a sphere inside a larger sphere (i.e. the deformed plane). The adhesive force is then proportional to the root square of the maximum load applied, that becomes a key parameter. The theoretical study of adhesion under plastic conditions goes beyond the LISA-PF system-oriented scope of this thesis, however it is worth underlying the influence of the maximum load on the pull-off force once the plastic regime is entered.

\subsection{Derjaguin-Muller-Toporov (DMT)}

In 1975 Derjaguin-Muller-Toporov \cite{DMT,Maugis} formulated a different theory in which adhesion does not modify the contact profile inside the Hertzian area, but only acts on a ring outside the sphere.

The adhesive force is:
\begin{equation}
	F_c = -2 \pi \gamma R
\end{equation}
that is added to the external force F.
The resulting model is then summarized by the following:
\begin{equation}
	\frac{a^3}{R} = \left(F + 2 \pi \gamma R \right) \left(\frac{3}{4 E^*} \right)
\end{equation}
and
\begin{equation}
	\delta = \frac{a^2}{R}
\end{equation}
where the overload due to adhesion is independent from the penetration.

\begin{figure}[!ht]
\includegraphics[width=\columnwidth]{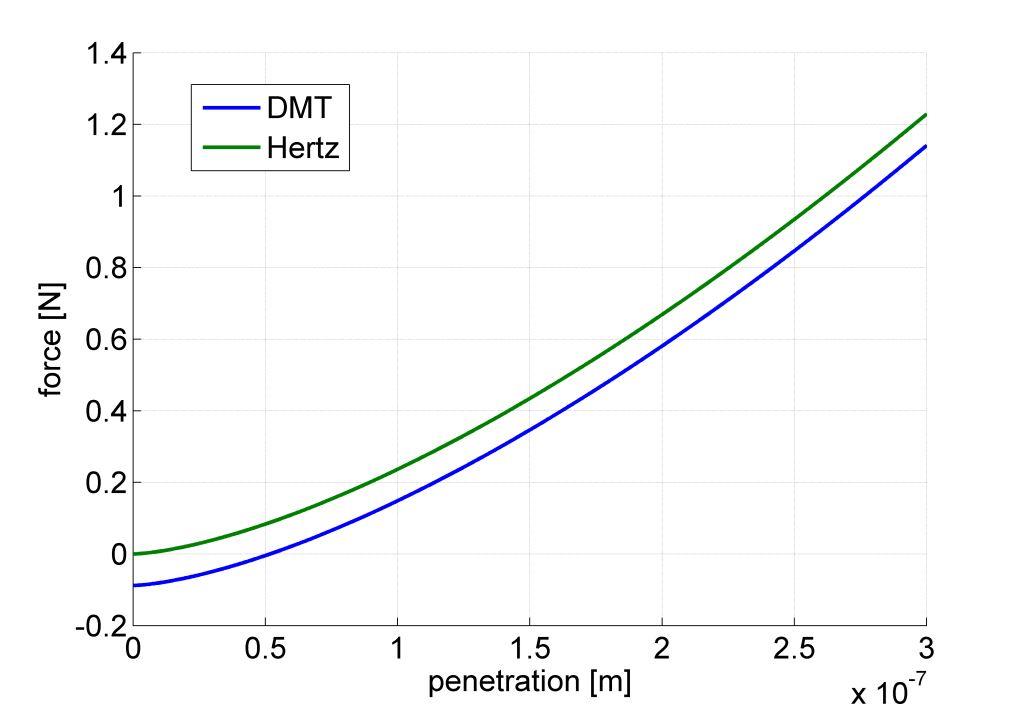}
\caption{Force-to-penetration law in the DMT model.}
\label{fig:DMT}
\end{figure}

Fig.~\ref{fig:DMT} shows the force-to-penetration law for a sphere on a plane according to the DMT theory ($R = 0.01 m$, $\gamma = 2.8 J/m^2$, $E^*=59 GPa$).

\subsection{JKR vs. DMT}

In 1976 Tabor highlighted how JKR model neglects tensile forces outside the contact area, while DMT neglects the deformations. The two theories are both good descriptions, but for different extreme cases.

The discriminant quantity is the Tabor parameter:
\begin{equation}
	\mu = \frac{\gamma^2 R}{E^2 Z_0^3}
\end{equation}
where $Z_0$ is the the equilibrium distance between atoms.

The DMT theory applies for $\mu << 1$ (high Young's modulus, small sphere radius and low surface energy), and that of JKR for $\mu >> 1$ (low Young moduli, large radius, high surface energy). The nominal surfaces and materials planned for LISA-Pathfinder should determine a contact in the JKR field.

\section{The Large Uncertainty on Adhesion}

\begin{figure}
\includegraphics[width=\columnwidth]{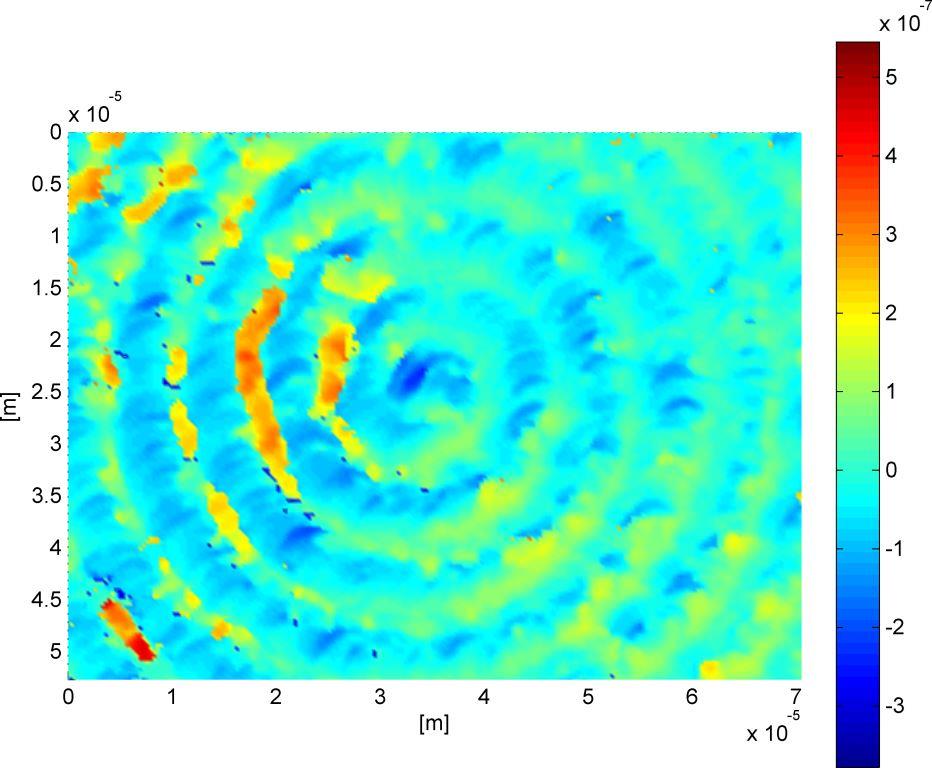}
\caption{View of a TM-like surface as measured by a profilometer.}
\label{fig:profile}
\end{figure}

Already from these models it is possible to demonstrate that the knowledge on the adhesion between the RT and the TM has a high uncertainty. At least two effects may determine a real behaviour significantly different from the nominal one:
\begin{enumerate}
	\item In reality, the TM and RT surfaces are rough, Fig.~\ref{fig:profile}. Roughness makes modeling of adhesion much harder. In principle each asperity can be considered as a sphere. However, each virtual sphere would have a different radius and height position. Therefore some spheres will be compressed while others will be pulled.
	\item Locally the surfaces may not deform in an elastic way. Rough and deformable surfaces are brought together until they coalesce and, as a consequence, the asperities may produce local plasticity. The effect of plasticity can be estimated by considering that in a sphere-plane contact the plane will change its shape into a spherical recess whose radius is larger than that one of the sphere. In such a case one can obtain a pull-off force as high as 1.5 N\footnote{assuming $R_{plane} = -0.011 m$ and the JKR model}, against a nominal one of 0.135 N. Similarly, assuming the plastic contact produces a complete welding of two objects, the Ultimate Tensile Strength of gold applied on the sphere-plane contact area would determine a peak force of about 0.4 N.
\end{enumerate}

These two points suggest that a theoretical approach alone would be affected by an excessive uncertainty, especially if a worst-case estimation is needed. The main source of data on the adhesion between RT and TM is then collected by means of a set of experimental campaigns.

\chapter[Adhesion: Facility and Results]{Adhesion: Experimental Facility for Dynamic Testing and Results}

This chapter overviews the experimental facility, the procedures and the results of the on-ground qualification activity of the GPRM release.

\section{The Transferred Momentum Measurement Facility}

The Transferred Momentum Measurement Facility (TMMF) is designed to study the adhesion phenomenon between metallic surfaces under dynamic separation. Indeed, great emphasis is given to the LISA-Pathfinder Grabbing Positioning and Release Mechanism (GPRM) case. The TMMF is one of the experiments performed at the University of Trento for on-ground preparation of LISA-Pathfinder, \cite{Ciani}. This experiment has been run and upgraded in parallel to the design and development of the release mechanism.

The concept of the experiment is the minimization of the effect of gravity by means of a pendulum suspension of an object representative of the TM. The pendulum is enclosed in a high vacuum chamber\footnote{$10^{-7}$ mbar} in order to reproduce a space-like environment. The suspended object is also called test mass, but will be here defined TM mock-up (TMmu) for the sake of clarity. In the same chamber, a tip (RTmu) is attached to a linear actuator. The geometry and the surface finishing of this piece are equal to the flight hardware RT. The RTmu is slowly actuated to engage the suspended TMmu, load and unload the contact. It is then quickly retracted, in order to pull the TMmu with the adhesive bond. The measurement of the momentum transferred to the TMmu by this retraction is provided by measuring the TMmu law of motion.

It is not feasible to include the 1.96 kg Au/Pt cube in this facility, three different masses have thus been used (0.0096 kg, 0.089 kg and 0.844 kg). This allows, if needed, to extrapolate the results to the actual mass value.

\begin{figure}
\includegraphics[width=\columnwidth]{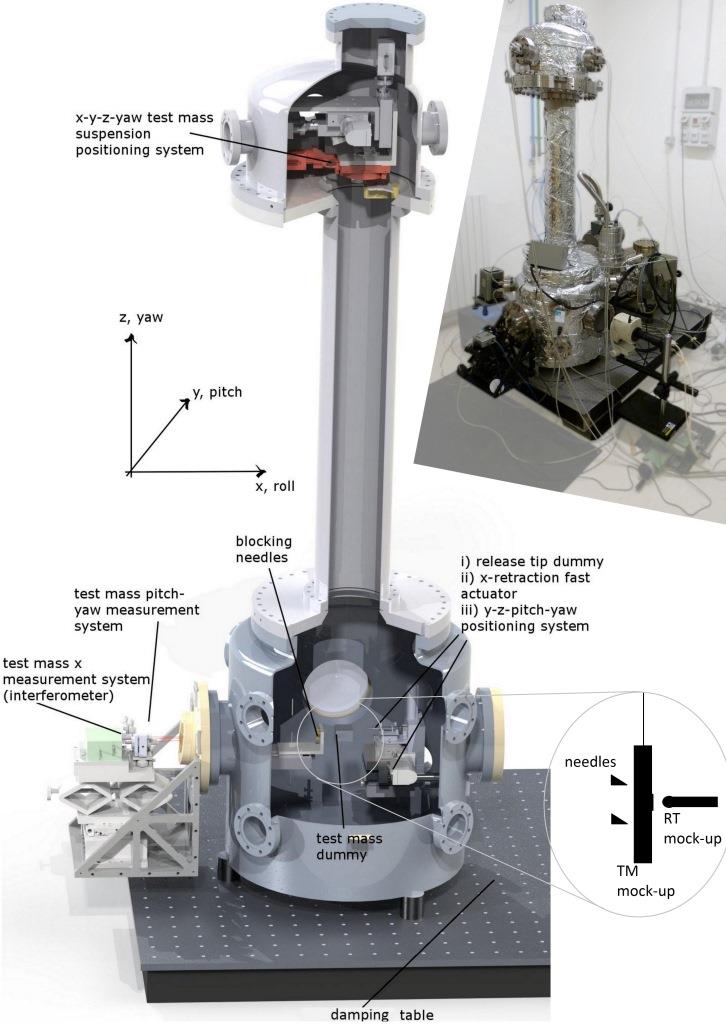}
\caption{Cut view of the TMMF, with a picture of the actual system in the top right corner.}
\label{fig:TMMF}
\end{figure}

\begin{figure}
\includegraphics[width=1.1\columnwidth]{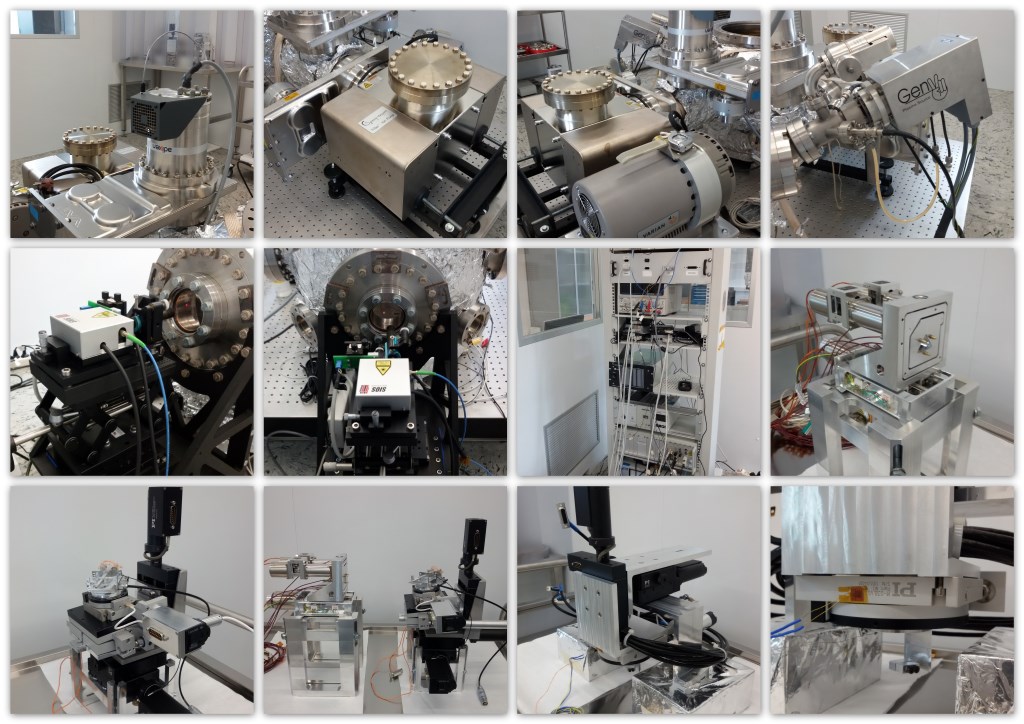}
\caption{From the top left: turbo-molecular pump, ionic pump, ionic and scroll pumps, plasma, 2 views of the interferometer, electronics rack, support system (needles), RTmu and the actuators that allow to change its position/attitude, support system together with the RTmu, actuators of the pivot of the pendulum with a detailed view.}
\label{fig:TMMFcollage}
\end{figure}

\begin{figure}
\includegraphics[width=\columnwidth]{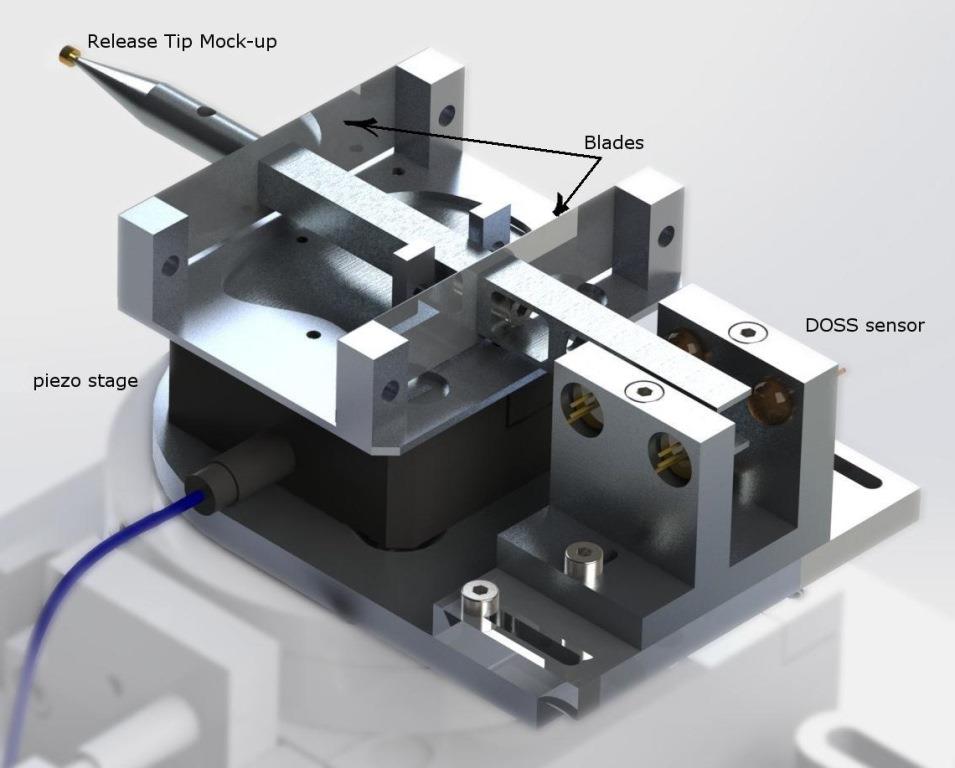}
\caption{Rendered view of the RT mock-up and the ultrasonic piezo actuator that moves it.}
\label{fig:RTmu}
\end{figure}

\begin{figure}
\includegraphics[width=\columnwidth]{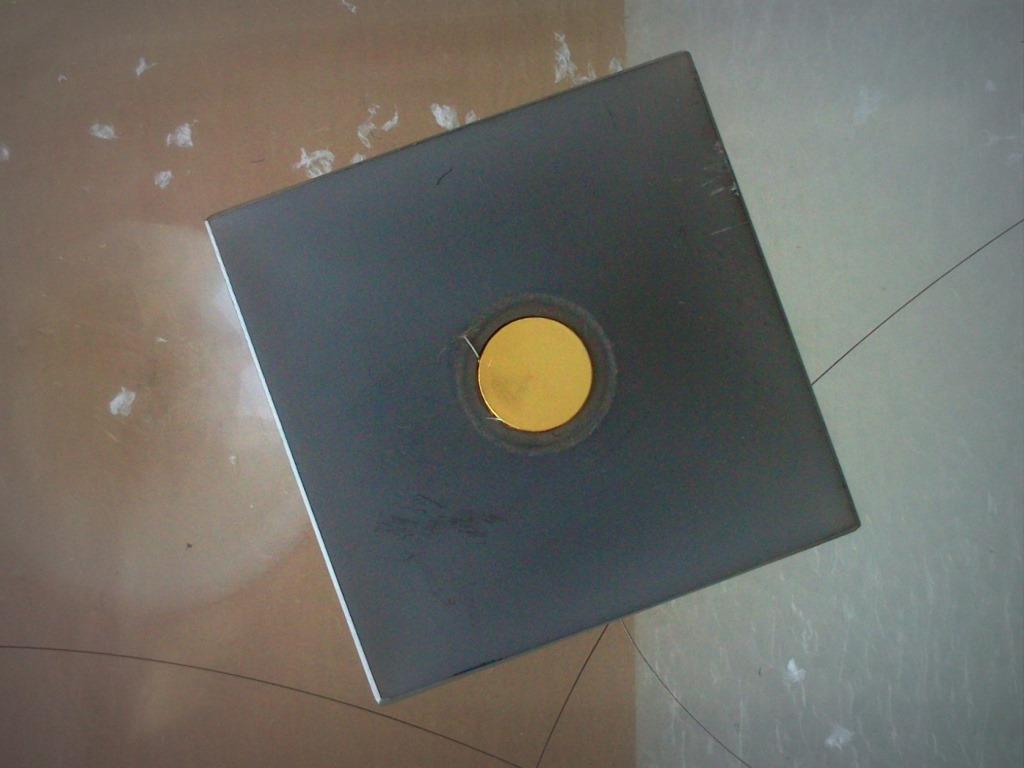}
\caption{View of a TM mock-up (0.0096 kg). The central part is the Au/Pt gold coated insert.}
\label{fig:TMmu}
\end{figure}

\begin{figure}
\includegraphics[width=\columnwidth]{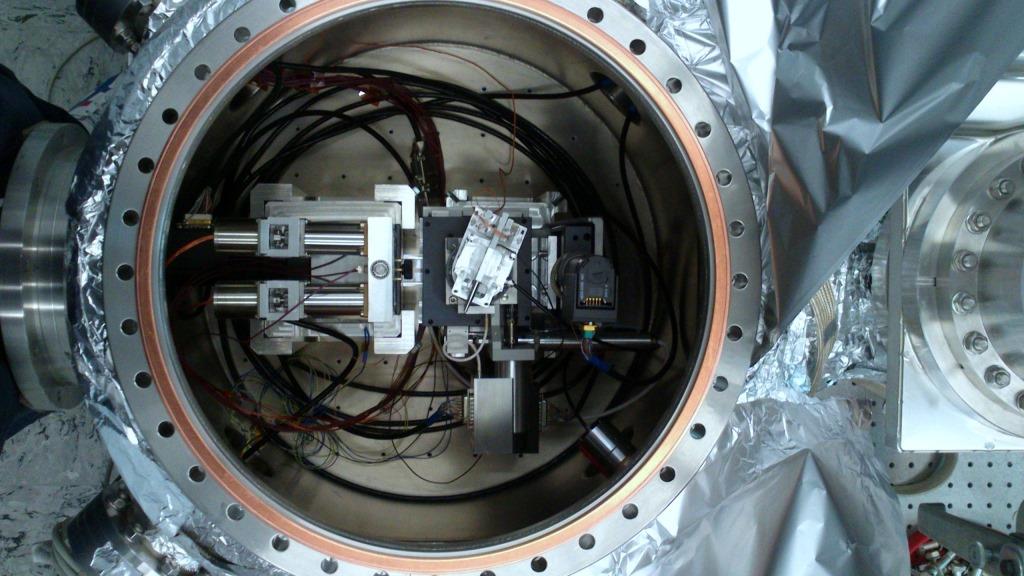}
\caption{View of the inside of the TMMF.}
\label{fig:TMMFinside}
\end{figure}

\begin{figure}
\includegraphics[width=\columnwidth]{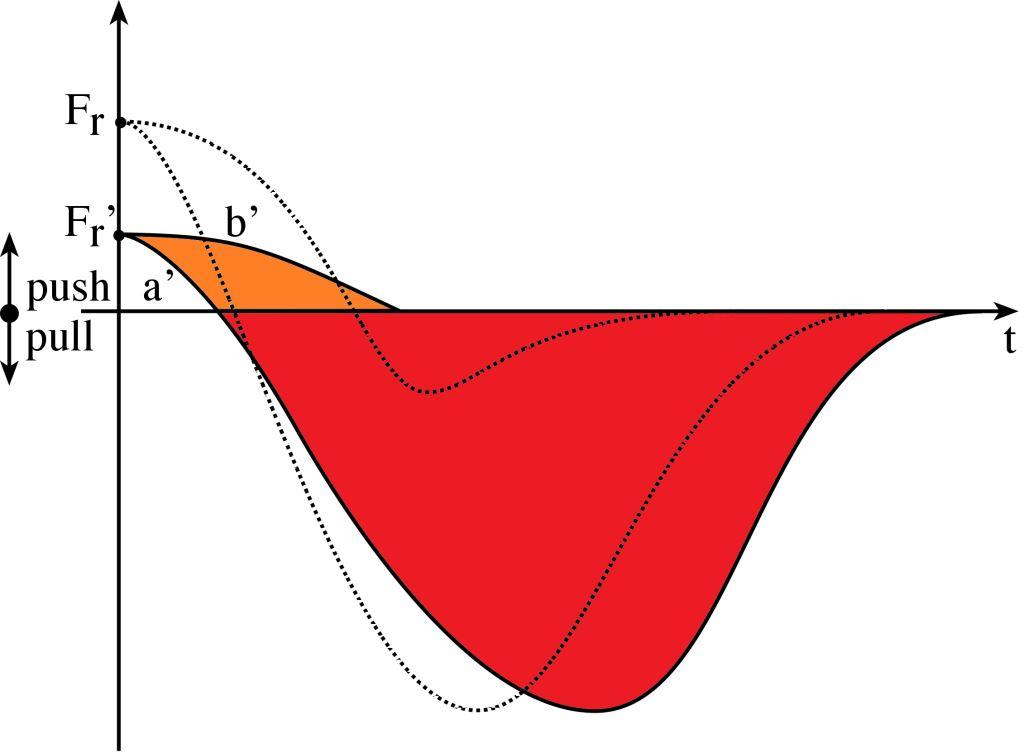}
\caption{Contact forces between TMmu, release tip (\textit{a'}) and blocking needles (\textit{b'}) in the TMMF. Dotted lines represent force profiles in the in-flight release (Fig.~\ref{fig3})}
\label{fig:forceprof}
\end{figure}

In order to measure forces in the order of few tens of mN applied for a timeframe of a ms and minimize effects that could hinder the arise of adhesion, the TMMF has to fulfill several requirements. It has to reproduce gravity-free conditions, at least in a plane, a space-like environment and a clean surface. Besides, vibrations isolation and the capability to accommodate slow ground tilting are required as well. At the same time, the TMMF has to be able to hold in a stable way the TM during the application of loads and measure both the TMmu position and attitude and the RTmu retraction. Details of the TMMF design are specified in reference \cite{RSI}. A schematic description of its main subsystems is here provided (Fig.~\ref{fig:TMMF}):
\begin{itemize}
	\item the TMmu is a Ti frame suspended by a 1.15 m long tungsten wire, hosting a cylindrical insert at the center of one of the faces. The bulk material of this insert is Au-Pt; its face pointing the tip is micro-milled and then gold coated. The rear side of the insert is mirror finished to reflect the incoming laser beams of the optical readout system;
	\item the RTmu (Fig.~\ref{fig:RTmu}) is a 2 mm disk lens-shaped with a 10 mm radius of curvature. It is actuated by an ultrasonic linear piezo positioner through a set of calibrated blade-springs, in order to control the normal force applied to the contact patch;
	\item a turbo molecular pump is used to evacuate the chamber to the pressure of $10^{-7}$ mbar, while an ion pump is used to maintain such a vacuum level avoiding the vibrations produced by the turbo pump during the measurement phase; 
	\item a plasma source able to perform ion etching (Ar+ and O- ions) guarantees the cleanliness of the mock-up surfaces;
	\item an active 6-dof anti-vibration table minimizes the micro-seismic vibration transmitted from the ground to the chamber;
	\item a laser interferometer is pointed to the TM insert rear surface to measure its position through an optical window;
	\item an optical lever is pointed to the TM insert rear surface to measure its attitude in terms of yaw and pitch angles;
	\item a Differential Optical Shadow Sensor (DOSS, Figure 2) measures the tip position \cite{DOThesis};
	\item several actuators are used to position the TM suspension point and to adjust the tip position and pitch/yaw attitude;
	\item 3 needle-edged screws mounted on an L shaped frame are used to hold the TM while it is engaged by the tip. The edges of these screws and the rear part of the TM are covered with an anti-adhesive coating (ZrN-based and CrN-based respectively). Both the coatings are conductive, minimizing effects due to triboelectrification phenomena. The set of needles can be translated and tilted (pitch and yaw) in order to comply with the TM attitude at the equilibrium;
	\item a high speed camera monitors the quick retraction of the tip and provides a visual feedback to the operator in charge of commanding the actuators;
	\item most of the vacuum chamber is covered with aluminum foils to minimize heat dissipation during baking. 
\end{itemize}

The operational procedure of the test is declined in two main ways as detailed in the next sections. However, its main focus is the representativeness of the in-flight case. Therefore there is a common background, summarized by the following points:
\begin{enumerate}
	\item the RTmu is put in contact with the TMmu. The TMmu is supported by the needles.
	\item the RTmu-TMmu contact patch is loaded up to a few hundreds of mN (300-400 mN).
	\item the RTmu is retracted from the contact first reducing the mutual push to 0 mN, then applying a pull by means of adhesion. The quickness and timing of these actions is the main operational difference among the test campaigns.
\end{enumerate}

Several experimental configurations have been tested in order to find a good compromise between experiment sensitivity, usability and representativeness of the in-flight conditions as explained in \cite{RSI} and \cite{PL3023}. Two aspects mark the difference between the current setup and the in-flight release conditions.
\begin{enumerate}
	\item The tip mock-up is actuated by an ultra sonic piezo drive, which can produce a far larger stroke (18 mm) than the flight piezo-stack, but with a reduced velocity in the first 10 $\mu m$ ($\approx$4 mm/s instead of $\approx$40 mm/s).
	\item The TM mock-up is released from a caged configuration between the tip and a set of blocking needles instead of the two opposed RTs. The needles apply a residual contact force of about 10 mN before the retraction and are kept still (1-sided release), while the residual contact force applied in flight by the RTs is substantially equal to the maximum and both RT are actuated.
\end{enumerate}
As a consequence of point 1, adhesion force accelerates the TM mock-up for a larger time than in-flight, producing an over-estimation of its time integral yielding the momentum transfer. Conversely, adhesion strength could be reduced by a slower elongation rate of the adhesion patch. However, tests performed in the past with our facility show that the transferred momentum can be described by a simple mathematical model in which the adhesion force is assumed to be conservative, i.e. independent of the quickness of  retraction of the tip \cite{EXME1}. This means that the force-to-elongation function can be assumed rate independent and can be transferred from the on-ground system to the in-flight environment. Consequence of point 2 is a distortion of the total force time profile applied to the TM in the testing configuration. As long as the maximum contact force is representative, adhesion strength is preserved and the extrapolation to the in-flight case can be performed as follows.

Two aspects of the experiment need further discussion. First, the introduction of the DOSS for the measurement of the displacement signal of the tip constitutes a valuable improvement of the facility with respect to what presented in \cite{MSSP13,RSI}, because it makes it possible to simplify the mathematical model of the system dynamics adopted to analyze the tests results. Second, a limiting effect on the experiment representativeness is the presence of the needles, that determines a non symmetrical action on the TMmu during release. As the needles do not move during RTmu retraction, the elastic energy stored in them converts into impulse on the TMmu, according to the following:

\begin{equation}
	I_{needles} = F_{pre}\sqrt{\frac{m}{k_{nd}}}
	\label{eq:catapult}
\end{equation}
		
where $F_{pre}$ is the load at the contact patch, $m$ is the TMmu mass and $k_{nd}$ is the stiffness of the needles, assumed linear. A large value of $m$ enhances the contribution of the needles push. It is worth noting that this equation is a good approximation as long as the RTmu exits the contact indentation in a time-frame much shorter than the needles-TMmu period of oscillation. This effect is sometime called \textit{catapult effect}.

For each release tip retraction experiment a few quantities were measured. The derivation of these quantities from data does not require any fitting procedure and is thus model independent. The quantities are:
\begin{enumerate}
	\item The total transferred momentum.
	\item The peak acceleration of the TM. From this value the peak, apparent pull-off force $F_{peak}$,  is derived by multiplication with the TM mass.
	\item The event time length, $\Delta t$. Time length is defined as the length of the time interval wherein the TM acceleration is above 5\% of the peak value.
\end{enumerate}

\begin{figure}
\includegraphics[width=1.1\columnwidth]{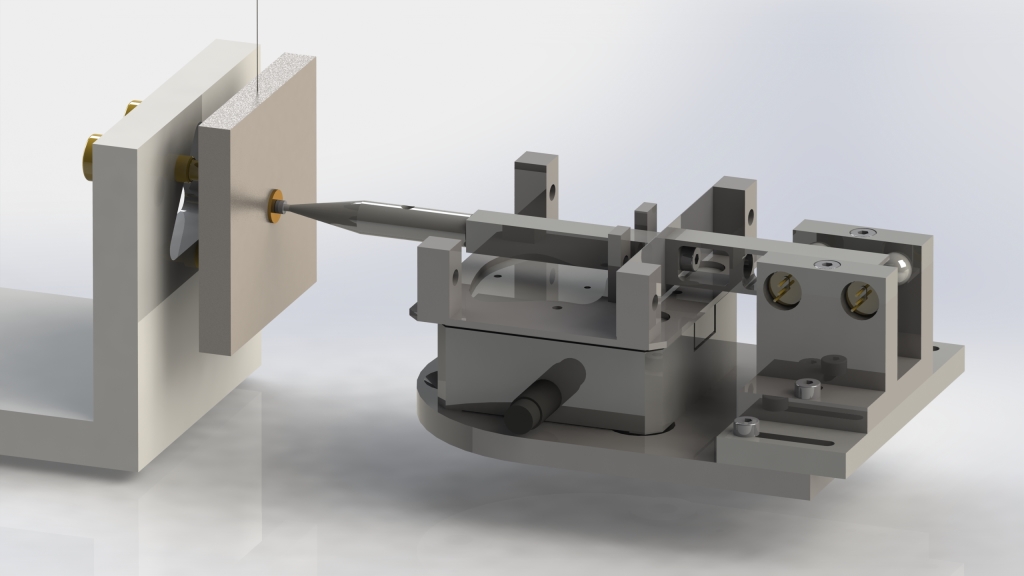}
\caption{Rendered view of the core of the TMMF, as used in the repeated light TM and high-load test campaigns.}
\label{fig:1blade}
\end{figure}

Five test campaigns where performed:
\begin{enumerate}
	\item A campaign with a TMmu of mass 0.0096 kg (light TM). The release tip in this campaign was supported by a pair of soft blade springs. High-resolution high speed videos showed that, during release, the tip actuated by the soft blade springs was retracted from the contact with a component of velocity parallel to the contact plane.
	\item A campaign with a TMmu of mass 0.089 kg (intermediate TM) and the DOSS sensor. In order to fix the problem of the tangential motion, when moving to the measurement campaign with the intermediate mass TM, blade springs where replaced by a stiffer set in order to minimize the shear action.
	\item A campaign with a TMmu of mass 0.844 kg (heavy TM) and the DOSS sensor. The set up was the same as in 2) above.
	\item The campaign with the light TMmu has then been repeated to harmonize the results as a consequences of the DOSS introduction, of few other modifications to the test facility and of the use of a single stiff blade, Fig.~\ref{fig:1blade}. This is done to further mitigate the transverse motion of the RTmu, which is due to a lack of performance of its actuator. This choice follows a test campaign dedicated to solving this issue, see the appendix for further details.
	\item A campaign with intermediate TM, the DOSS sensor and high (300 mN) initial preload. Also in this campaign, the RTmu is supported by one blade alone.
\end{enumerate}

\section{Test Campaigns with Low Residual Load}

The experiments of these campaigns are performed with the following operational procedure. The peculiar feature is the nominal slow reduction of the preload before the quick retraction.
\begin{enumerate}
	\item the TM swing and yaw modes are stabilized by means of a feedback loop. Both the tip and the blocking needles are not in contact with the TM;
	\item the TM is slowly approached by the blocking needles until a displacement of roughly 1 $\mu m$ is detected;
	\item the tip slowly engages the TM applying a load of roughly 400 mN, similarly to the in-flight procedure;
	\item the load is kept constant for about 10 s, then is reduced. A residual load is left in order to guarantee the contact against residual spurious vibration of the chamber;
	\item the tip is quickly retracted. The interferometer measures the TM swing motion, the optical lever measures the TM attitude and the DOSS measures the tip retraction. All these measurements are synchronously sampled at 200 kHz in a time window of 2 s;
	\item this procedure is repeated setting different yaw and pitch angles of the tip searching the maximum impulse. In order to increase the statistical significance of the results, at least 5 tests are repeated for each direction.
\end{enumerate}

The residual load of pt.~4 is unavoidable, as it is required in order to keep the TMmu-RTmu contact patch joined against spurious seismic vibrations and to account positioning errors for avoiding any detachment. This load is the cause of a net impulse on the TMmu that is summed to the effect of adhesion, Eq.~\ref{eq:catapult}. Such an extra momentum would not be critical, if all the quantities in Eq.~\ref{eq:catapult} were known and the systematic effect quantified. The residual preload $F_{pre}$ is equal to a (minimum) intentionally left value, left to avoid detachments, plus an unknown quantity given by the product of the positioning error and the constraining stiffness of the caged TM.

\subsection{Results}
\begin{figure}
\includegraphics[width=\columnwidth]{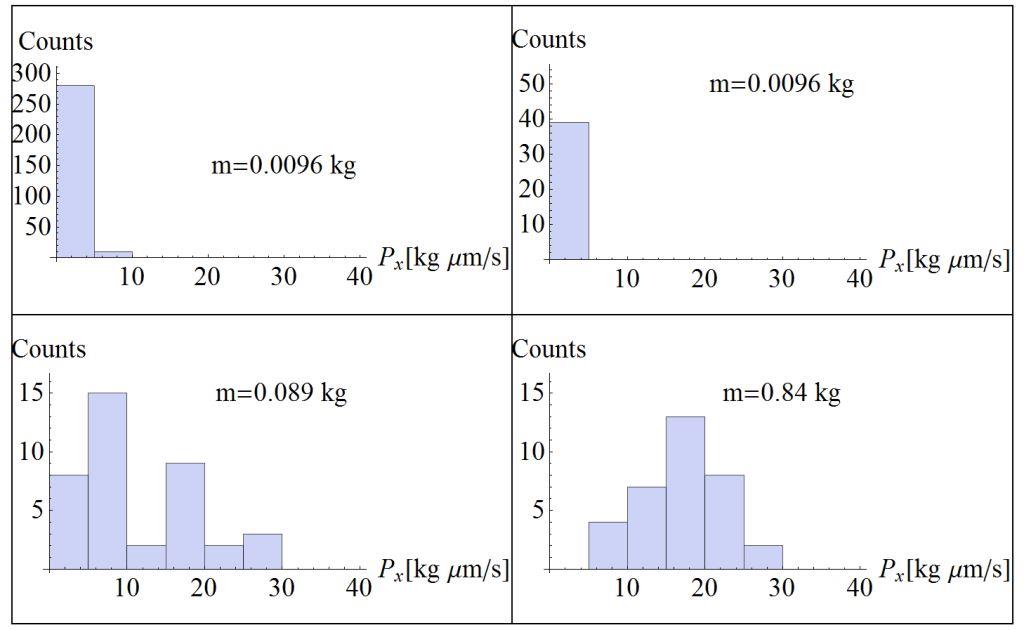}
\caption{Histograms of transferred momentum.}
\label{fig:momentum}
\end{figure}

\begin{figure}
\includegraphics[width=\columnwidth]{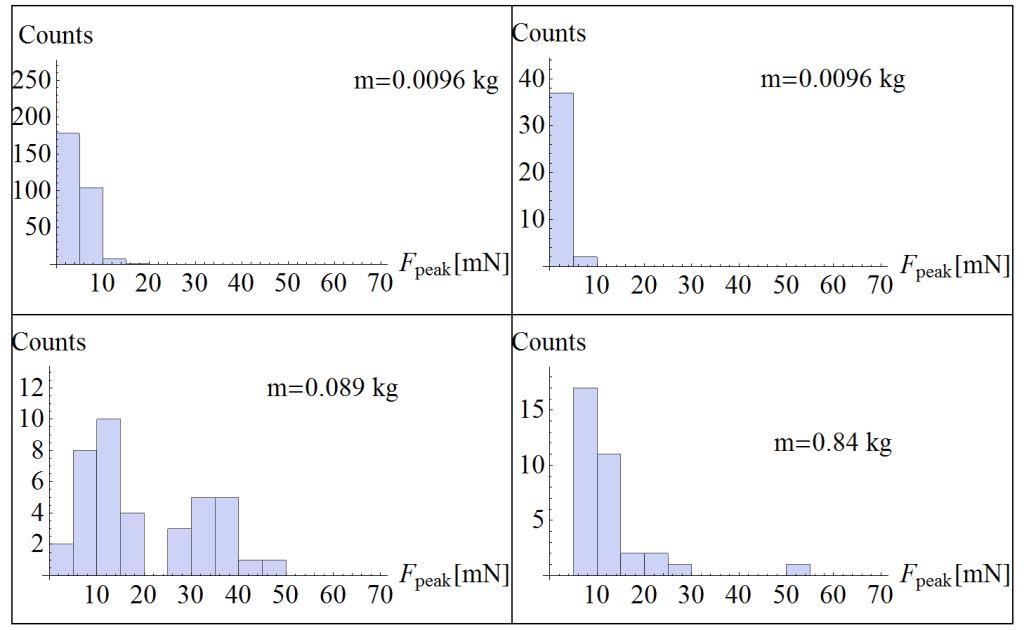}
\caption{Histograms of peak pull-off force.}
\label{fig:pulloff}
\end{figure}

Histograms of the measured quantities \cite{TestSummary} for the four different experimental campaigns are reported in  Fig.~\ref{fig:momentum} and Fig.~\ref{fig:pulloff}.

The key parameter values of the histograms are summarized in Tab.~\ref{tab:0.01kg}, Tab.~\ref{tab:0.01kgb}, Tab.~\ref{tab:0.1kg} and Tab.~\ref{tab:1kg}.

\begin{table}[!ht]
\small
\renewcommand{\arraystretch}{1}
\caption{Key parameters values for the light TM (old campaign)}
\label{tab:0.01kg}
\centering
\begin{tabular}{c|c|c|c}
\hline
 & \textbf{$P_x [\mu N s]$} & \textbf{$F_{peak}[mN]$} & \textbf{$\Delta t [ms]$} \\
\hline
Mean & 2.4  & 4.9 & 1.2 \\
\hline
Standard Deviation & 1.1  & 2 & 0.2 \\
\hline
Standard Deviation of Mean & 0.06  & 0.1 & 0.01 \\
\hline
Minimum & 0.34  & 0.37 & 0.38 \\
\hline
Maximum & 8  & 17 & 1.9 \\
\hline
\end{tabular}
\end{table}

\begin{table}[!ht]
\small
\renewcommand{\arraystretch}{1}
\caption{Key parameters values for the light TM (2014)}
\label{tab:0.01kgb}
\centering
\begin{tabular}{c|c|c|c}
\hline
 & \textbf{$P_x [\mu N s]$} & \textbf{$F_{peak}[mN]$} & \textbf{$\Delta t [ms]$} \\
\hline
Mean & 1.38  & 3.1 & 0.9 \\
\hline
Standard Deviation & 0.58  & 1.13 & 0.18 \\
\hline
Standard Deviation of Mean & 0.09  & 0.18 & 0.03 \\
\hline
Minimum & 0.22  & 0.72 & 0.48 \\
\hline
Maximum & 2.54  & 5.38 & 1.52 \\
\hline
\end{tabular}
\end{table}

\begin{table}[!ht]
\small
\renewcommand{\arraystretch}{1}
\caption{Key parameters values for the intermediate TM}
\label{tab:0.1kg}
\centering
\begin{tabular}{c|c|c|c}
\hline
 & \textbf{$P_x [\mu N s]$} & \textbf{$F_{peak}[mN]$} & \textbf{$\Delta t [ms]$} \\
\hline
Mean & 11  & 20 & 1.45 \\
\hline
Standard Deviation & 8  & 13 & 0.4 \\
\hline
Standard Deviation of Mean & 1  & 2 & 0.06 \\
\hline
Minimum & 1.9  & 3.4 & 1.0 \\
\hline
Maximum & 27  & 48 & 2.6 \\
\hline
\end{tabular}
\end{table}

\begin{table}[!ht]
\small
\renewcommand{\arraystretch}{1}
\caption{Key parameters values for the heavy TM}
\label{tab:1kg}
\centering
\begin{tabular}{c|c|c|c}
\hline
 & \textbf{$P_x [\mu N s]$} & \textbf{$F_{peak}[mN]$} & \textbf{$\Delta t [ms]$} \\
\hline
Mean & 17.5  & 12 & 3.1 \\
\hline
Standard Deviation & 5  & 9 & 1.2 \\
\hline
Standard Deviation of Mean & 0.9  & 2 & 0.2 \\
\hline
Minimum & 8  & 5 & 0.8 \\
\hline
Maximum & 27  & 54 & 5.6 \\
\hline
\end{tabular}
\end{table}

\begin{figure}
\includegraphics[width=\columnwidth]{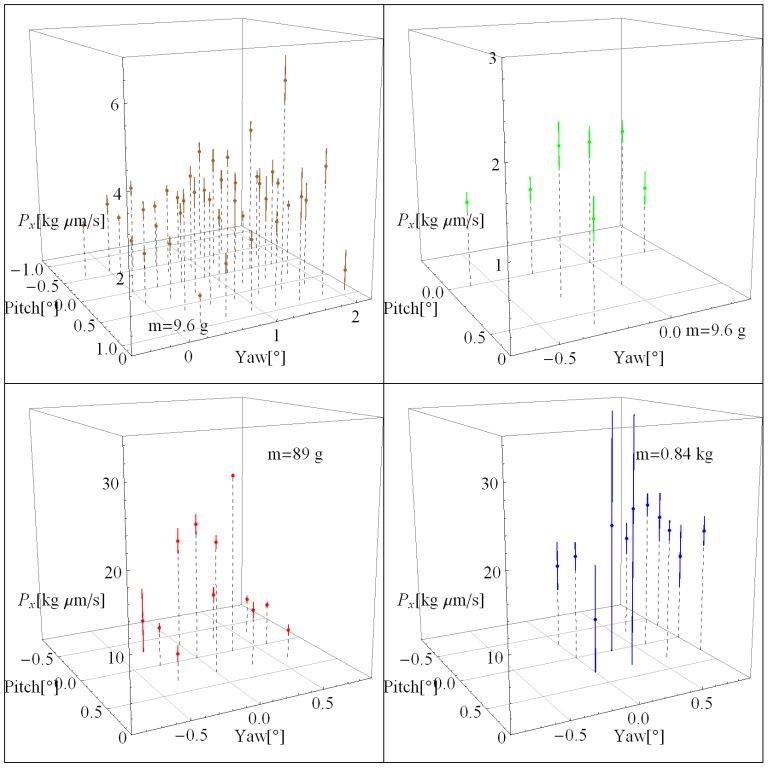}
\caption{Angular dependence of transferred momentum.}
\label{fig:anglemom}
\end{figure}
\begin{figure}
\includegraphics[width=\columnwidth]{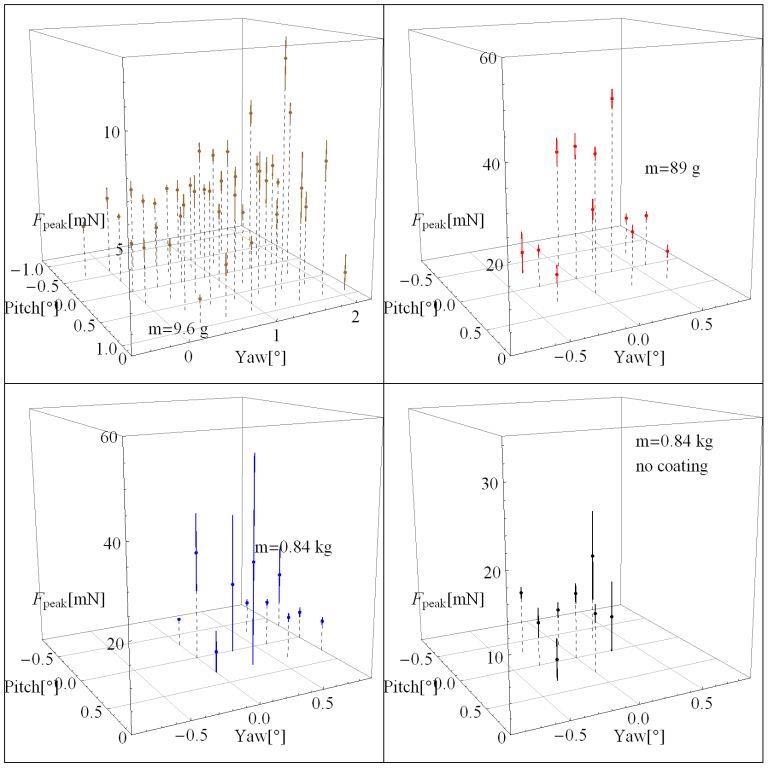}
\caption{Angular dependence of the pull-off force.}
\label{fig:angleforce}
\end{figure}

The dependence of transferred momentum and peak pull-off force on the pitch and yaw angles of the trajectory of the release tip are shown in Fig.~\ref{fig:anglemom} and Fig.~\ref{fig:angleforce}.

The comparison of Tab.~\ref{tab:0.01kg} and Tab.~\ref{tab:0.01kgb} with Tab.~\ref{tab:0.1kg} and Tab.~\ref{tab:1kg} is more complex. In particular:
\begin{itemize}
	\item Even if the maximum value in Tab.~\ref{tab:1kg} is of the same order of that in Tab.~\ref{tab:0.1kg}, it is evident from the mean values and from the histograms in Fig.~\ref{fig:pulloff}, that this is more of a coincidence due to a single outlier in the distribution of the data collected with the heavy TM. If the outlier is removed, the peak force appears to have been reduced by almost a factor 2 moving from the intermediate to the heavy TMmu.
	\item The mean event duration is larger for the heavy TM than for the intermediate. In addition $\Delta t$ is found to be significantly correlated with $F_{peak}$ (linear correlation coefficient ~0.6)
	\item The light TMmu always shows a low pull-off force, both with the old campaign (soft blades) and the new one (single blade). This suggests that dynamic effects (e.g. rotation) of the TMmu may influence the result.
	\item There is a slight reduction in the pull-off and momentum from the old to the new light TMmu campaigns. Such a reduction can be due to a better position accuracy of the needles actuator. However, this demonstrates that the spurious transverse motion of the RTmu was not critical.
\end{itemize}

The preload $F_{pre}$ applied by the blocking needles on the TMmu, generates part of the transferred momentum (catapult effect). $F_{pre}$ can be calculated from the compression of the blade springs and from their stiffness. $F_{pre}$ is estimated to be 5÷10 mN with the stiff springs (campaigns with the intermediate and the heavy TM) and 1÷3 mN with the soft springs (light TMmu). The uncertainty on $F_{pre}$ derives mainly from the corresponding uncertainty on the identification of the position at which the tip completely engages the TM against the blocking system. 
In the absence of any adhesion, the release of $F_{pre}$ would impress a momentum to the TMmu given by \ref{eq:catapult}. Under the same hypotheses, the duration of the event, and the peak pull-off force are expected to be:
\begin{equation}
	\Delta t \approx \frac{\pi}{2}\sqrt{\frac{m}{k_{nd}}}
\end{equation}
\begin{equation}
	F_{peak} \approx F_{pre}
\end{equation}
Tab.~\ref{tab:catapult} lists the values calculated from the previous equations using the measured value for k, $k \approx 5\div8\times10^5$ N/m.

\begin{table}[!h]
\small
\renewcommand{\arraystretch}{1}
\caption{Contribution of the effect of the residual preload}
\label{tab:catapult}
\centering
\begin{tabular}{c|c|c|c|c}
\hline
\textbf{m [kg]} & \textbf{Blades} & \textbf{$F_{pre} [mN]$} & \textbf{$P_x [\mu N s]$} & \textbf{$\Delta t [ms]$} \\
\hline
Light TMmu & Soft & $1\div3$ & $0.1\div0.4$ & $0.2$ \\
\hline
Light TMmu & Stiff & $3\div6$ & $0.4\div1.3$ & $0.2\div0.4$ \\
\hline
Intermediate TMmu & Stiff & $5\div10$ & $2\div4$ & $0.5\div0.7$ \\
\hline
Heavy TMmu & Stiff & $5\div10$ & $5\div13$ & $1.6\div2.0$ \\
\hline
\end{tabular}
\end{table}

For all the campaigns, the measured momentum values appear to be larger or equal to these “threshold” momenta, within the uncertainty of the estimates. For the light and intermediate TMmu, the threshold is barely significant compared to the measured values and a large fraction of those are significantly higher that the threshold. For the heavy TMmu a significant fraction of the measured values are compatible with just the action of the catapult. In these events, the peak pull-off force tends to be comparable with just the preload. The analysis of the duration values gives instead no clear picture. 

The angular plots of $P_x$ and $F_{peak}$ reported in Fig.~\ref{fig:angleforce} and Fig.~\ref{fig:anglemom}, show a marked dependence on both pitch and yaw only for the intermediate TM. Some less pronounced dependence may also be visible in the campaign with the heavy TM, in the presence of gold coating.

Putting together the evidence above:

\begin{itemize}
	\item A dominant contribution of adhesion has been observed with the intermediate TMmu. This adhesion shows a peak value as high as $\approx$50 mN and can transfer a momentum as high as $\approx$30 $\mu N s$ in a time of the order of $\approx$1 ms.  
The pronounced angular dependence of pull-off force and transferred momentum is also a sign of adhesion contact with its dependence on the nature of the surfaces in contact. Simple vector projection effects that scale as the cosine of angles, are negligible within the investigated angular range.
	\item Both the momentum and the peak force are expected to be independent of the mass. The much-reduced values measured with the light TMmu relative to those obtained with the intermediate TMmu, show then that some spurious effects must have intervened to suppress or reduce the adhesion.
	\item The contribution of adhesion to the transferred momentum for the case of the heavy TMmu is less evident. Given the large uncertainties in the prediction of Tab.~\ref{tab:catapult}, the possibility exists that adhesion has played only a minor role in in this case. 
	\item The points above are not inconsistent with the following further observations:
\begin{itemize}
	\item In the campaign with the light TMmu, the rotations of the TMmu upon load reduction and retraction may have contributed to break the adhesion, thus reducing the transfer of momentum in the horizontal direction.
	\item During the 400 mN preload application and subsequent release with the heavy TMmu, rotations of the TM by 30 $\div$ 50 $\mu rad$ have been observed in both pitch and yaw. This motion may have stressed the contact patch and consequently weakened the adhesion. The high mass of this TMmu makes the system apparently more stable, however, the low loads applied makes it hard to fully constrain the pendulum on the needles with the tip and therefore load it properly.
 \end{itemize}
\end{itemize}

\begin{figure}
\begin{center}
\includegraphics[width=7cm]{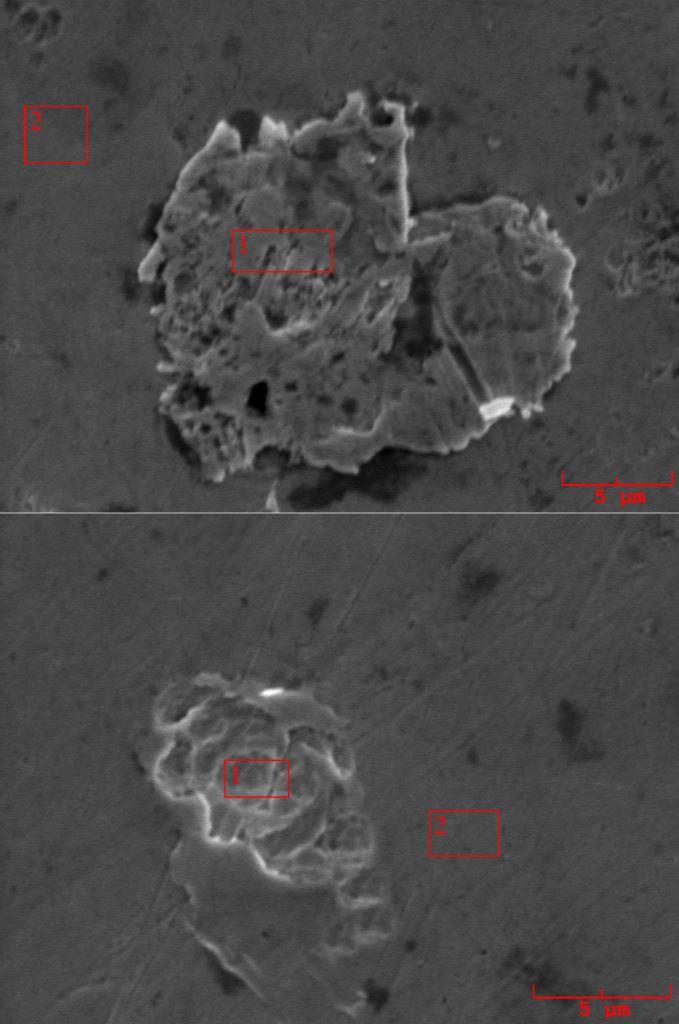}
\end{center}
\caption{SEM images of the release tip after the measurement campaign showing spots of gold (points 1, 95\% top, 89\% bottom), whereas its bulk concentration (points 2) is 78\% and 80\%.}
\label{fig:spots}
\end{figure}

The contact between soft materials is expected to be dominated by ductile adhesion. However, the observed values of maximum pull-off force are not compatible with a full ductile contact model for gold on gold, upon a preload of 400 mN. The pull off force for such a contact is expected to be in excess of the applied preload. According to \cite{AdhesionClean} a force up to 1 N could be expected for a preload of  400 mN. This is much larger that the observed 50 mN upper bound.

The contact between a hard material and bulk gold, as is nominally the case in our system, has been observed \cite{AdhesionClean} to behave approximately as a gold to gold contact, thus even this kind of contact appears to be hardly compatible with the observations. 
In order to independently assess the presence of ductile adhesion, SEM and optical inspections have been performed of the contact. Fig.~\ref{fig:spots} shows SEM pictures of the tip that was used for the intermediate TM campaign. 

The pictures show well visible spots with concentration of gold much larger than the concentration of the bulk alloy (Hafner Orplid keramik PF 77.7\% Au – 19.5\% Pt). The size of these spots (5÷10 $\mu m$) is a sizeable fraction of the expected contact area of $\approx$35 $\mu m$. This would indicate that indeed some ductile contact was established at some point and the contact has been fractured during one of the maneuvers performed during the experiment.  

As the maximum pull-off force in a ductile contact is just the product of the gold tensile strength by the contact area, a 5 $\mu m$ radius spot will show a maximum pull-off force reduced by a factor $\approx$50 relative to that of a uniform contact of $\approx$35 $\mu m$, something of the order of the observed peak forces. Though this picture is suggestive, in reality there is no way of identifying when this gold has been transferred, if during one of the retraction events, during preload release, or finally during some other manipulation of the contact. 

\begin{figure}
\includegraphics[width=1.1\columnwidth]{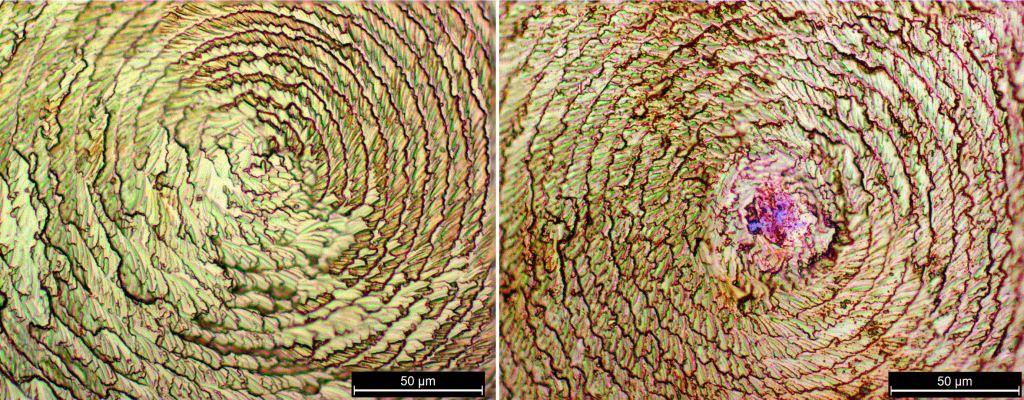}
\caption{Microscope view of the contact areas between the release tip and the TM. Left a native contact. Right the area of contact after two measurement campaigns. The radius of the damaged surface is on the order of that expected from the Hertzian contact (~35 $\mu m$).}
\label{fig:insert}
\end{figure}

If for each retraction, some gold would be removed from the TMmu, the contact area would rapidly become free of gold. Though the contact area is indeed damaged (Fig.~\ref{fig:insert}), the SEM analysis indicates that the contact is still covered with gold.

The observed values of pull-off forces with the intermediate TMmu would be compatible with a fully elastic model \cite{ductileadhesion}, which is able to describe, for instance, the adhesive contact between smooth mica surfaces coated with a nanometer rough gold layer. However, the adhesion in this kind of contacts rapidly decreases in order of magnitude if the surfaces are not pristine and very smooth. This is not the case for the heavily handled surfaces of tip and TM, which anyhow have a native roughness of about 100 nm. Thus, the measured value seems to be quite high for such kind of adhesion, which, in addition, would be hardly compatible with the observed transfer of gold to the RTmu.

Finally it is worth mentioning that the data would still be compatible with an ideal ductile contact formed at a preload of about 10 mN. This last scenario may happen if, during the operation of releasing the 400 mN pre-load, the contact is broken and re-established at the pre-release preload. However such a clean scenario does not seem compatible with the observed damage to the contact interface. 

\begin{figure}
\includegraphics[width=1.1\columnwidth]{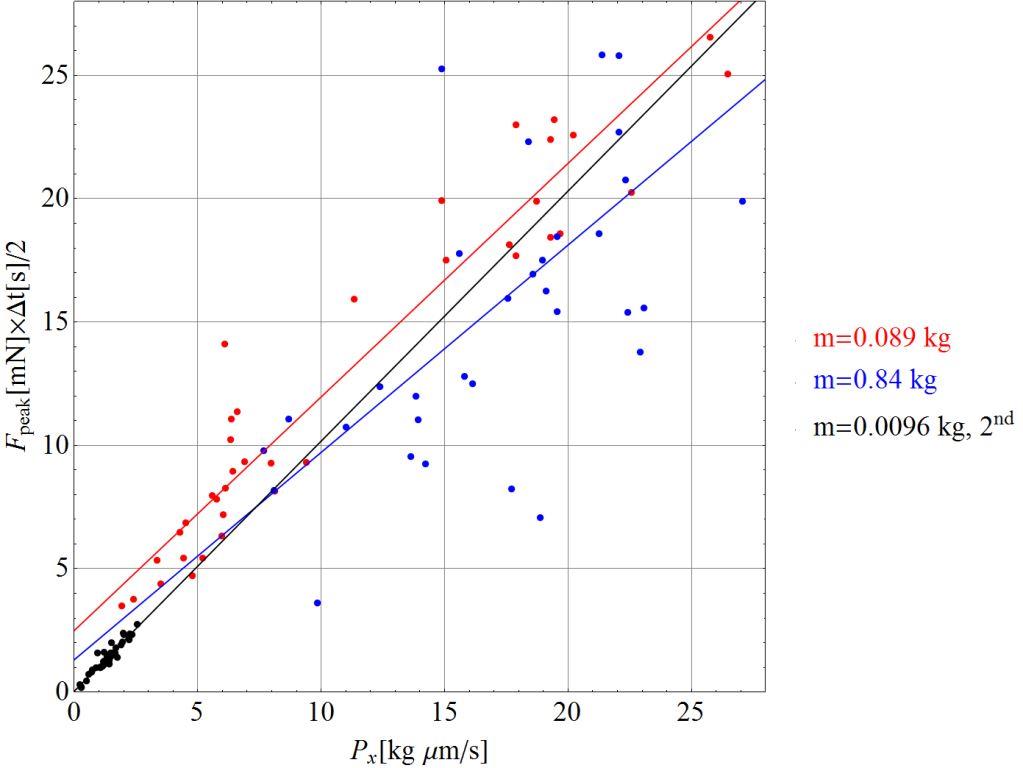}
\caption{Momentum calculated from Eq.~\ref{eq:approxmom} as a function of the measured momentum.}
\label{fig:approxmom}
\end{figure}

It is important to underline however that the measured values refer to a contact between surfaces that have been damaged during large set of tests. This may differ from the conditions of the contact in-flight.

In addition, as the observations do not match well the previous knowledge on related kinds of contact, one must be aware that the observed contacts might be damaged by the manipulation required by the test itself, and that the contact in flight might then be of a rather different nature. It is also worth underlying that, while the maximum contact load is equal in-flight and on-ground, the residual load at the time of retraction on-ground is smaller by a factor 30 than what is now expected in-flight. 

\section{Test Campaign with High Residual Load}

During the hand-over phase the GPRM must remain in contact with the TM, \cite{Handover} as an intermediate detachment may determine extra TM misalignment. In order to fulfill this requirement, the bad positioning performance of the actuators forces to leave a contact load as high as the maximum one ($\approx$300 mN). 

In theory, adhesion when the contact locally plasticize is a function of the maximum load experienced. Thus, the load right before the quick retraction should not be a key parameter, except for having a negative effect on the momentum transferred (Eq. \ref{eq:catapult}).
The on-ground experimental qualification always reproduces the expected maximum load expected in flight. However, it is extremely challenging guaranteeing that the reduction of the contact load in the test facility does not result in an assessment motion that reduces the effective maximum load. In order to exclude such a detrimental effect in the experiment, a specific test campaign has been performed.

Chronologically, this test campaign is the last performed (mid 2014). The spurious transverse motion of the RTmu is eliminated by using only a single blade as a support of the RTmu on top of its piezo stage, Fig.~\ref{fig:1blade}. Besides, instead of exploring different pitch-yaw angles, priority is given to the contact patch position y-z, in order to perform the contact on un-damaged TMmu surface patches. It is worth noting that the change in the y-z coordinates constitutes also an uncontrolled change in the local retraction angle. The roughness and the surface finishing determines that the local vector normal to the average surface might be misaligned with respect to the nominal normal.

\begin{figure}
\includegraphics[width=0.73\columnwidth]{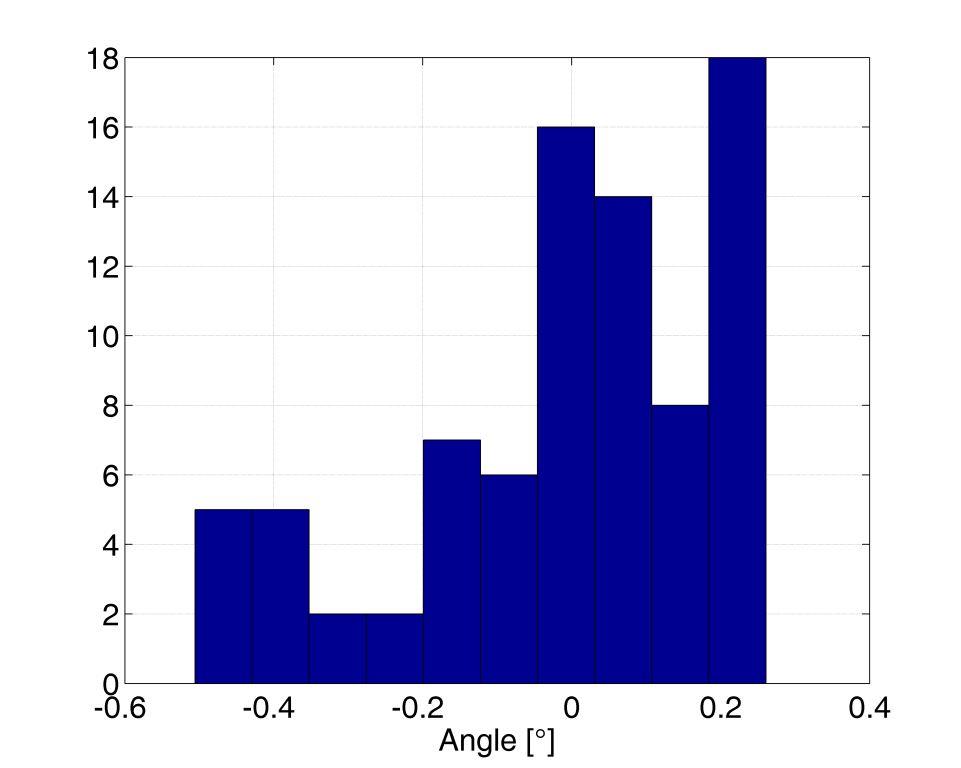}
\caption{Local misalignment angle.}
\label{fig:angle1}
\end{figure}

\begin{figure}
\includegraphics[width=0.73\columnwidth]{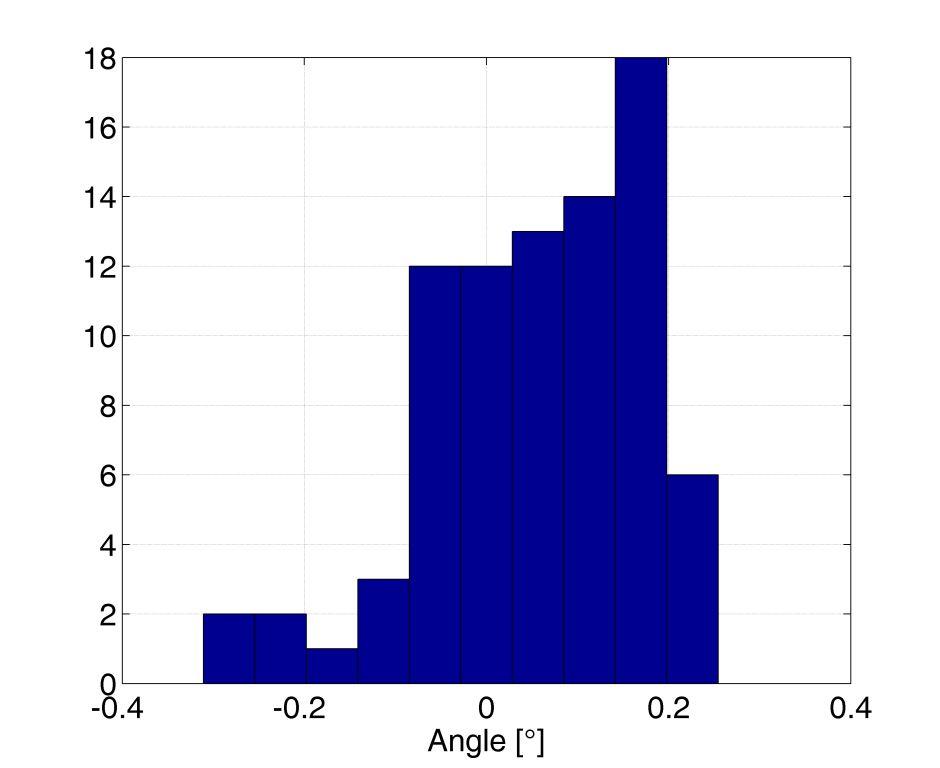}
\caption{Local misalignment angle. This angle is orthogonal to that of Fig.~\ref{fig:angle1}}
\label{fig:angle2}
\end{figure}

The estimations of the local angles with respect to the nominal normal vector are shown in Fig.~\ref{fig:angle1} and \ref{fig:angle2}. They are obtained by fitting a plane into a fraction of the surface as big as the contact patch. The coefficients of the plane are transformed in angles assuming the inclinations are small.


Indeed, the uncontrolled change in the real angle can be a source of uncertainty on the adhesion peaks and elongations.

The experimental operations for this campaign are summarized by the following points:
\begin{enumerate}
	\item the TM swing and yaw modes are stabilized by means of a feedback loop. Both the tip and the blocking needles are not in contact with the TM;
	\item the TM is slowly approached by the blocking needles until a displacement of the TMmu of roughly 1 $\mu m$ is detected;
	\item the tip slowly engages the TM applying a load of about 300 mN, similarly to the in-flight procedure. 300 mN are applied by moving the ultrasonic piezo by 47 $\mu m$;
	\item the load is kept constant for about 10 s;
	\item the RTmu is quickly retracted. The interferometer measures the TMmu swing motion, the optical lever measures the TMmu attitude and the DOSS measures the tip retraction. All these measurements are synchronously sampled at 200 kHz in a time window of 2 s;
	\item this procedure is repeated following a grid of y and z RTmu coordinates.
\end{enumerate}

\begin{figure}
\begin{center}
\includegraphics[width=1\columnwidth]{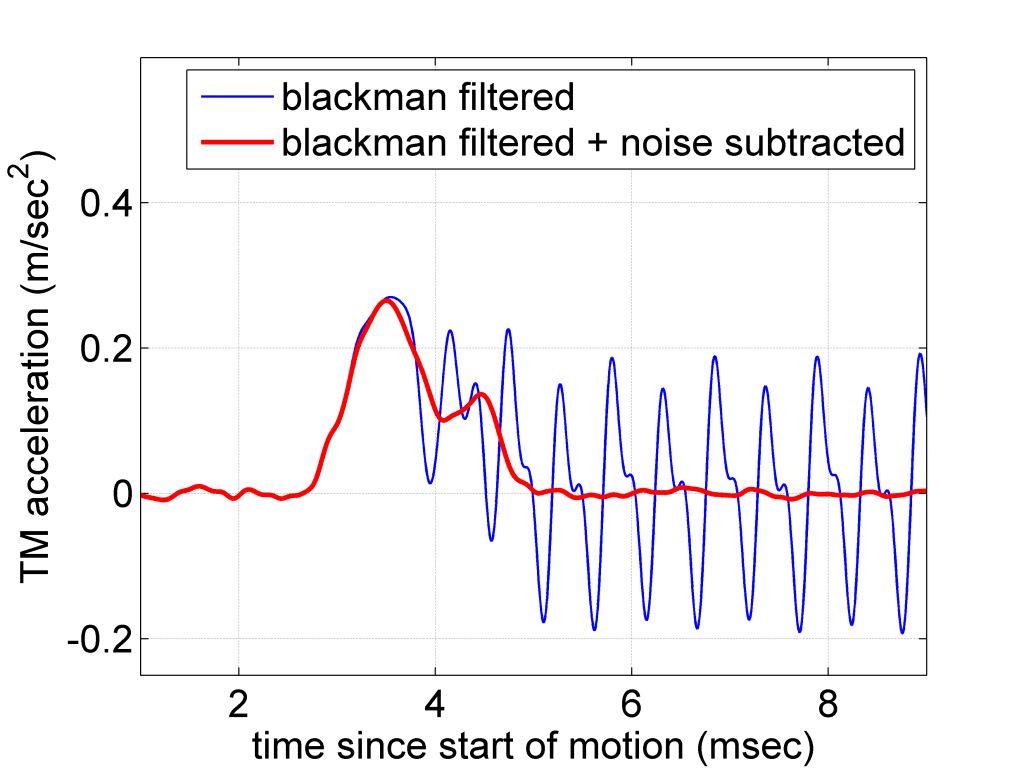}
\end{center}
\caption{Example of TM acceleration as measured and after correction in the high preload test campaign.}
\label{fig:acchl}
\end{figure}

Indeed, the choice of leaving a high residual load enhances the catapult effect. With a few hundreds of mN load, the RTmu retraction time is comparable to the needles-TMmu natural oscillation period. Eq.~\ref{eq:catapult} is therefore not valid. The measured acceleration is shown in Fig.~\ref{fig:acchl} and it is clear that the peak force is lower than 300 mN (as it would be if the RTmu disappeared instantaneously).

Unfortunately, the measurement of the TMmu motion does not provide an immediate accurate force profile as it is affected by a spurious disturbance attributed to the interferometer electronics. This is critical in this campaign in which the portion of signal of interest is long, as a consequence of the push of the needles.

The disturbance removal procedure is discussed in the Appendix. Fig.~\ref{fig:acchl} shows the results in terms of acceleration before and after the removal.

Qualitatively one can assume the first peak in Fig.~\ref{fig:acchl} is due to the needles push while the second is due to adhesion, however this must be demonstrated.

The approach followed for disentangling needles push and adhesion is different from that one of the previous sections. In particular, an adhesion model -which is indeed an assumption itself- is avoided. A model is proposed only for the needles. The needles push is derived from this model in which the TM motion is the input. The difference between the measured acceleration and the needles push is then the force applied on the contact patch by the RTmu. If this force becomes positive adhesion is present.

The force applied on the contact patch between TMmu and RTmu is derived by:
\begin{equation}
	F_{TM,RT} = m_{TMmu}\ddot{x}_{TMmu} - F_{ND,TM}\left(x_{TMmu}\right)
\end{equation}
where $\ddot{x}_{TMmu}$ is the measured acceleration, $m_{TMmu}$ is the mass (0.089 kg) and $F_{ND,RT}$ is the force that the needles exert on the TM and is a function of the TMmu law of motion, according to the model of the blocking system.

For instance, the simplest and straightforward model is that one in which the needles push is:
\begin{displaymath}
	F_{pl} = \left\{
     \begin{array}{lr}
       k_{ND}(x_{TM} - x_{0}) & : x_{TM} - x_{0} < 0 \\
       0 & : x_{TM} - x_{0} \geq 0
     \end{array}
   \right.
\end{displaymath} 
where $k_{ND}$ is the stiffness of the needles, $x_{TM}$ is the TMmu measured motion and $x_{0}$ is the initial position that allows the residual load.

\begin{figure}
\begin{center}
\includegraphics[width=0.7\columnwidth]{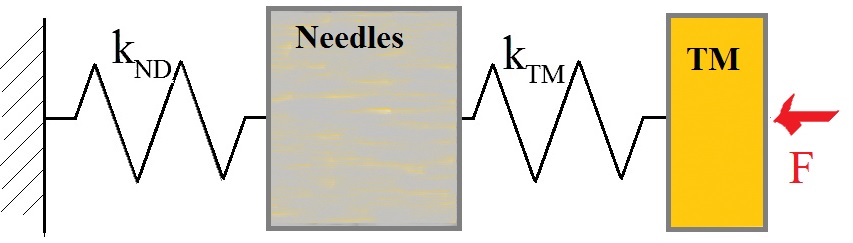}
\end{center}
\caption{Model of the system supporting the TM.}
\label{fig:NDmodel}
\end{figure}

\begin{figure}
\includegraphics[width=1.1\columnwidth]{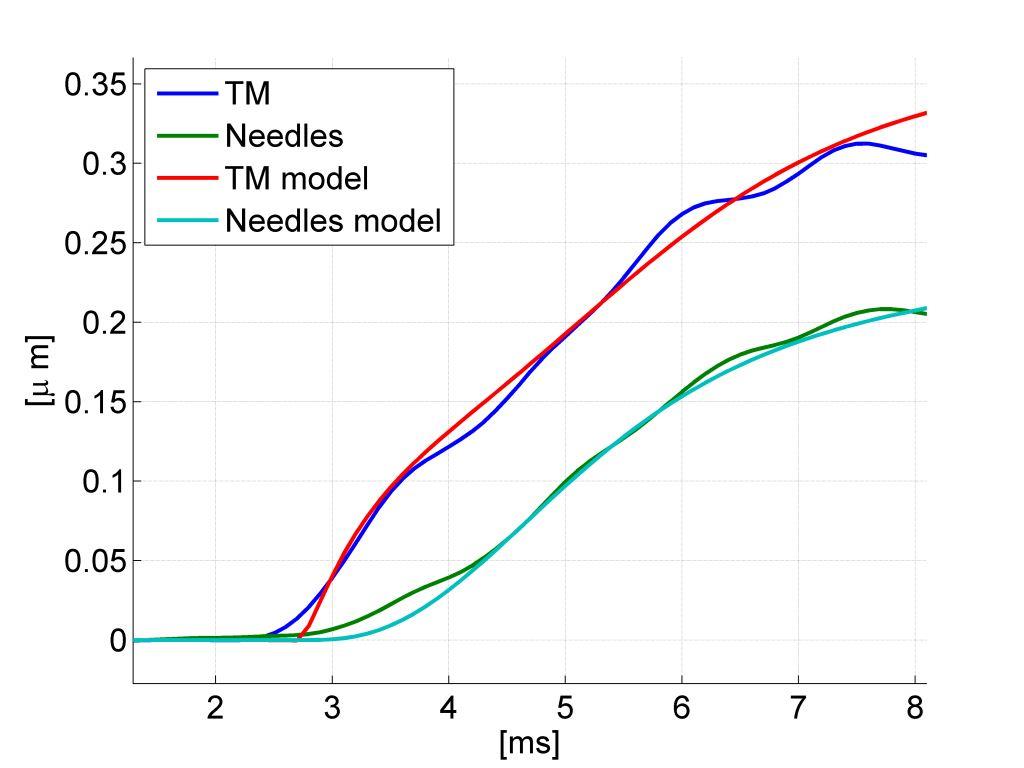}
\caption{TM/needles measured and identified motion.}
\label{fig:NDid}
\end{figure}

The model of the blocking system is derived from specific tests in which the RT has been loaded on the TM with the same velocity, acceleration and jerk of the release tests. Both the TM and the needles motion have been measured, although not synchronized. The needles motion is measured by pointing the interferometer on the back of their supporting platform and is clearly different from that one of the TM, Fig.~\ref{fig:NDid}. A simple stiffness is not enough for describing the needles and a mass is needed for taking into account dynamic effects. The motion equations are then:
\begin{align}
	m_{ND} \ddot{x}_{ND} & = - k_{ND}x_{ND} - k_{TM}(x_{ND} - x_{TM}) \\
	m_{TM} \ddot{x}_{TM} & =  = \left\{
     \begin{array}{lr}
       k_{TM}(x_{ND} - x_{TM}) & : (x_{ND} - x_{TM}) < 0 \\
       0 & : (x_{ND} - x_{TM}) \geq 0
     \end{array}
   \right. - F_{ext}
\end{align}
where $m_{ND}$ is the mass of the needles, $k_{ND}$ is the stiffness between the mass of the needles and ground, $k_{TM}$ is the stiffness between the TMmu and the mass of the needles and $m_{TM}$ is the mass of the TMmu (0.089 kg in this test campaign). $F_{ext}$ is the external force applied on the TMmu by the RTmu. $x_{TM}$ and $x_{ND}$ are the motion of the TMmu and of the needles respectively. They have been measured, although not synchronized as only one interferometer is available. $m_{ND}$, $k_{ND}$ and $k_{TM}$ are the unknowns. In order to avoid continuous oscillations, a damper has been included in parallel to both $k_{ND}$ and $k_{TM}$.

8 measurements are available for both TMmu and needles motion. The applied force $F_{ext}$ has been assumed repeatable and has been measured by acquiring the motion performed by the ultrasonic actuator that moves the RTmu and calibrating the stiffness of the blade. The tests used for identification are characterized by a pushing force (from 0 to 300 mN) instead of the usual release (i.e. pull force). The TMmu and needles datasets are not synchronized, hence 64 combinations are possible. The parameters identified are listed in Tab.~\ref{tab:parnd}.

\begin{table}[!h]
\small
\renewcommand{\arraystretch}{1}
\caption{Parameters of the TMmu supporting system.}
\label{tab:parnd}
\centering
\begin{tabular}{c|c|c|c}
\hline
\textbf{Name} & \textbf{Units} & \textbf{Mean} & \textbf{Standard} \\
\textbf{} & \textbf{} & \textbf{} & \textbf{Deviation} \\
\hline
$k_{TM}$ & N/m & $2.03\times10^6$ & $1.65\times10^5$ \\
\hline
$k_{ND}$ & N/m & $1.12\times10^6$ & $2.40\times10^4$ \\
\hline
$m_{ND}$ & kg & $1.74$ & 0.28 \\
\hline
\end{tabular}
\end{table}

\begin{figure}
\includegraphics[width=\columnwidth]{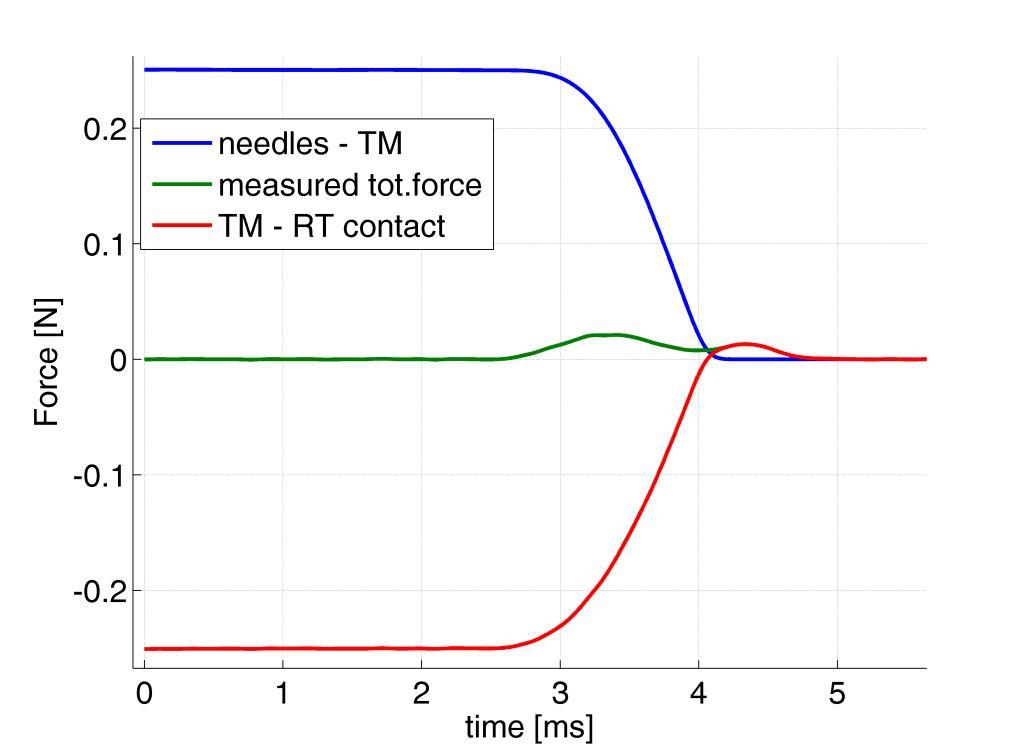}
\caption{Needles push derived from the model. The contact behaviour between TMmu and RTmu is derived by the difference between the push and the measured force.}
\label{fig:diff1}
\end{figure}

\begin{figure}
\includegraphics[width=1\columnwidth]{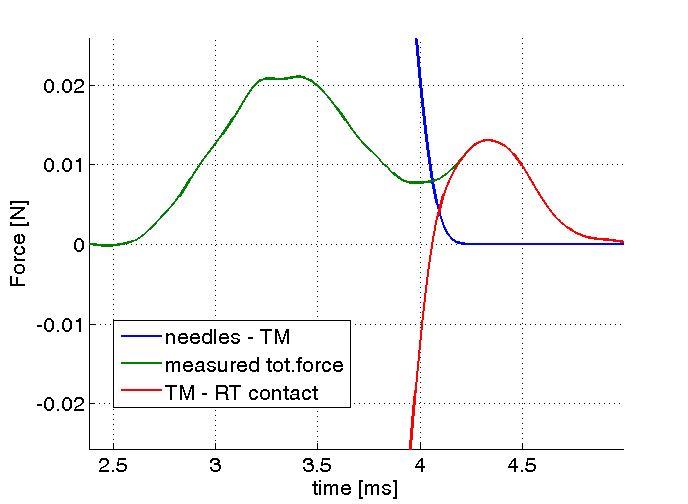}
\caption{Detail. Needles push derived from the model. The contact behaviour between TMmu and RTmu is derived by the difference between the push and the measured force.}
\label{fig:diff2}
\end{figure}

\begin{figure}
\begin{center}
\includegraphics[width=\columnwidth]{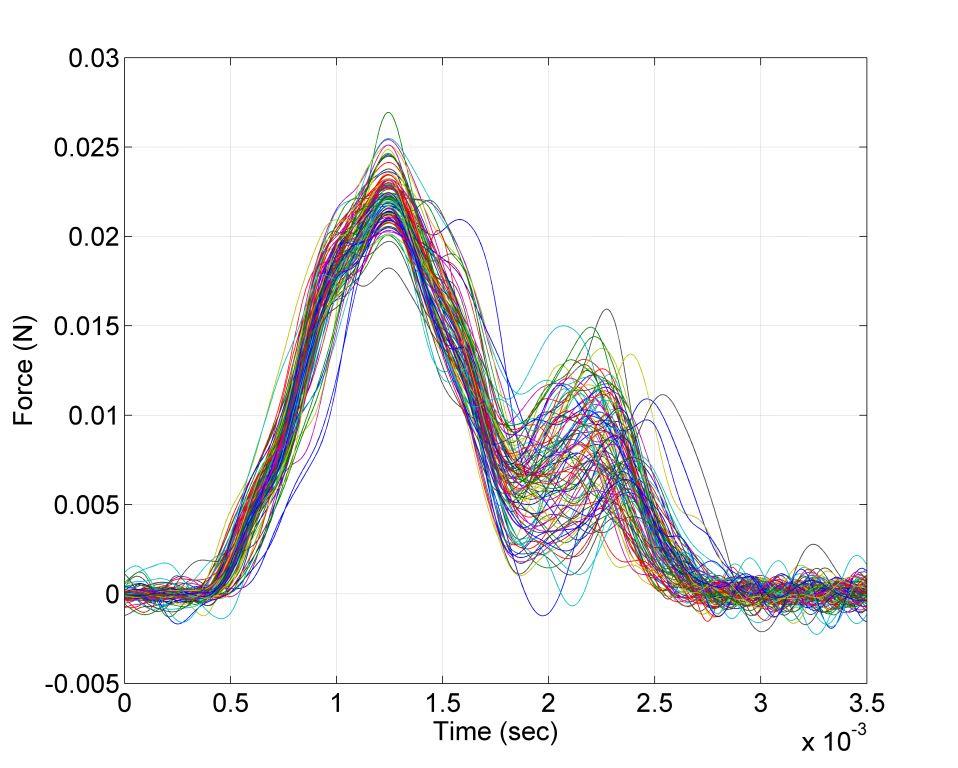}
\end{center}
\caption{Set of forces measured on the TM. The datasets are synchronized with respect to the first peak.}
\label{fig:forces}
\end{figure}

These parameters are used with each retraction data-set (Fig.~\ref{fig:forces}) in order to derive the needles-TMmu push force and the contact force between TMmu and RTmu.

An example of this results is shown in Fig.~\ref{fig:diff1} and \ref{fig:diff2}. It is evident that adhesion is roughly coincident with the second peak of the acceleration profile, as it was qualitatively expected. Tab.~\ref{tab:res01kghl} summarizes the results in terms of pull-off force and transferred momentum, due to adhesion.

\begin{table}[!h]
\small
\renewcommand{\arraystretch}{1}
\caption{Results of the test campaign with high residual preload.}
\label{tab:res01kghl}
\centering
\begin{tabular}{c|c|c|c}
\hline
\textbf{Name} & \textbf{Units} & \textbf{Mean} & \textbf{Standard} \\
\textbf{} & \textbf{} & \textbf{} & \textbf{Deviation} \\
\hline
Pull-off & mN & $9.5$ & $2.5$ \\
\hline
Momentum & Ns & $4.68\times10^{-6}$ & $1.32\times10^{-6}$ \\
\hline
\end{tabular}
\end{table}

Fig.~\ref{fig:ftoelhl} shows a few of the force-to-elongation functions.

\begin{figure}
\includegraphics[width=1\columnwidth]{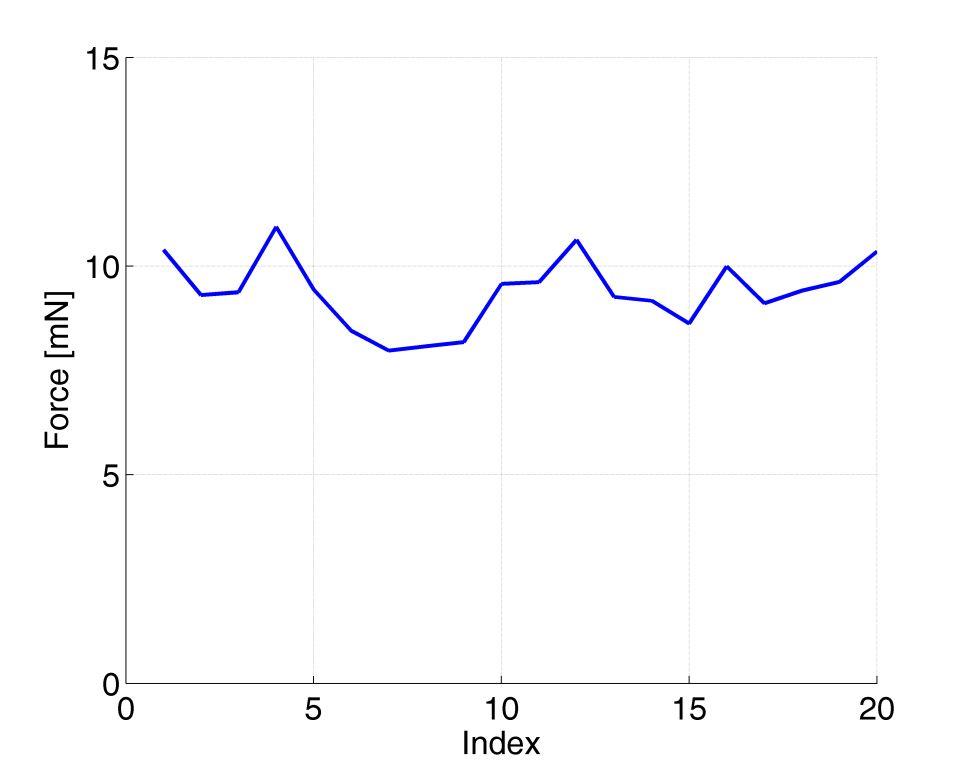}
\caption{Peak force as a function of the test execution order.}
\label{fig:fadhtime}
\end{figure}

\begin{figure}
\includegraphics[width=1.1\columnwidth]{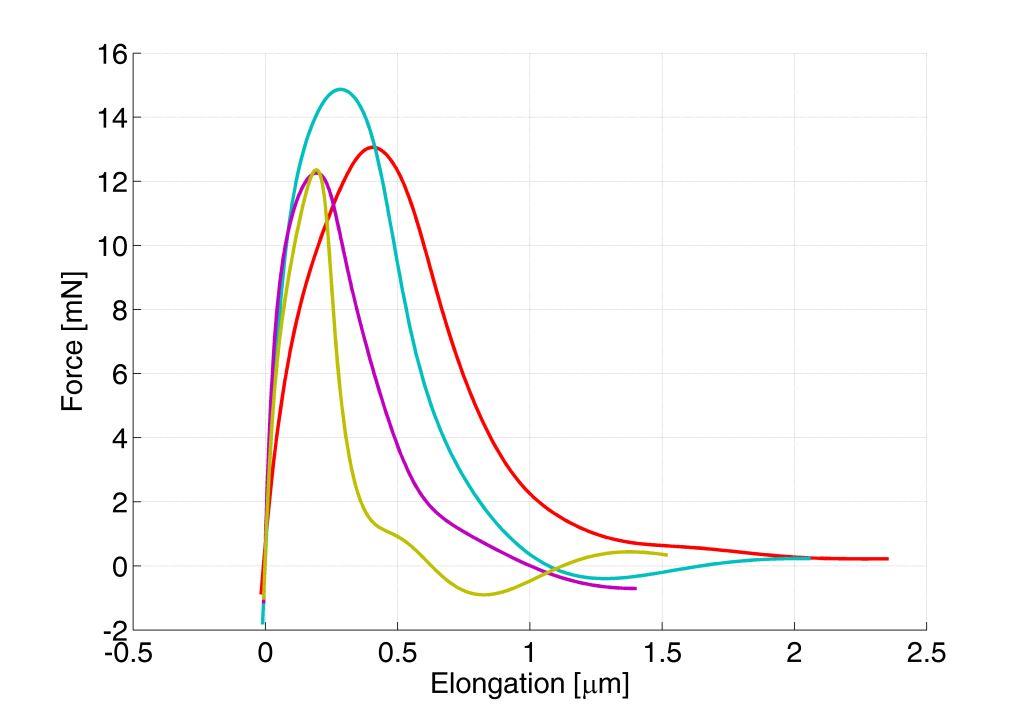}
\caption{Force-to-elongation functions (3 out of the 107 available). Both force and elongation signals are filtered with a Blackman with a 3 kHz cut-frequency.}
\label{fig:ftoelhl}
\end{figure}

One of the major concerns is that by repeating the test on the same position one would ruin the surface determining an underestimation of adhesion. However, repeating the tests on the same y-z coordinate suggests that this effect does not influence adhesion, at least in the tens of tests, Fig.~\ref{fig:fadhtime}. This observation is of key importance not only for the experimental procedure, but also for the flight release. In case the release fails, for any reason, a re-grabbing and re-release is planned. This result suggest that the mechanical behaviour will not differ strongly from one release to another one. 

The results of this test campaign are much more repeatable than the results of the campaigns with low residual load, despite the uncontrolled chenge of the contact angle described at the beginning of this section. At the same time, the results of this campaign along with the previous ones never show a pull-off force higher than 50 mN. This strengthens the idea that adhesion in-flight will be about a few tens of mN or less.

\section{Surface Analysis Before and After the Tests}

In the scope of this thesis, adhesion is mainly analyzed in terms of its macroscopic effects. Indeed, precious information are obtained with a microscopic approach as well. In the high-residual-load campaign the surface profile is measured both before and after the tests. It is worth reminding that each y-z position sees only one or two contacts with the RTmu.

\begin{figure}
\includegraphics[width=\columnwidth]{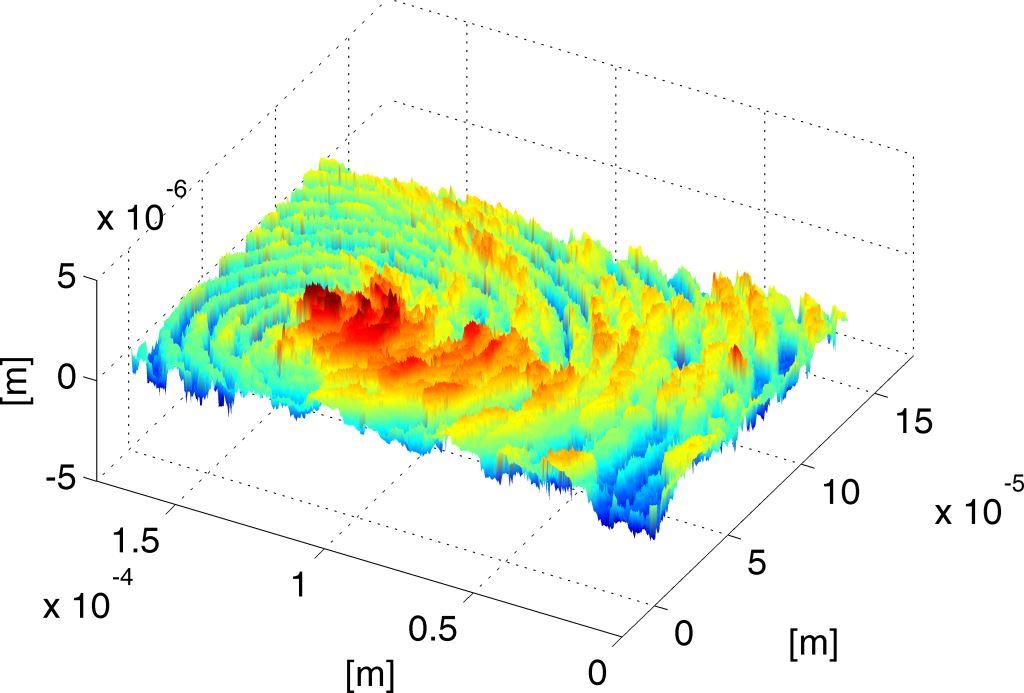}
\caption{Surface profile before the experimental campaign.}
\label{fig:before}
\end{figure}

The 3D profile of the surface is measured right before and after the tests. Fig.~\ref{fig:before} shows the profile before the contact. The effect of the application of 300 mN is not clear and requires some analysis.

\begin{figure}
\includegraphics[width=\columnwidth]{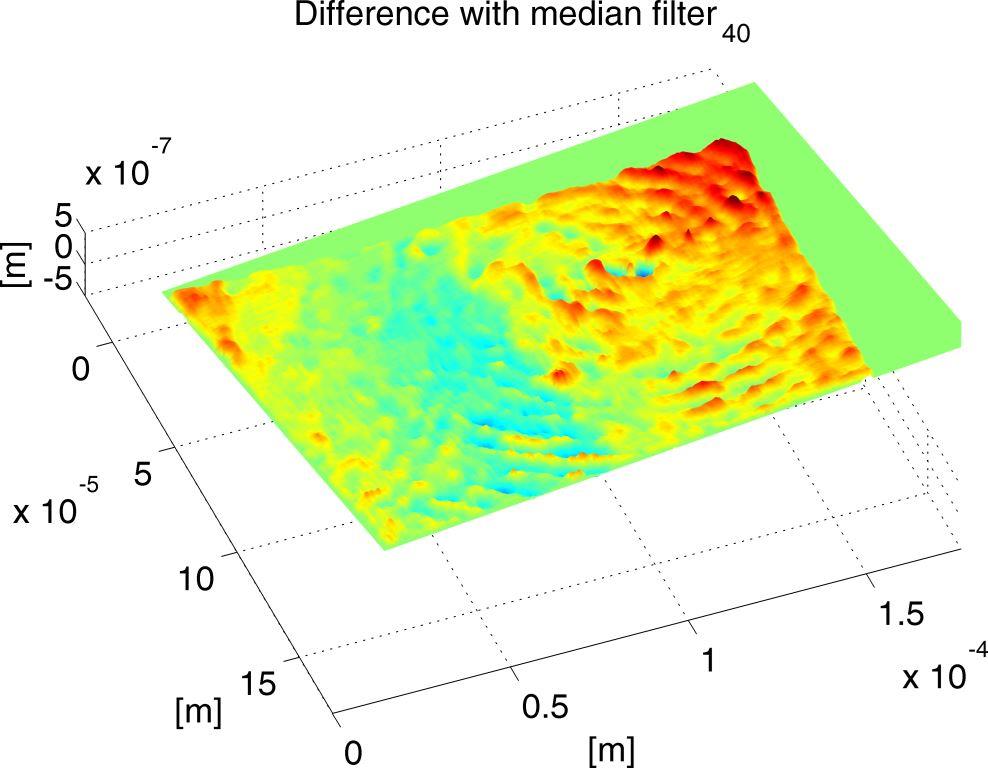}
\caption{Difference between the surfaces profiles.}
\label{fig:diff}
\end{figure}

\begin{figure}
\includegraphics[width=\columnwidth]{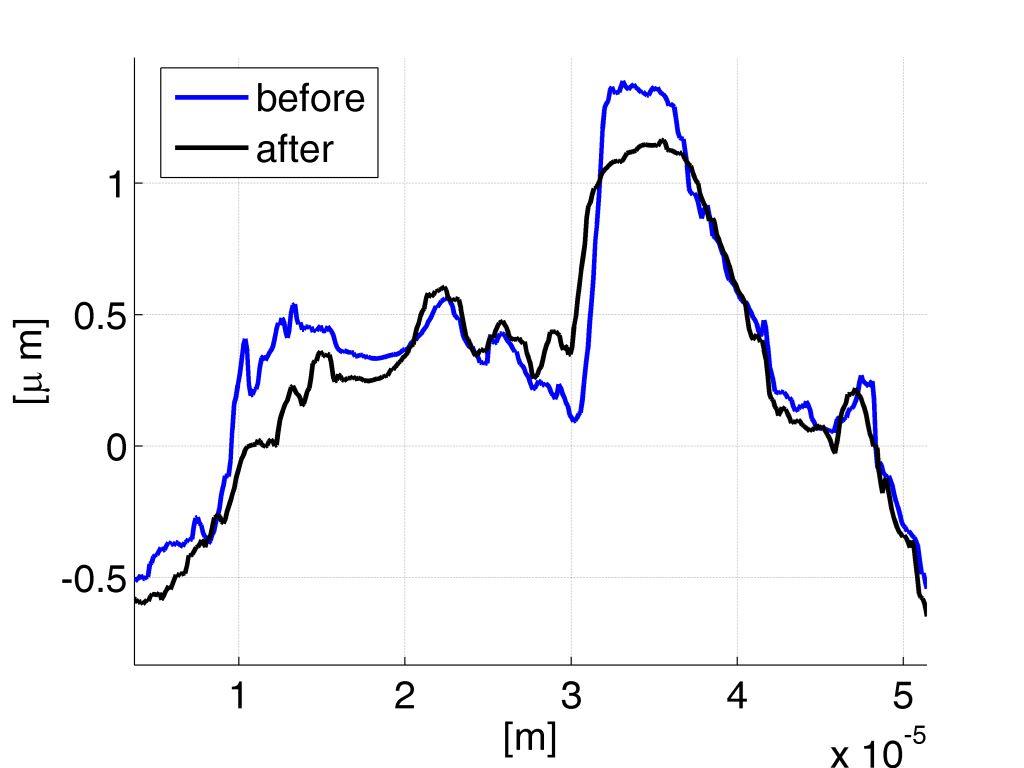}
\caption{Section of the surfaces along a line that highlights the differences.}
\label{fig:sect}
\end{figure}

First of all, the sensor noise is estimated by repeating the profile measurement 5 times. The comparison of the surfaces before and after the contact requires that the same analysis is performed on the same patch. Unfortunately, there are no reference points on the surface that allows an automatic positioning. This makes a post-processing matching mandatory. The script and a few details are available in the Appendix. The main idea is to minimize the root mean square of the difference of the bottom parts (i.e. valleys) of the two surfaces. The idea is that the valleys should be less affected by the contact. Fig.~\ref{fig:diff} shows the difference between the two surfaces after the matching procedure. Fig.~\ref{fig:sect} shows the two surfaces in a section that highlights the differences on the top of the mountain at the TMmu center. It is worth highlighting that a filter is needed in order to visually understand the shape change. A 2D median filter is chosen. Each output value contains the median value in the $m$-by-$n$ neighborhood around the corresponding value in the input image. $m$ and $n$ have been chosen equal and with value 10, 20 and 40. This type of filter is particularly useful to mitigate the effect of outliers. An analysis of the profilometer performance shows that the standard deviation produced has a low value (about 40 nm) that becomes much higher (about 500 nm) in certain positions where the illumination and/or the surface local orientation produce a detrimental effect.

The shape of the surface difference, Fig.~\ref{fig:diff}, shows a behaviour that is qualitatively in agreement with a contact with a nominal 0.01 m hemisphere and shows that a shape change occurred, which means the contact locally entered the plastic domain.

The non negligible difference of the two surfaces is demonstrated by the results of Tab.~\ref{tab:surfstd} as well. The two distributions of the surface coordinates have different standard deviation. Such a difference is confirmed at various median filter levels. 

\begin{table}[!ht]
\small
\renewcommand{\arraystretch}{1}
\caption{Surfaces Standard Deviation and Noise}
\label{tab:surfstd}
\centering
\begin{tabular}{c|c|c|c}
\hline
\textbf{Median filter} & \textbf{STD [$\mu m$]} & \textbf{STD [$\mu m$]} & \textbf{Noise [$\mu m$]} \\
\textbf{dimensions} & \textbf{before} & \textbf{after} & \textbf{} \\
\hline
0 & 0.433 & 0.387  & 0.079\\
\hline
10 & 0.403 & 0.369  & 0.074\\
\hline
20 & 0.377 & 0.346  & 0.072\\
\hline
40 & 0.331 & 0.308  & 0.071\\
\hline
\end{tabular}
\end{table}

Lastly, the two distributions do not belong to the same population also according to standard hypothesis test, such as the Kolmogorov-Smirnov, that fails for every filter dimension. The test fails even assuming the mean value is the same, which is unlikely.

This observation indicates, once again, that a plastic deformation happened on the TMmu and will likely happen in-flight. As a consequence, the maximum contact load is effectively a key parameter and the JKR/DMT theories are not sufficient in modeling adhesion.

\begin{figure}
\includegraphics[width=\columnwidth]{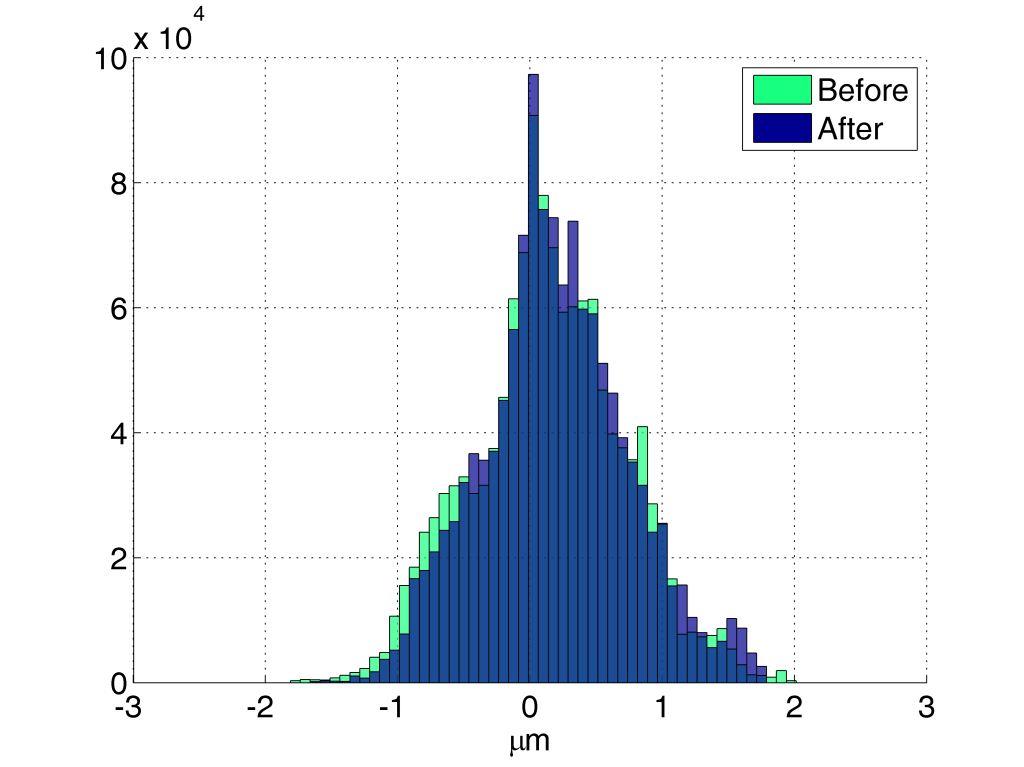}
\caption{Surface coordinates distributions. The number of points is very high (1286575), thus even a small visual difference determines a failure of the hypothesis test.}
\label{fig:dist}
\end{figure}

\section{Estimation of Analytical Force-to-Elongation Functions}

A very rough estimation of the risk of the measured level of adhesion is based on the pull-off force values obtained in the TMMF. The momentum transfer can be derived, assuming a nearly triangular acceleration time profile, by using the formula:
\begin{equation}
	P_x = \frac{1}{2}F_{peak}\Delta t
	\label{eq:approxmom}
\end{equation}
where $\Delta t$ is the release event time length. The formula works reasonably well in our experiments (Fig.~\ref{fig:approxmom}). As $\Delta t$ is expected to be of 200 $\mu s$ with the GPRM (time needed by the tip to travel about 7 $\mu m$), the requirement of $5 \mu m/s$ requires that $F_{peak} < 100 mN$, a number well above the maximum measured value of ~50 mN. From these quantities a rough estimate of 2.4 $\mu m/s$ worst-case residual velocity is obtained. 

A finer extrapolation requires the knowledge of the adhesion force-to-elongation functions. Such functions can be derived both from the test campaigns with low and high initial load. The focus will be on the first one and on the intermediate TM campaign (0.089 kg) whose results are worst-case. The analytical approach is also needed in order to run comprehensive simulations.

\begin{figure}[!ht]
\includegraphics[width=\columnwidth]{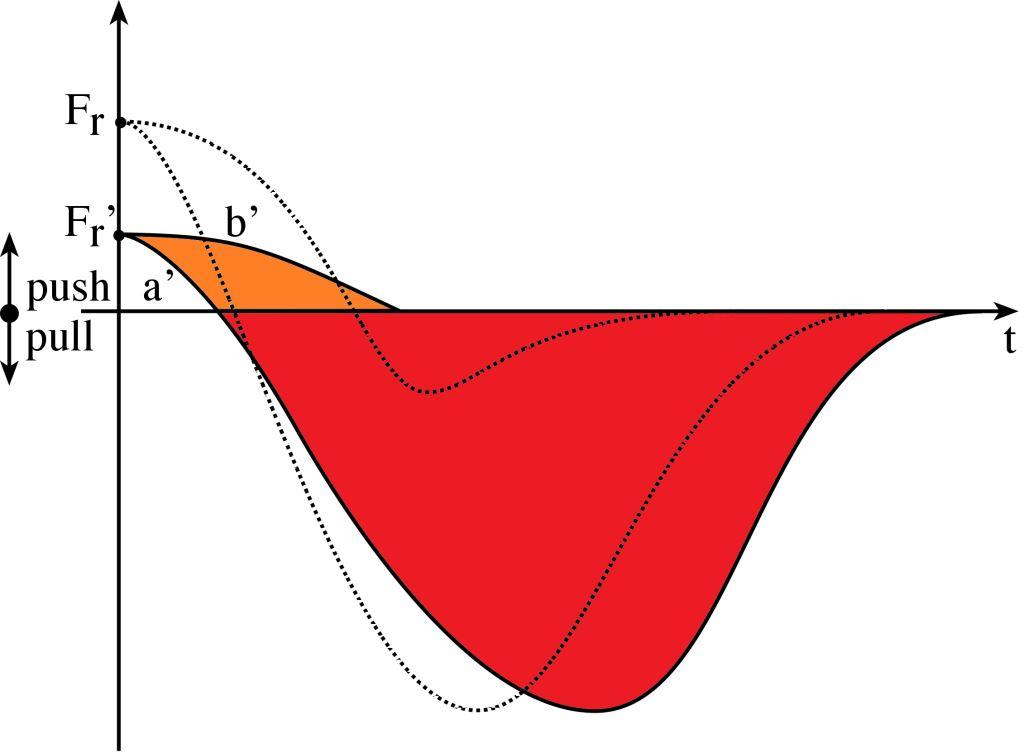}
\caption{Contact forces between TM, release tip (\textit{a'}) and blocking needles (\textit{b'}) in the TMMF. Dotted lines represent force profiles in the in-flight release (Fig.~\ref{fig3})}
\label{fig9}
\end{figure}

The contact force profile, Fig.~\ref{fig9}, shows a peak that is ruled by the maximum contact load experienced \cite{Gane}, that is a constant in all the experiments. After the peak the force tends to zero in a larger timescale. The blocking needles instead work like a non-actuated RT (line \textit{b'}): they initially balance the load applied by the RT mock-up \textit{a'} and then they push the TM away with a cosine profile like a preloaded oscillator. Their contribution to the momentum transfer (nearly equal to the orange area in Fig.~\ref{fig9}) can be calculated with Eq.~\ref{eq:catapult}

Clearly, the transferred momentum in the testing configuration (equal to the sum of the 2 red and orange areas in Fig.~\ref{fig9}) differs from the in-flight one. However, the force-to-elongation profile of curve \textit{a'} (Fig.~\ref{fig9}) in the testing configuration can be estimated and included in the mathematical model of the in-flight release, in order to predict the effect of the contact force in the actual release.

\begin{figure}[!ht]
\includegraphics[width=\columnwidth]{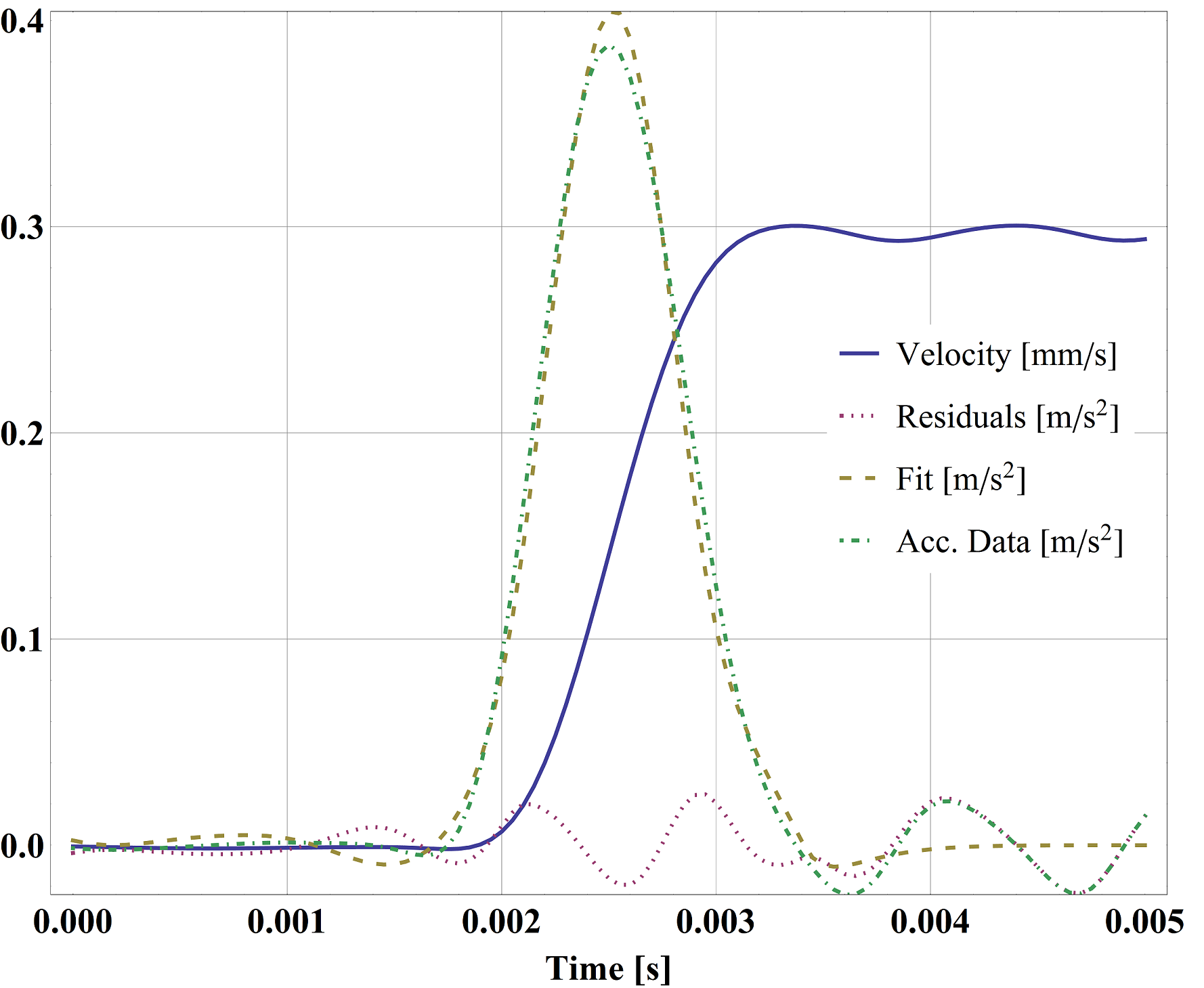}
\caption{TM measured velocity, acceleration, best fit and residuals of the adhesion estimation procedure.}
\label{fig10}
\end{figure}

In the experimental configuration, the TM is accelerated by the force exerted by the blocking needles, proportional to the TM displacement, and by adhesion, which is described by an exponential function of the elongation of the contact \cite{MSSP13}, i.e. the difference between the tip and TM displacement signals. 
The adhesion mathematical model has the following shape:
\begin{equation}
F_{adh} = A_1 (\Delta l) e^{-\frac{\Delta l}{\lambda}}
\label{eqadhX1X2}
\end{equation}

where $\Delta l$ is the elongation, $A_1$ the initial stiffness and $\lambda$ the elongation constant. The fitting model, which considers also a possible adhesion on the needles, is then:
\begin{equation}
\begin{split}
a_{mod} = \omega^2_{nd} \delta x(t) e^{-\frac{\delta x(t) \text{unitstep}(\delta x(t))}{\lambda_{nd}}} & \\
+ \omega^2_{Tmu} \delta y(t) e^{-\frac{\delta y(t)\text{unitstep}(\delta y(t))}{\lambda_{Tmu}}} &
\end{split}
\label{eqadh}
\end{equation}
with:
\begin{equation}
\delta x(t) = x_{TM}(t)-x_{0,TM}
\end{equation}
and
\begin{equation}
\delta y(t) = x_{Tmu}(t)-\frac{\omega^2_{nd} x_{0,TM}}{\omega^2_{T}}
\end{equation}
where $a_{mod}$ is the modeled acceleration, $\omega^2_{nd}$ is the contact stiffness between needles and TM normalized with respect to $M_{TMmu}$, $x_{TM}$ is the TM mock-up displacement, $x_{0,TM}$ its initial position (proportional to the residual load), $\lambda_{nd}$ is the adhesion elongation constant at the needles, $\omega^2_{Tmu}$ is the normalized contact stiffness between the tip mock-up and TM, $x_{Tmu}$ is the difference between the displacements of the experimental RT and TM (i.e. adhesion elongation) and $\lambda_{Tmu}$ is the elongation constant of adhesion between TM and tip mock-up. Eq.~\ref{eqadh} yields the non-linear fitting model of the TM acceleration signal, where $x_{Tmu}$ is the measured input to the system. The TM acceleration signal is calculated by means of a finite-difference Euler method applied to the sampled displacement signal after the application of a low-pass Blackman filter with a 2.5 kHz cutoff frequency. Fig.~\ref{fig10} shows the measured and the fitted acceleration profiles, together with the fit residuals.

\begin{figure}[!ht]
\includegraphics[width=\columnwidth]{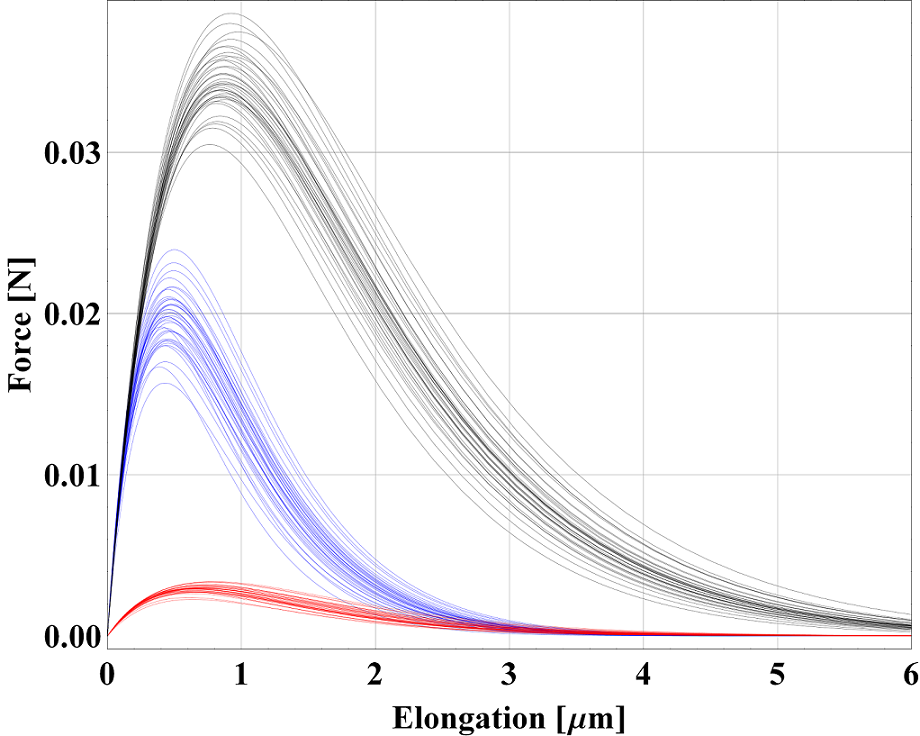}
\caption{Adhesion force-to-elongation curves estimated from 3 best-fits and their covariance matrix with uncertainty ($\pm \sigma$). More than 40 are available.}
\label{fig11}
\end{figure}

The negligible $\lambda_{nd}$ best fit parameter excludes that a measurable adhesion arises between the needles and the TM mock-up frame, thanks to the limited contact surface and the presence of an anti-adhesive coating (CrN). Conversely, $\lambda_{Tmu}$ and $\omega^2_{Tmu}$ best fit parameters yield for each test the adhesion force-to-elongation profile. A selection of these curves is depicted in Fig.~\ref{fig11}, which shows the low repeatability of adhesion produced at the contact between the rough gold-dental gold  surfaces. It is worth underlying that the limited repeatability is not determined by the accuracy of the estimation procedure, but is dominated by the reduced repeatability of the phenomenon.

The main reason for the high dispersion of the adhesion profiles is related to the fact that it is ruled by microscopic properties of the contact surfaces \cite{Maugis}. The surfaces involved in this contact are engineered to withstand the contact pressure, and an ultra-smooth surface finish is avoided because it would enhance adhesion \cite{FullerTabor, Israelachvili}. However, the surface roughness produces a variation of the local properties from contact to contact, because a different portion of the surfaces is involved each time. 

In the experimental tests with low residual load, the behaviour of the contact forces between RT and TM is mainly explored in the pulling regime (adhesion) and just a low initial force $F_{r'}$ is left before the retraction in order to limit the static push of the blocking needles. This means that the estimated force to elongation curves need to be extended to the compressive regime from 10 mN up to 350 mN ($F_r'$ and $F_r$ respectively, Fig.~\ref{fig9}). Such an extension is performed according to Hertz theory \cite{Maugis} the two opposed contacts are therefore expected to behave differently mainly if the radii of curvature of the RTs are different. Such a radius has a nominal value of 10 mm (Fig.~\ref{fig:GPRMcut}) with a tolerance ISO 2768-f, that is $\pm 1$ mm. This tolerance value is associated with the $3\sigma$ of a normal distribution from which the curvature radius is extracted.

Performing a qualitative comparison between the dispersion of the RT retraction profiles in Fig.~\ref{fig:retractions} and of the adhesion forces plotted in Fig.~\ref{fig11}, it is expected that the limited repeatability of adhesion dominates the expected dispersion of momentum transfer contribution produced by pulling forces (red area of Fig.~\ref{fig3}). At the same time, the dispersion of the RT retraction profiles dominates the dispersion of the momentum transfer contribution of pushing forces (orange area of Fig.~\ref{fig3}), which results much more repeatable.


\chapter{Simulations and Discussion}

This chapter derives the results of the simulations of the release, based on the data of the previous chapters. Such results estimate the performance of the GPRM. Moreover, qualitative guidelines for improving such performance are eventually provided.

\section{Simulation of the Release and Results}

The mathematical model of the electro-mechanical dynamics of the release mechanism (Section \ref{sec:RTmodel}) is completed by including the adhesion force profiles presented in Section \ref{sec:Adhesion} and the equation of motion of the TM.

In order to take into account all the model uncertainties, a large number of releases must be simulated by extracting each model parameter from an appropriate probability density function. The sources of uncertainty are grouped: those related to the mechanism, to the system initial condition and to the adhesion force. Each source is here discussed in order to clarify the overall budget affecting the Montecarlo simulation.
\begin{enumerate}
	\item The GPRM release mechanism is modeled with equations \ref{eqel}, \ref{eqMg} and \ref{eqpiez}. In order to simulate a release phase, two sets of parameters are extracted from Gaussian distributions defined according to their estimated covariance matrix, in order to describe the two opposed release mechanisms. During the retraction, the force profile on both sides also depends on the actual radii of curvature of the RTs, which are extracted from two independent normal distributions according to the end of Section \ref{sec:Adhesion}. 
	
Finally, a normal distribution is used to describe the statistic behaviour of the slope of the command voltage time profile. Such a distribution is defined according to the estimated variance as described in Section \ref{sec:RTmodel}.
	\item The relevant initial condition of the system at the release is described by the compressive preload exerted by the RTs on the TM. As described in Section \ref{sec:Caging}, during the passover procedure from the plungers to the RTs, an average contact force of 300~mN is exerted, which is substantially unrelaxed and present as initial condition at the release. This makes the time profile of the pushing contact forces different from what tested on ground (branch 1-2 of the curves plotted in Fig.~\ref{fig3} and Fig.~\ref{fig9}), while the pulling forces due to adhesion are assumed unaffected. The accuracy of the force control loop allows us to estimate that the contact force before the release is normally distributed with a standard deviation equal to 16.6 mN. 
	\item Random adhesion force profiles to be simulated at the two contacts are generated according to the experimental results discussed in Section \ref{sec:Adhesion}. 
	\item The misalignment of the direction of retraction of the RT with respect to the orthogonal to the TM surface is neglected. On one side, this quantity is limited thanks to design requirements. On the other side, it produces shear stress on adhesion patch that reduces its strength and its criticality. The assumption is therefore conservative.
\end{enumerate}

The combination of these effects is modeled by a set of 7 differential equations: 3 for each release mechanism and the TM equation of motion along the direction of retraction of the RTs:
\begin{equation}
\begin{split}
M & \ddot{x}(t) = \\
	&
	\begin{cases}
	F_{adh}\left(x_{RT,1}(t)-x(t)\right), & \text{if } x_{RT,1}(t)-x(t) > 0 \\
	F_{hertz}\left(x_{RT,1}(t)-x(t)\right), & \text{if } x_{RT,1}(t)-x(t) \leq 0
	\end{cases}
	\\
	+ &
	\begin{cases}
	F_{adh}\left(x(t)-x_{RT,2}(t)\right), & \text{if } x(t)-x_{RT,2}(t) > 0 \\
	F_{hertz}\left(x(t)-x_{RT,2}(t)\right), & \text{if } x(t)-x_{RT,2}(t) \leq 0
	\end{cases}
\end{split}
\end{equation}
where $M$ is the TM mass (1.96~kg) and $x_{T,i}$ is the i-th RT retraction. It is worth noting that adhesion and Hertz forces in Eq.~\ref{eqpiez} are represented by the quantity $F_c$. Adhesion force is described by the semi-empirical model of Eq.~\ref{eqadhX1X2}, where $\Delta l = x_{RT,1}(t)-x(t)$.

\begin{figure}[!ht]
\includegraphics[width=\columnwidth]{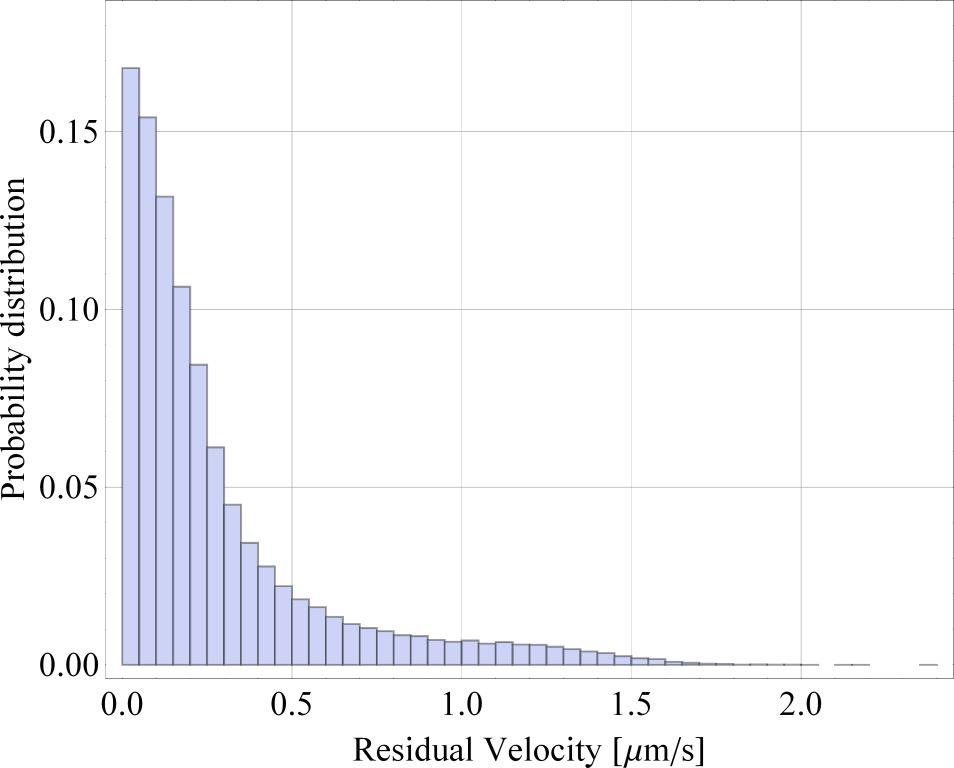}
\caption{Release velocity probability distribution obtained from Montecarlo simulations.}
\label{fig:MontecarloTot}
\end{figure}

100000 simulations are run with this model. Fig.~\ref{fig:MontecarloTot} shows the distribution of the release velocities obtained. 99.73~\% of the simulations (i.e. $3\sigma$ for a normal distribution) are below 1.50~$\mu$m/s where the requirement is 5.

\begin{figure}[!ht]
\includegraphics[width=\columnwidth]{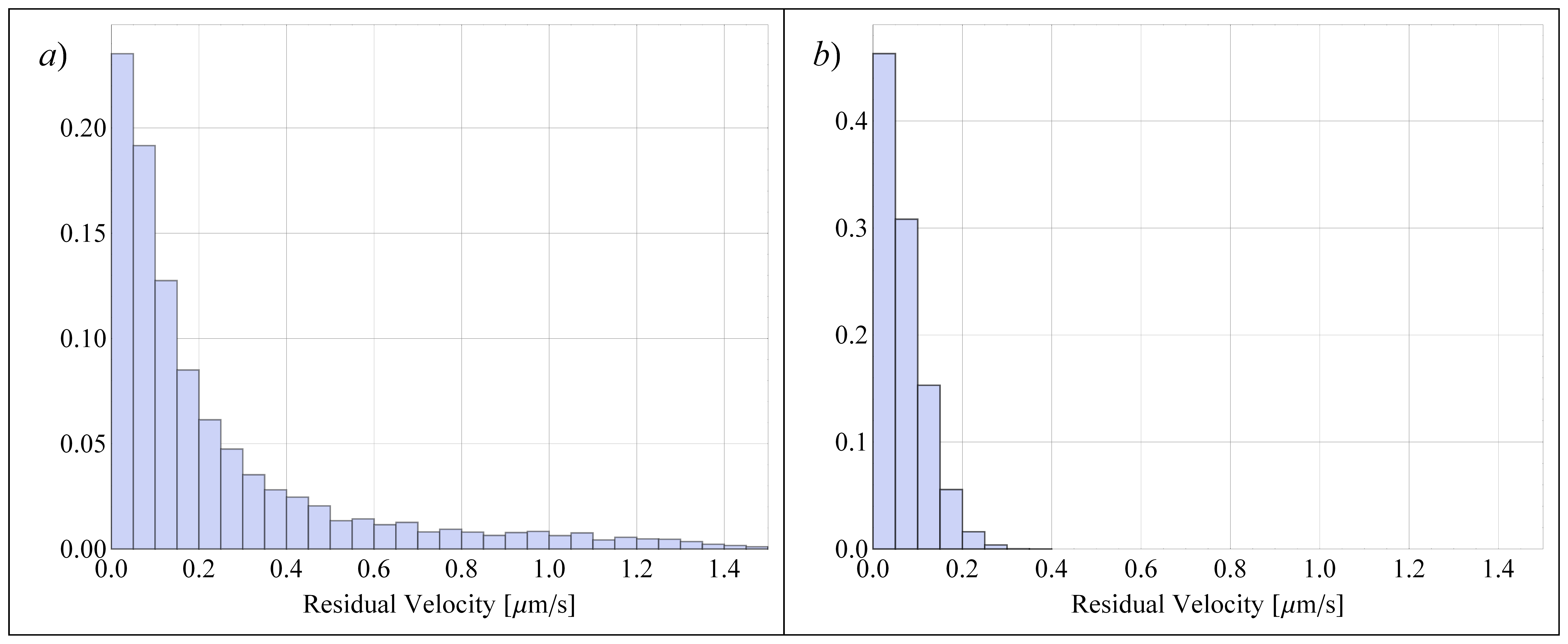}
\caption{Release velocity distributions. (a) adhesion effect (b) other effects (mechanism and initial load)}
\label{fig:Montecarlo2}
\end{figure}

This overall distribution is the combined result of several effects. Fig.~\ref{fig:Montecarlo2}a shows the contribution of random adhesion with 300~mN residual load and equal nominal mechanisms. The 99.73~\% velocity is 1.36~$\mu$m/s. Meanwhile, Fig.~\ref{fig:Montecarlo2}b shows the TM velocity with random mechanisms and preload with the same nominal adhesion at the two RTs. The $3\sigma$ velocity is here 0.24~$\mu$m/s. These results suggest that the TM momentum is mostly due to adhesion. 

Although this analysis is more advanced and complete, the preliminary in-flight predictions presented in \cite{Bortoluzzi2013} and \cite{ESMATS13} are here confirmed.

\begin{figure}[!ht]
\centering
\includegraphics[width=0.7\columnwidth]{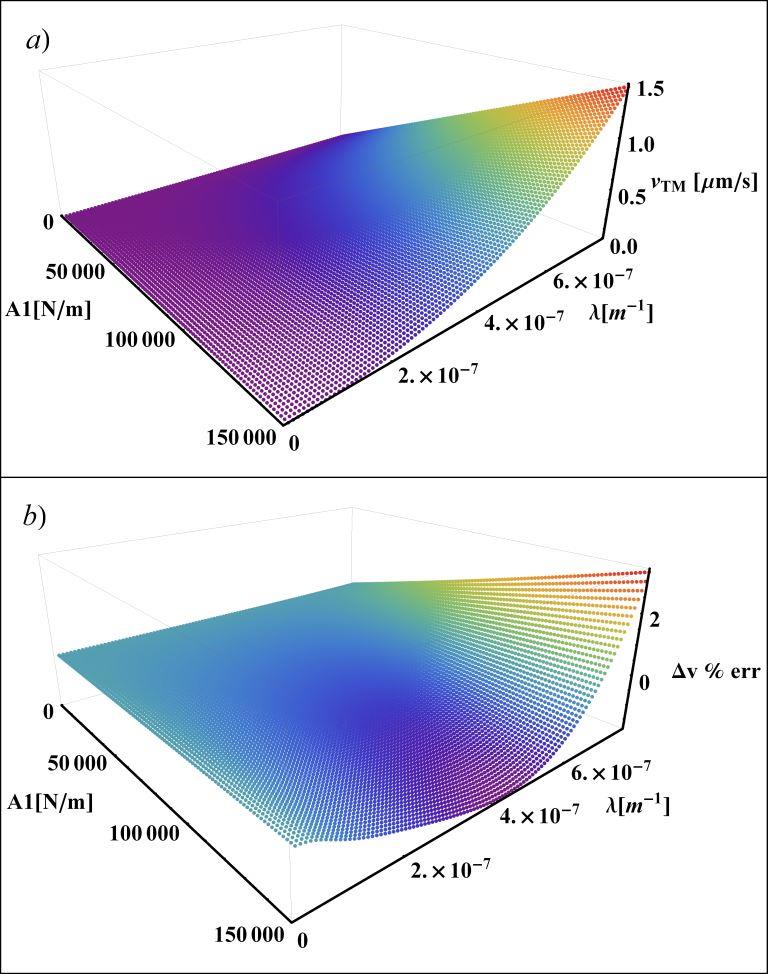}
\caption{a) Simulated release velocity as a function of $A_1$ and $\lambda$, b) best fit error.}
\label{fig:X1X2}
\end{figure}

The effect of the adhesion parameters $A_1$ and $\lambda$ of Eq.~\ref{eqadhX1X2} (that are proportional to the stiffness and the elongation of the adhesive bond) is evaluated by means of a set of simulations with adhesion active on only one TM side and no other asymmetry effects. In these simulations $A_1$ ranges between 0 and $1.5 \times 10^{5}\:N/m$ and $\lambda$ between 0 and $8\times10^{-7}\:m$, that are the boundaries experienced in experimental data. The simulated release velocity is shown in the 3D plot of Fig.~\ref{fig:X1X2}a as a function of the two adhesion parameters.

The energy of the adhesive bond can be easily obtained integrating Eq.~\ref{eqadhX1X2}:
\begin{equation}
	\Delta U = A_1 \lambda^2
\end{equation}

In the simplified model of the release dynamics presented in \cite{JAMBenedetti} it is shown that energy, $\Delta U$, and the TM final velocity are proportional, with a proportionality factor that depends on the RT retraction velocity in unloaded conditions, $v_{RT}$:
\begin{equation}
	v_{TM} = \frac{\Delta U}{M v_{RT}}
	\label{velpred}
\end{equation}
The adhesion energy is substituted in Eq.~\ref{velpred} and then fitted into the simulated release velocities (Fig.~\ref{fig:X1X2}a). Assuming the tip velocity as a fit parameter, its optimal value results in $v_{RT} = 34.3$~mm/s (residuals below 2.5 \% of the maximum, Fig.~\ref{fig:X1X2}b). The best fit parameter is compatible with the unloaded RT velocities measured on-ground, that have mean value 33.9 mm/s and standard deviation 4.7 mm/s. This confirms the capability of Eq.~\ref{velpred} in estimating the adhesion contribution to the final TM velocity.

\section{Lesson Learned}

Several parameters influence the performance of the GPRM in releasing the TM. Indeed, the most critical is the load experienced between TM and RT. A substantial reduction of the residual load before the release would mitigate any effect due to asymmetry of the two mechanism. At the same time, a strategy in which the RT detaches and in a short time re-grabs the TM would reduce the maximum load experienced at the contact patch and also adhesion. 

The synchronization between the two mechanism is not a requirement in LISA-PF. Although it is not the main drive of momentum transfer, its effect can still be further minimized. Besides, as this quantity is easily testable, a proper test campaign would reduce the uncertainty that still partially covers the amount of asymmetry.

At the same time, the overall release would greatly benefit by approaching the release in a more system-like way, i.e. considering it as a unique phase instead of an interface between two subsystems (GPRM and DFACS). A concurrent design of the two systems, also in light of the challenging release would greatly improve the performance or allow better science parameters (e.g. larger gaps). For instance, the axis of retraction -z- is the most critical (i.e. larger disturbance, \cite{PaulThesis}), but it is also the one with less control authority.

One parameter that could improve the performance without a deep redesign of mechanism, control system or central control unit is the tip radius. As a first order approximation, adhesion is proportional to the contact area. In the Hertz model \cite{Maugis}, the radius $a$ of the contact area and the force $F$ are:
\begin{equation}
	a = \frac{3^{1/3} \sqrt[3]{\frac{F R}{E_m}}}{2^{2/3}} 
\end{equation}
\begin{equation}
 F=\frac{4}{3} d^{3/2} {E_m} \sqrt{R}
\end{equation}	
where $R$ is the radius of the hemisphere in contact, $d$ the indentation and $E_m$ the equivalent elastic modulus. A larger $R$ means a larger $a$ and contact area. At the same time a small $R$ increases the indentation $d$. The larger the contact area, the larger will be adhesion. Conversely, a larger indentation means a longer contact time and an enhancement of the effects of asymmetrical mechanisms actions. A trade-off value can therefore be found and it can be shown that a radius of about 1 mm would provide much better worst-case performance than the current one (10 mm).

\begin{figure}[!ht]
\includegraphics[width=0.85\columnwidth]{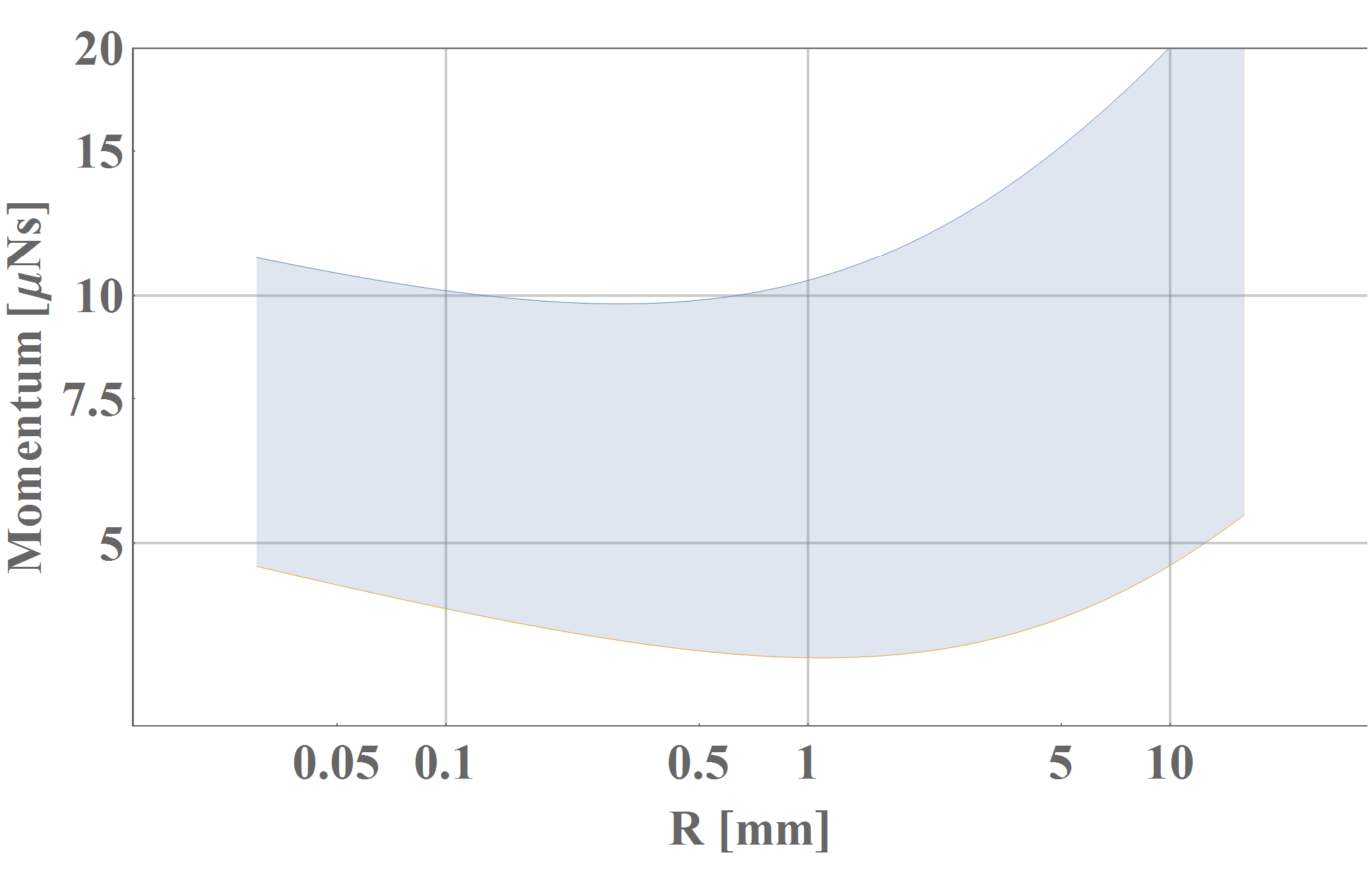}%
\caption{\label{fig:radius}Estimation of the worst case velocity as a function of the tip radius. The two curves are an upper and bottom estimation of the worst-case.}
\end{figure}

Fig.~\ref{fig:radius} shows a range of worst-case transferred momentii. The pushing (catapult) effect is obtained by multiplying a force for the time needed by one mechanism to exit the contact (the other one is assumed being much faster). This time depends on the hertzian indentation and the jerk of the RT motion. Adhesion effect is due to the pull-off force multiplied by a time, where the pull-off is equivalent to the gold rupture strength on all the contact area. The time is estimated with an adhesion elongation and a constant RT velocity. Thus the RT is assumed moving with constant jerk in the first tens of nm and with constant velocity in the $\mu m$ range.

Two effects determine the contribution of the push:
\begin{enumerate}
	\item the synchronization of the mechanisms motions.
	\item the symmetry of the surface properties in the contact patch.
\end{enumerate}
In LISA-Pathfinder, no requirement and neither test campaigns are defined for these two features. If such a specification is made available for eLISA, an optimal or quasi-optimal radius can then be determined.

However, in order to put some numbers in the idea, the following values are used along with very simplified models:
\begin{itemize}
	\item bottom limit estimation: contact force = 0.2 N, RT constant velocity = 180 mm/s
	\item upper limit estimation: contact force = 0.4 N, RT constant velocity = 45 mm/s
\end{itemize}
Jerk, Young's modulii and mass are kept always constant. 

Fig.~\ref{fig:radius} shows that in most cases the current value of 10 mm is large. 10 mm would be optimal if the momentum due to the asymmetry of the mechanisms is by far the large contribution. However, this would be alone a good news, because the estimations of this chapter show that -without adhesion- the residual velocity would be less than 1/10 of the requirement.

Finally, one effect that may help reduce the effect of adhesion is a shear stress in the contact patch. However, such an effect has to be considered carefully and tested on-ground. As a matter of facts, assuming the same force (50 mN) once in the direction normal to the TM surface and once in the tangential, it can be shown that it is more critical in the second case:

\begin{equation}
	\Delta \omega = \frac{\Delta t F a}{2 I} = \frac{0.0001s\:\:0.05N\:\:0.046m}{2\:\:0.000706kg m^2} = 330 \mu rad/s > 100 \mu rad/s
\end{equation}

\begin{equation}
	\Delta v = \frac{\Delta t F}{2 M} = \frac{0.0001s\:\:0.05N}{2\:\:1.96kg} = 2.5 \mu m/s < 5 \mu m/s
\end{equation}

Still, qualitatively speaking, the use of shear stress in the contact patch would help breaking adhesion. It would also have the effect of distributing the TM velocity along the 3-axis of the control system, while now there's no reason to assume significant velocities along $x$ and $y$ (the RT retraction is performed along $z$).



\chapter{After the Release: critical factors}\label{ch:after} 

The most critical part of the release is the mechanical action of the GPRM. However, the whole procedure cannot be declared a success until the TM is nominally centered in the EH with zero nominal velocity. Once the TM is set free of the mechanical contact, the DFACS provides actuation, as introduced in Sec.~\ref{DFACSintro}. A success operation of the DFACS depends both on its authority and on the amount of disturbance effects. Moreover, the release is critical because it defines the initial condition to the charge control system.

The following sections will address both topics in order to complete the picture of the TM release phases.  

The section on the DFACS is intended only as an overview and the punctual results are avoided\footnote{more can be found in the relevant Airbus Space notes both from the author of this work and from others}, except where strictly needed.

\section{Capturing the TM with the DFACS}

The success of the release depends on the mechanical actions described in the previous chapters, on the phenomena influencing the TM motion right after the RT retraction and on the worst-case initial states that the DFACS is able to control. The two latter topics are here over-viewed.

\begin{figure}[!ht]
\includegraphics[width=\columnwidth]{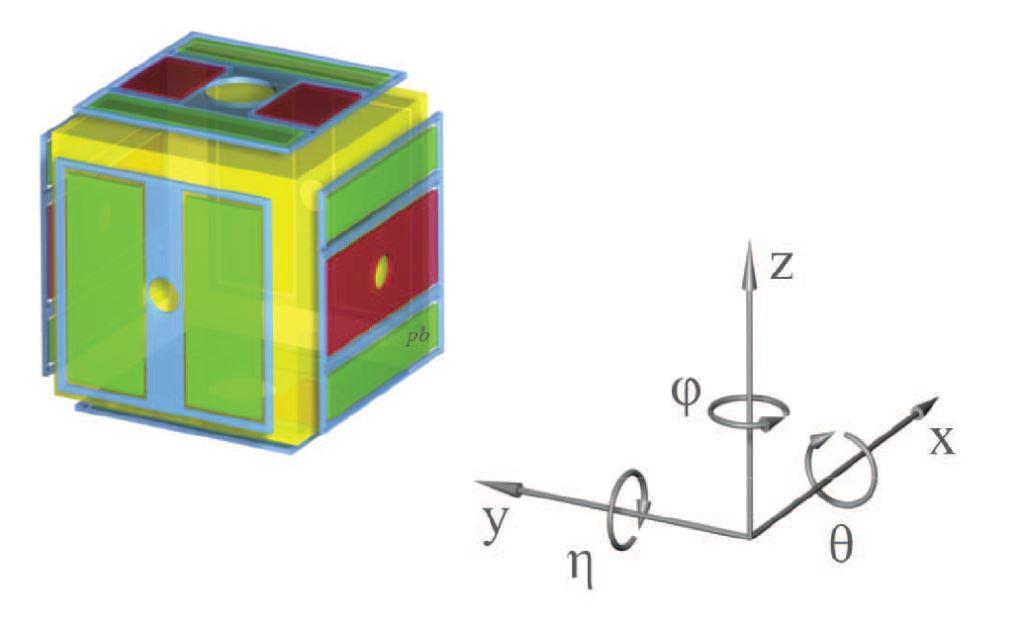}
\caption{Electrodes surrounding the TM and standard reference system (courtesy of Airbus Defence and Space).}
\label{fig:electrodes}
\end{figure}

The DFACS acts by means of the electrodes surrounding \cite{TMactuation} the TM and shown in Fig.~ref{fig:electrodes} where the GPRM axis is z.

As mentioned, the procedure after RT retraction (assumed here at t = 0) provides:
\begin{enumerate}
	\item t = 0+: RT retraction;
	\item 0 < t < 15 s: the Plungers move from the initial distance (equal to the Tip length, ~15 µm) to 500 µm;
	\item t = 15 s: start of observation and control;
	\item 15 s < t < 25 s: the Plungers remain at their position;
	\item t = 25 s: first estimation of the TM velocity with a less than 5 \% error;
	\item t > 25 s: if the velocities are assumed controllable the Plungers move in a safe position and the control keeps guiding the TM.
\end{enumerate}

\subsection{TM-Plunger voltage and its effect on the TM motion}
The general equation for the electrostatic force/torque on the TM, derived from standard energetic considerations of the electrostatic field of conductors, is also given in \cite{Nico, StiffnessM, ElAnalysis, HRose}. Including the Plungers as additional conductors, it becomes:

\begin{equation}
	\begin{split}
	F_{q} = & \frac{1}{2} \sum_{i=1}^{18} \frac{\partial C_{Eli,TM}}{\partial q}(V_i - V_{TM})^2 \\
	+ & \frac{1}{2} \sum_{i=1}^{18} \frac{\partial C_{Eli,H}}{\partial q}(V_{i})^2 \\
	+ & \frac{1}{2} \frac{\partial C_{TM,H}}{\partial q}(V_{TM})^2 \\
	+ & \frac{1}{2} \frac{\partial C_{Pl1,Pl2}}{\partial q}(V_{Pl1} - V_{Pl2})^2 \\
	+ & \frac{1}{2} \sum_{i=1}^2 \frac{\partial C_{Pli,TM}}{\partial q}(V_{Pli} - V_{TM})^2 \\
	+ & \frac{1}{2} \sum_{i=1}^{18} \sum_{j=1}^2  \frac{\partial C_{Pli,j}}{\partial q}(V_i - V_{Pli})^2 \\
	& \frac{1}{2} \sum_{i=1}^2 \frac{\partial C_{Pli,H}}{\partial q}(V_{Pli})^2
	\end{split}
\end{equation}

\begin{figure}[!ht]
\includegraphics[width=\columnwidth]{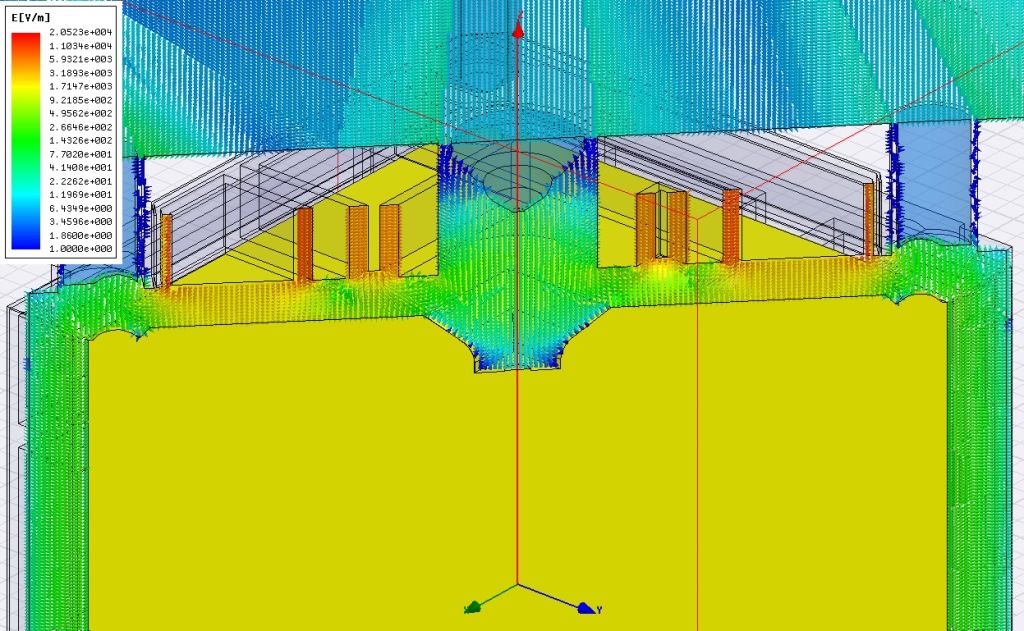}
\caption{Electrical field around the TM, with the Plungers retracted.}
\label{fig:E3}
\end{figure}

where the contributions are summarized in Tab.~\ref{tab:capacitances}.
\begin{table}[!h]
\small
\renewcommand{\arraystretch}{1}
\caption{Force Contributions}
\label{tab:capacitances}
\centering
\begin{tabular}{p{6cm}|p{6cm}}
$\frac{1}{2} \sum_{i=1}^{18} \frac{\partial C_{Eli,TM}}{\partial q}(V_i - V_{TM})^2$ & force contribution due to TM-electrodes effect \\
\hline
$\frac{1}{2} \sum_{i=1}^{18} \frac{\partial C_{Eli,H}}{\partial q}(V_{i})^2$ & force contribution due to housing-electrodes effect \\
\hline
$\frac{1}{2} \frac{\partial C_{TM,H}}{\partial q}(V_{TM})^2$ & force contribution due to TM-housing effect \\
\hline
$\frac{1}{2} \frac{\partial C_{Pl1,Pl2}}{\partial q}(V_{Pl1} - V_{Pl2})^2$ & force contribution due to plunger-plunger effect \\
\hline
$\frac{1}{2} \sum_{i=1}^2 \frac{\partial C_{Pli,TM}}{\partial q}(V_{Pli} - V_{TM})^2$ & force contribution due to TM-plunger effect \\
\hline
$\frac{1}{2} \sum_{i=1}^{18} \sum_{j=1}^2  \frac{\partial C_{Pli,j}}{\partial q}(V_i - V_{Pli})^2$ & force contribution due to plunger-electrodes effect \\
\hline
$\frac{1}{2} \sum_{i=1}^2 \frac{\partial C_{Pli,H}}{\partial q}(V_{Pli})^2$ & force contribution due to housing-plungers effect \\
\end{tabular}
\end{table}


Assuming that all but the TM voltages are zero, the previous formula is simplified in:

\begin{equation}
	\begin{split}
	F_{q} = & \frac{1}{2} \sum_{i=1}^{18} \frac{\partial C_{Eli,TM}}{\partial q}(V_i - V_{TM})^2 \\
	+ & \frac{1}{2} \frac{\partial C_{TM,H}}{\partial q}(V_{TM})^2 \\
	+ & \frac{1}{2} \sum_{i=1}^2 \frac{\partial C_{Pli,TM}}{\partial q}(V_{Pli} - V_{TM})^2 \\
	\end{split}
\end{equation}

In a previous activity \cite{ZnnAstrium}, the capacitance of both the Plungers have been analyzed and modeled with a Finite Element software\footnote{Ansoft MAXWELL}(Fig.~\ref{fig:E3} and fitted with proper functions. The FE analysis shows that in the first seconds after the release, when the plungers are close to the TM (relative distance < 500 $\mu m$) the contributions of housing and electrodes are negligible. It is worth reminding that the electrodes are not actuated for the first 15 s. The TM motion depends only on the release velocity and on the attraction due to the plungers. As the effect of the plungers is not included in the E2E\footnote{End-to-end} simulator, the estimation of the effect of such a disturbance has to be performed with an autonomous simulation. Worst-case results are obtained extending the simulation to all the first 25 s, that include also the first instants of sensing and weak actuation.

Of course, the plungers effect is not the only force acting on the TM. Other external disturbances are present, as shown in Tab. [omitted here on \textit{arXiv}, see \cite{PaulThesis}], where $a_{TM,DC}$ is the DC component of the disturbance, $a_{TM,max}$ is the peak value and $u_{max}$ the maximum actuation available.

\begin{figure}[!ht]
\includegraphics[width=\columnwidth]{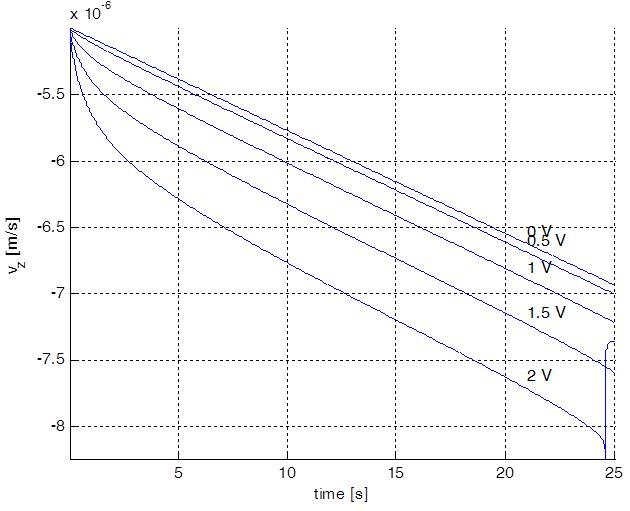}
\caption{TM velocity along z with different voltages. The 2 V line singularity is the effect of the contact and does not have physical sense.}
\label{fig:afterrelease}
\end{figure}

The worst-case increase in the TM velocity due to a TM-Plunger voltage is shown in Fig.~\ref{fig:afterrelease}. The initial TM state is assumed equal to the requirement.

According to \cite{TriboStefano}, the expected voltage is about 1 V.

\begin{figure}[!ht]
\includegraphics[width=\columnwidth]{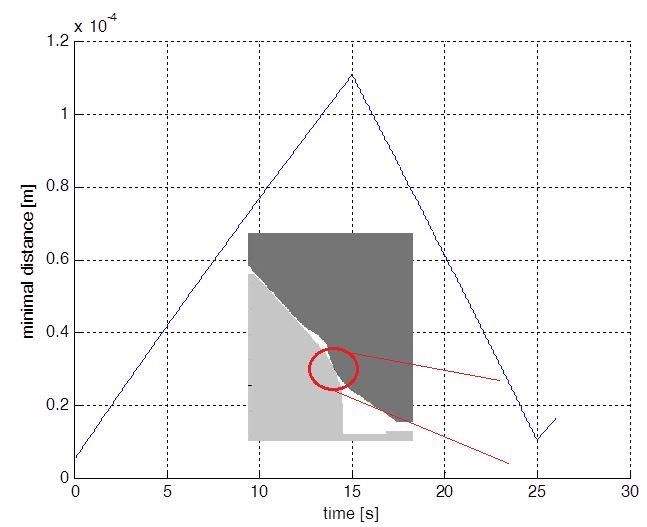}
\caption{Minimal distance between TM and Plunger estimation. The small picture shows the closest couple of points.}
\label{fig:mindistance}
\end{figure}

The worst-case analysis of the Plunger attraction on the TM shows that the Plunger effect is expected to be about 5 \% of the TM velocity, however it also shows that the Plunger may get very close to the TM (Fig.~\ref{fig:mindistance}). As a consequence of this concern, the retractions of the Plungers have been increased from the baseline of 500 $\mu m$ to $1800 \mu m - 2400 \mu m$ (the actual distance depends on step size). With such a procedure, the Plunger force on the TM is below 1 \% of the worst-case actuation after 4.3 s and the extra velocity is less than 0.1 $\mu m/s$. The worst-case actuation is that one in which the TM is as far away as allowed by the requirement from the EH center.

\subsection{DFACS worst-case initial conditions}

As in any space project, the subsystem requirement is established such that it has a margin at a system level. For example, the 5 $\mu m/s$ figure required in the design of the GPRM, becomes 7.5 $\mu m/s$ for DFACS design purposes.

Eventually, the real margin of the DFACS becomes a key parameter for the evaluation of the overall release performance. Three strategies have been followed to estimate the hyperspace of the initial conditions that the DFACS is able to deal with. All of them require the use of the E2E simulator \cite{E2Evitelli} with the hypothesis that it represents the actual mission behavior.


Three ways of describing the allowable initial conditions for the control system are explored. It is worth noting that the initial condition in this case is intended under an operational point of view. This means that these initial states are the states of the TM right after the mechanical release, not the states when the control system is effectively initialized. 
\begin{itemize}
	\item First of all, assuming that only one TM initial state differs from zero, the maximum value of that state is derived. This is repeated for each of the 12 states. The simulations show that the control system is able to control the TM even if its after-release velocity along the retraction axis z is equal to 12 $\mu m/s$, assuming that that is the only non-zero state of the TM. Although this approach may give a hint on the DFACS performance, it does not take into account the cross-coupling between the states. 
	\item In order to obtain a more comprehensive result, the maximum scale factor that it is possible to apply on the GPRM requirement is derived. This means that the hyper-volume defined by the requirement is scaled by a certain factor until a further increase would mean that a state determines a failed action of the DFACS.  Such a procedure estimates a scaling factor equal to 1.4.
	\item The final criterion derives a trade off between the previous two extreme cases and is built in such a way that one can use it to know \textsl{a priori} if a certain state will result in a successful release or not. Thus, the aim of this criterion is the definition of a dis-equation describing the set of allowable initial states. The scaling factor on the requirement is a possible criterion, but would be too much conservative. The desired dis-equation should take into account both the 1.4 margin factor and the fact that a single initial state is more easily controlled than all together.
According to this criteria, a full set of TM initial states can be controlled if:
\begin{equation}
	\sum_{i=1}^{12} \left( \frac{x_{0,i}}{x_{LIM,i}} \right)^4 < 1
\end{equation}
Where $x_{0,i}$ are the initial states and the values of $x_{LIM,i}$ are estimated looking at the results of the simulations with the E2E simulator. For internal coherence $x_{LIM,i}$ must comply with the following constraint:
\begin{equation}
	\sum_{i=1}^{12} \left( \frac{1.4\:req_{i}}{x_{LIM,i}} \right)^4 = 1
\end{equation}
\end{itemize}

All these three approaches are checked by running  a large number of simulations in the E2E simulator. For each of the descriptions of the initial states it is verified that all the initial states belonging to the relevant hyper-volume produce a captured TM. The initial states are generated both with a 12-dimensional grid, both randomly.

\section{Electrification of the TM during the Release}

One issue intimately linked to the release is the amount of charge that accumulates on the TM once it is freed from the mechanical constraints. A specific system \cite{Tobias} has been designed in order to control the TM charge. However, in case of failure of such a system, a re-grabbing is the only available way to bring charges into acceptable limits.

The TM will develop an initial charge immediately after the release due to the breaking of the contact. In facts, part of the charge remains trapped on each body as a consequence of a phenomenon called contact electrification \cite{Tribo}. This value is critical for both the TM dynamics (as seen in the previous section) and of course for the charge control.

The detailed description of the estimation of the initial TM charge and voltage is in \cite{TriboStefano}. The main points are here summarized. Besides, the terms included in the equations are here updated. The final result remains the same.

The charge that accumulates on two bodies in separation is well described by:
\begin{equation}
	Q_{TM}=C_{0}V_{0}
\end{equation}
where $C_0$ is the capacitance between the bodies at the very beginning of the separation (nominally, when the resistance of the contact tends to $\infty$) and $V_0$ is the potential difference at the same time.

As the TM release involves more than two bodies the exact relation is:
\begin{equation}
	Q_{TM}=C_{TM,EH}\Delta V_{EH}+C_{TM,Pyr}\Delta V_{Pyr}+C_{TM,Cyl}\Delta V_{Cyl}+C_{TM,RT}\Delta V_{RT}
\end{equation}
where $Pyr$ indicates the pyramidal Plunger, $Cyl$ the cylindrical and $\Delta V$ a voltage difference when mechanical contact with the TM is lost.
The same charge, when the Plunger are completely retracted becomes:
\begin{equation}
	Q_{TM}=C_{TM,EH}\Delta V_{EH}
\end{equation}

From this equation, the TM voltage can be estimated in:
\begin{equation}
	V_{TM}=(C_{TM,Pyr}\Delta V_{Pyr}+C_{TM,Cyl}\Delta V_{Cyl}+C_{TM,RT}\Delta V_{RT})/C_{TM,EH}
\end{equation}

where $C_{TM,Pyr}=35\:pF$, $C_{TM,Cyl}=10\:pF$, $C_{TM,RT}=2.5\:pF$ and $C_{TM,EH}=34.4\:pF$. The values of the Plunger and EH capacitances have been derived with FE analysis, $C_{TM,RT}$ is computed analytically between the RT nominal spherical cap and a plane. \cite{TriboStefano} indicates 14 pF for this quantity, but computes the total capacitance between a whole sphere and a plane. Here, only the real cap is considered and its value is estimated by:
\begin{equation}
\int_{0}^{\theta_f} \frac{\epsilon_0 R_{RT}^2 2 \pi sin(\theta)}{d+(R_{RT}-R_{RT} cos(\theta))} d\theta
\end{equation}
where $\epsilon_0$ is the vacuum permittivity, $R_{RT}$ is the radius of curvature of RT, $\theta_{f}$\footnote{$0.4\times10^{-3}$ is the radius of the release tip} is $atan(0.4\times10^{-3}/R_{RT})$ and $d$ is the mutual distance between RT and TM, assumed equal to 100 nm. 

\cite{TriboStefano} overestimates $C_{TM,RT}$, but also neglects one of the plungers (the cylindrical, that has a lower capacitance). The effect of both plungers is here considered. It is true that the very last contact will be only on one side, but still both plungers will be close to the TM.

Both bibliographical sources \cite{Tribo} and experimental data\footnote{performed at the University of Modena, unpublished.}, further analyzed in \cite{TriboStefano}, suggest that the Au-Au voltage difference is a random variable whose mean is zero and the 2$\sigma$ value is $\delta V=$1 V. Such an estimation is based on the work functions of the two metallic patches.

Propagating the voltage distribution in $Q_{TM}$ and $V_{TM}$, the following is obtained:
\begin{equation}
	Q_{TM} < \sqrt{C_{TM,EH}^2+C_{TM,Pyr}^2+C_{TM,Cyl}^2+C_{TM,RT}^2}\times \delta V \approx 3\times 10^8e
\end{equation}

\begin{equation}
	|V_{TM}| < \frac{\sqrt{C_{TM,Pyr}^2+C_{TM,Cyl}^2+C_{TM,RT}^2}}{C_{TM,EH}}\times \delta V \approx 1.1 V
\end{equation}

Cosmic rays provide continuous positive charging of the TM when the spacecraft is in orbit. Therefore, in case of failure of the charge system, a $-3\times 10^8e$ charge would mean that about 50 days are required to bring the TM voltage to zero. Instead, if the charge is positive, two month of operations with the allocated error budget would require at least two grabbing procedures and subsequent releases, \cite{TriboStefano}. 


\chapter{The Drag-Free CubeSat and its Housing}\label{ch:Stanford} 

This chapter describes the activities related to the development of the housing for the TM in the scope of a low-cost drag-free project conceived at Stanford University.

The Drag-Free CubeSat mission has been proposed to demonstrate the feasibility of a Gravitational Reference Sensor (GRS) with an optical readout for a 3 units (3U) CubeSat. The Drag-Free CubeSat is designed to shield a 25.4 mm spherical TM from the external non-gravitational forces and to minimize the effect of internal generated disturbances. Several of the disturbances are passively reduced by the design of the TM housing. 

The housing has an effect on the mechanical, thermal and magnetic environment around the TM. All of them have been analyzed. The mechanical vibrations have to fit the launch environment and the modes have to be outside of the measurement range ($10^{-4} – 1$ Hz). The magnetic field has to be reduced by a 0.01 factor from the outside to the inside. The temperature difference between internal opposing surfaces, determining pressure on the TM, has to be below $10^{-3}(1 mHz/f)^{1/3} K Hz^{-1/2}$.  

The housing, together with the TM, the sensors and the UV LEDs for charging control, constitutes the GRS, which would then fit into a 1U. The other 2Us are occupied by the caging mechanism that constraints the TM during launch, the thrusters, the Attitude Determination And Control System (ADACS) and the electronics. 

Next generation GRS technology for navigation, earth science, fundamental physics, and astrophysics has been under development at Stanford University since 2004. The Drag-Free CubeSat is a Stanford, University of Florida, KACST\footnote{King Abdulaziz City of Science and Technology} and NASA combined project. In 2014, it has been up-scoped to a Drag-Free mission within the limits of the KACST spacecraft bus. The design described here is referred to the initial CubeSat plan.

\section{Overview of the Project}

The heart of drag-free missions is the GRS, which measures the position of a free falling TM relative to the spacecraft. As mentioned in the Introduction, the TM position is used in a control system to command the thrusters and to constrain the spacecraft orbit to that of the TM. If the TM is freed of all forces but gravity, the hosting spacecraft also follows a purely geodesic orbit. The practical obstacle in obtaining a pure free-falling status is the presence of disturbances acting on the TM. Goal of the GRS is not only to measure the TM position, but also to shield and minimize these disturbances.

A CubeSat with the goal of pushing these technologies has been proposed by a team composed by Stanford, University of Florida, KACST, NASA and with international support. LISA-Pathfinder is going to become the state of the art in the drag-free field. Indeed then, in the scope of the Drag-Free CubeSat, the LISA-Pathfinder tight requirement is relaxed to $10^{-12}$ $m/s^{2} Hz^{-1/2}$ in the $10^{-4}-1$ Hz measurement range. The project has recently been upgraded to a full microsatellite hosted in a KACST bus. Nonetheless, this mission will be the first drag-free one with an optical readout and the first drag-free low cost project.

The performance requirement is a challenge for a spacecraft of reduced dimensions. The development of the Modular Gravitational Reference Sensor (MGRS), performed at Stanford since 2004 \cite{MGRSSun}, provides an invaluable background for this design. The primary components of the MGRS include a spherical TM, a LED based Differential Optical Shadow Sensor (DOSS), a caging mechanism based on the flight-proven DISCOS design \cite{Dan1976} and a charge control system. At the same time, one of the key systems of the GRS in shielding from the disturbances is the housing. This is a mechanical device designed to host the TM, physically hold the sensors, limit the effect of vibrations, passively reduce magnetic fields and thermal pressure and minimize the gravitational attraction induced by the spacecraft itself. 

In the next part of the thesis, an updated overview of the main planned components of the Drag-Free CubeSat is given. For a more detailed description of the mission concept the main reference is \cite{DFCubeSat}. The remaining part of the chapter is focused on the design of the housing with its performance and issues.

\subsection{CubeSat Main Components}

\begin{center}
\begin{figure}[ht!]
\caption{Open View of the CubeSat}
\label{fig:DFcubesat}
\includegraphics[width=1.1\columnwidth]{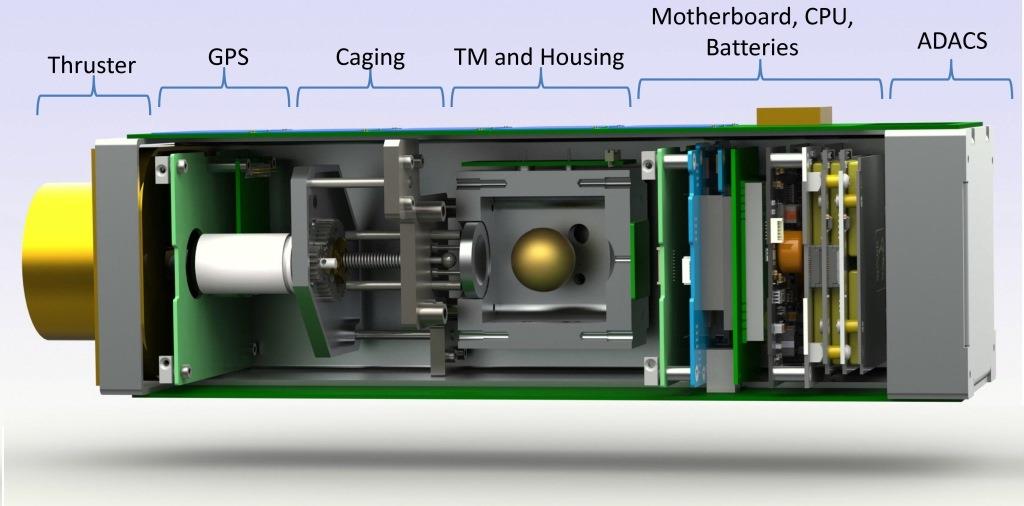}
\end{figure}
\end{center}

\begin{description}
	\item[Spacecraft Structure and Deployer:] The satellite (Fig.~\ref{fig:DFcubesat}) is a 3U CubeSat. All the systems are directly supported by the shell or by a Ti bulkhead normal to the satellite main axis. CubeSats'launch is almost always provided as a secondary payload in compliance with the Poly Picosatellite Orbital Deployer (P-POD) \cite{CubeSatRep} launcher (volume: $10\times10\times34$ cm, mass: < 4 kg).
	\item[Test Mass:] The TM is a sphere with 25.4 mm (1 inch) of diameter and 171 g of 70\%/30\% Au/Pt. This alloy is chosen because it is dense, can be machined, and has a low magnetic susceptibility \cite{DFCubeSat}. The TM surface is coated in SiC, which has high photo-emissivity, for charge control, high elastic modulus and hardness. It is hence unlikely to obtain large adhesion forces \cite{JKR1971}, such as those measured for LISA Pathfinder \cite{Bortoluzzi2013}, even after the high preloads required during launch.
	\item[Differential Optical Shadow Sensor:] The Differential Optical Shadow Sensor (DOSS) measures the position of the TM with respect to the housing. It consists of four sets of two LEDs, two photo diodes and the relevant electronic board. The four boards are mounted to the housing. A total of 8 beams are centered around the TM such that half of each beam is blocked when the TM is at its nominal position. When the TM moves with respect to the housing, an intensity change is detected with the photo diodes. In order to reduce common mode noise, the difference between the measurements of two diodes on opposite sides is taken. The design goal for the DOSS is a sensitivity of 1 nm at 1 mHz.
	\item[Caging System:] \cite{SUcage} The impact constant is defined as the product of launcher acceleration, unconstrained mass and gap length. The high value of this case (~0.002 kg m) makes the presence of a caging mechanism mandatory \cite{ESAobjectinjection}. This system (Fig.~\ref{fig:CagingSU}) is designed to restrain the TM during launch and release it afterward. It is loosely based on the design of the DISCOS system aboard the Triad satellite, which used a lead screw and plunger to hold the spherical TM \cite{Dan1976}. The TM is seated in a hemispherical recess on the interior wall of the housing. The system passively applies a force of at least 200 N to the TM, equivalent to the LISA requirement of 3000 N \cite{ESMATSThales}, scaled down by mass (1.96 kg to 0.171 kg).
\begin{center}
\begin{figure}[ht!]
\caption{Caging and housing (with a DOSS board). The ruler is in inches.}
\label{fig:CagingSU}
\includegraphics[width=1\columnwidth]{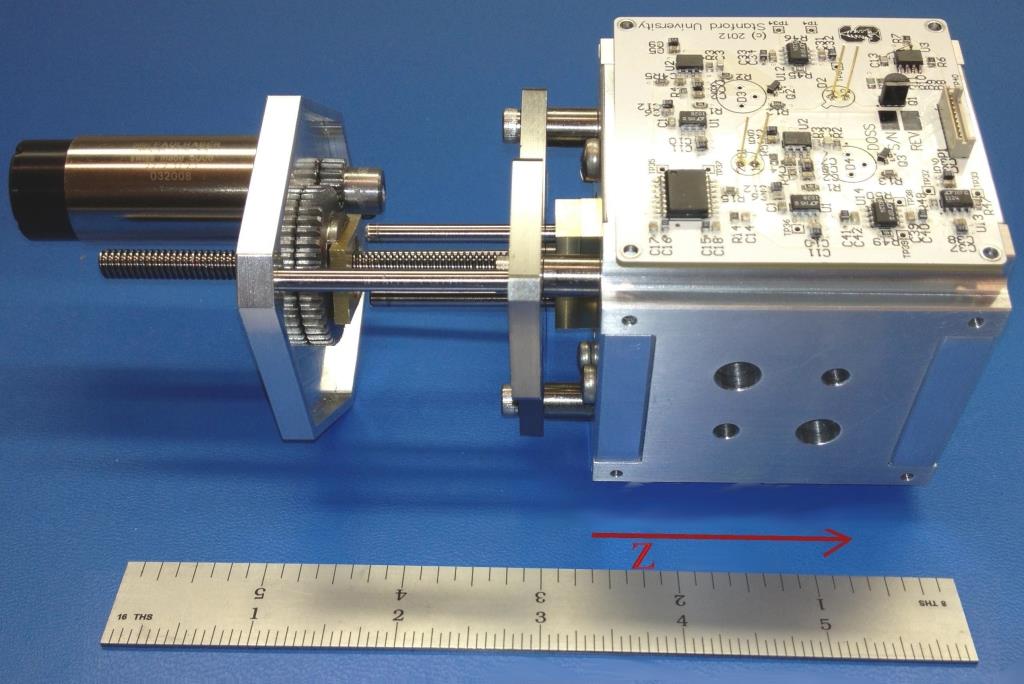}
\end{figure}
\end{center}	
The caging system is fixed to the spacecraft structure by a titanium bulkhead. Titanium is chosen because it is light and strong. A fine pitch 1/4-20 acme lead screw is driven by a DC motor through a bronze nut in the bulkhead. The total travel is approximately 25 mm, enough to move the plunger on the end of the lead screw from the locking position. Although the acme screw should prevent back-driving under most conditions, random vibration testing is needed to verify the preload remaining at or above the required load of 200 N while the caging system is unpowered.
	\item[UV Charge Control:] Charge imbalances between the housing and the TM caused by caging/uncaging as well the penetration of high energy particles leading to primary and secondary electron emission can cause an electrostatic disturbance force.  Typical charging rates for drag-free missions are on the order of 50 electrons per second \cite{charging}, with the exact rate depending on TM cross sectional area, thickness and composition of the housing, and orbit, among other factors.  TM charge control is achieved via UV photoemission.  Deep UV LEDs operating at 255 nm have been identified as an ideal candidate for charge control due to their small size, high dynamic range in power output, low power requirements, and the ability to be modulated at high frequency outside the drag-free control band \cite{LEDcharging}.  The TM is coated with SiC because of the material's relatively high quantum efficiency of 4.86 eV. Charge control is performed via bias-free charge control: UV light is shined onto the TM, and some light reflects back to the housing.  Photoelectrons are generated from both surfaces and the electrons will preferentially move such that the potential between the two surfaces reaches equilibrium.
	\item[Drag-Free and Attitude Control System:] The Drag-Free and Attitude Control System (DFACS) is the system that constraints the orbit of the spacecraft to that of the TM. 6 degrees of freedom (DOF) are sensed and 4 are actuated. The thruster applies its force along z. The actuation from a cold-gas thruster has to be an on/off one. With the on/off actuation, the motion of the satellite is a sequence of parabolic arcs ($~4\times10^{-4}$ m in amplitude). The limit of this algorithm is the minimum impulse bit of the thruster. A reduction of this parameter would allow better drag-free performance.
	
Indeed, atmospheric drag is not the only external disturbance force. On top of that, the total force will not be aligned with the direction of travel and attitude control is required in the compensation of these non-gravitational actions. Attitude actuation is a task of the Attitude Determination and Control System (ADACS) at the top (+z) of the satellite. Drag-free control in the transverse x and y directions is performed by adjusting the attitude such that the CubeSat opposes the drag force. The expected yaw and pitch angles are < 10 deg \cite{DFCubeSat}. 
 
Finally, as no differential measurement between test masses is used and the TM is a sphere, electrostatic actuation inside the housing is not required. This allows substantial savings in complexity, money and mass.
	\item[Micro-propulsion:] The preliminary requirements for the micro-thrusters are an impulse of between 10 mNs and 100 mNs with a precision of better than 1 mNs. As baseline the use of a micro propulsion cold gas thruster \cite{Propulsion,ColdGas} by VACCO is foreseen. This thruster has axial thrust capability only. A number of other options can be considered, including newer versions of the cold gas thrusters by VACCO and other companies, ionic fluid (also known as colloidal) thrusters and field emission thrusters.
\end{description}

\section{Preliminary Design of the Housing}
The mechanical part of the GRS is constituted by the housing, which is a box surrounding the TM. Its main functions are holding the sensors and other GRS systems and passively reduce the effects with a negative impact on the drag-free performance. Compared to LISA-Pathfinder, the GRS complexity is here reduced by the absence of electrostatic sensing and control inside the housing.

In order to comply with the drag-free performance goal, the housing has to: 
\begin{itemize}
	\item hold, together with the caging mechanism, the TM during launch;
	\item allow a safe release;
	\item	shield the TM thermally: maximum temperature difference between internal opposing surfaces below $10^{-3}$ $\left(\frac{1 mHz}{f}\right)^{1/3}$ $K$ $Hz^{-1/2}$;
	\item	shield the TM magnetically, with a 0.01 reduction factor;
	\item	hold the shadow sensors rigidly: the mechanical modes of the housing inside the spacecraft have to be clearly outside of the measurement range ($10^{-4} - 1$ Hz);
	\item	minimize the gravitational gradient;
	\item	make a re-caging possible;
	\item	hold the UV LEDs for charge control;
	\item	have a surface that allows charge control;
	\item	minimize patch effects;
	\item	fit both the DF CubeSat and the ShadowSat (the CubeSat for DOSS testing, \cite{DOSSCube,DOSS13}) with minimal re-design;
	\item	be machinable;
	\item	be vacuum compatible;
	\item	be easily assemblable for tests;
	\item	be compatible with the systems already designed (Caging, DOSS, CubeSatKit...).
\end{itemize}

Some of these requirements are guaranteed by deposing a coating inside the housing. More specifically:
\begin{enumerate}
	\item a safe release is allowed by minimizing the adhesion bonding between TM and housing. This means that the TM does not remain attached to the caging or the housing when the first one is retracted by a small amount. At the same time, the TM residual velocity is high after the release (it is an effect of the 200 N preload, as the catapult effect in LISA-Pathfinder). This velocity is then dissipated with a certain amount of bounces of the TM between the caging and the housing. Therefore, the geometry has also to survive these bounces and especially avoid small damages to the TM. The surfaces have then to be hard and chemically inert;
	\item charge control depends on the photo-emissivity of the TM and of the housing surfaces where the UV rays are reflected;
	\item patch effect is minimized by a short wavelength voltage variability on the inner surfaces.
\end{enumerate}

A material whose properties fit these guidelines is the SiC (Silicon Carbide) that covers the TM too. However, the use of a coating inside the housing imposes tight constraints on the geometry of the housing itself because the coating deposition process has to be taken into account. A uniform distribution is allowed only with an open geometry. The best would be the separation of all the sides in different pieces. On the other hand this would make the tolerancing and alignment much more critical.
It is then worth noting that the TM does not require position or attitude control in this case. At the same time, patch effect strongly depends on, and thus limited by, this gap. For these reasons, the housing dimensions can be and have to be larger than the TM radius allowing a big TM-to-housing gap. 
The minimization of adhesion force and charge control require the presence of coating only on the surface around the UV LEDs and where the TM is held during the launch. As a consequence, assuming that the housing is a cube, the surfaces primarily interested by the coating are only the two normal to the CubeSat main axis (z, in Fig.~\ref{fig:DFcubesat}). 

\subsection{Final Geometry}
It is highly preferable that the piece supporting the DOSS is not fractioned allowing better sensors alignment. A choice has to be made between uniform coating deposition on every surface and improved tolerancing for sensor alignment. Considering that the coating is not strictly required on the housing side surfaces, the priority is given to the sensor alignement.

\begin{figure}
\includegraphics[width=8cm]{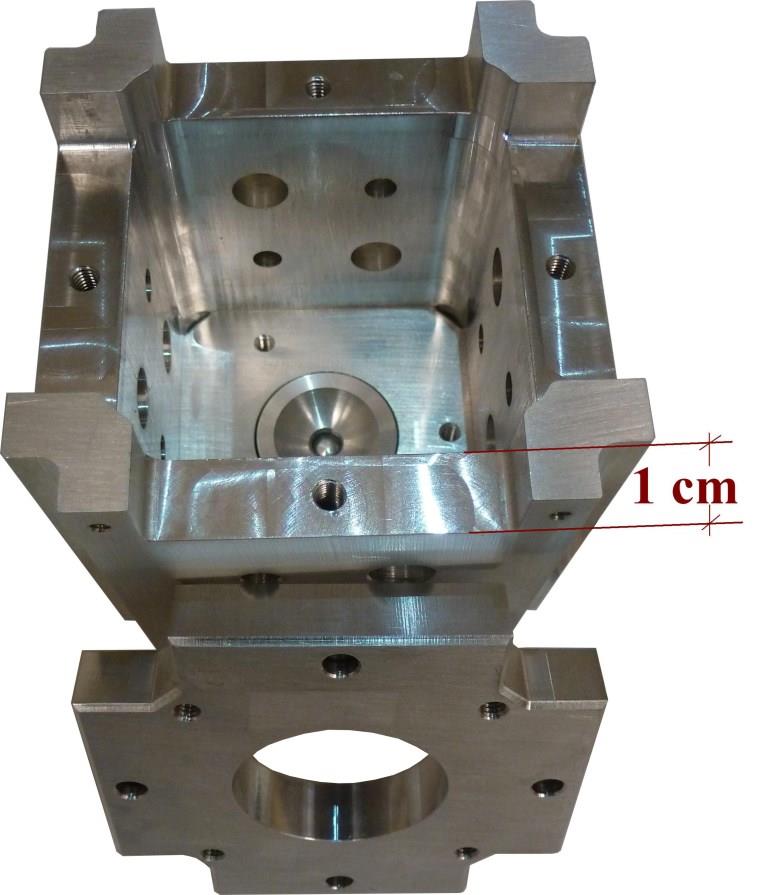}
\caption{Housing model for shaking test. The hemispherical recess on the bottom is here connected to a load cell.}
\label{fig:housing1}
\end{figure}

\begin{center}
\begin{figure}[ht!]
\caption{CAD model of the housing disassembled.}
\label{fig:housing2}
\includegraphics[width=1\columnwidth]{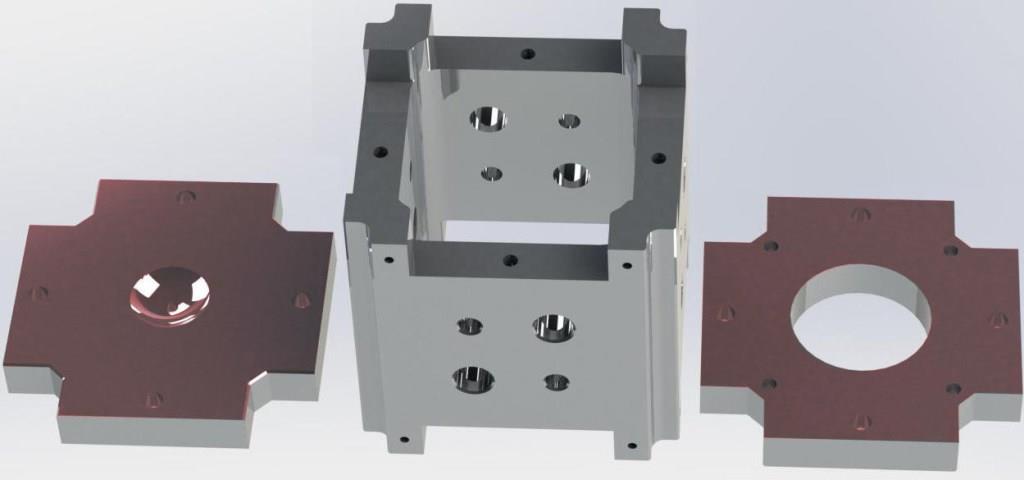}
\end{figure}
\end{center}

Taking all this into consideration, the final geometry is a thick-walled cube with 70 mm external and 50 mm internal side. The cube is divided in 3 parts (Fig.~\ref{fig:housing1} and \ref{fig:housing2}), +z and -z caps and the remaining sides (as a square tube). These two caps will be covered with SiC on the inside. To prevent extreme events (e.g. TM bonded to a side surface), it is still possible to depose coating on the other surfaces, but with less control on the thickness and reduced uniformity. The +z cap has to include a hemispherical recess that holds the TM during launch. The -z part has to allow the motion of the caging mechanism. The material chosen is Al6061, or Al7075, that has relatively low density, high yielding stress and good thermal conductivity.  The x and y sides of the housing have holes for DOSS LEDs and diodes. The outer edges along z are cut to host the linear guides of the caging mechanism. The 3 parts are then assembled together with a set of screws (M4). The resulting box is connected to the caging and is mounted on a titanium bulkhead perpendicular to the CubeSat z axis (Figure 2). 

Fig. \ref{fig:housing2} shows the modular structure of this design of the housing. The + and - z caps can be substituted with different pieces when required for testing purposes (e.g. a load cell mounted to the hemispherical recess during the shaking test). 
An important source of disturbance is the gravitational force induced by the presence of the spacecraft itself. The minimization of this source could impose slight changes in the final geometry. However, this effect is analyzed when the project is in a more advanced state because the final position and the mass properties of all the devices have to be precisely known. At that point, the spacecraft center of mass has to be located on the TM center. Also the gravitational gradient has an impact on the disturbance (the TM is not continuously in the center). In order to minimize this value, the inertia of the spacecraft has to be considered \cite{gravitationswank}. However, as a general guideline, a good GRS design is always the most symmetrical. That is why, for instance, all the corners of the housing are cut even if only two of them host the caging linear guides.  

\subsection{Mechanical Analysis}
The housing has to survive the launch environment, allow a safe caging and release of the TM and hold the sensor rigidly enough. While the best test to identify and qualify a complex mechanical system with several parts and screws is a shaking test, an estimation of the expected behavior has been performed with a FE analysis in COMSOL.
The modes of both the housing alone and the housing mounted in the spacecraft are considered. In order to bound the numerical error a convergence analysis is performed. The results (Table 1) do not present any possible issue. However, it will be useful to perform a comparison between numerical and measured values. As expected, the behavior of the housing alone is also symmetrical.

\begin{table}[!th]
\small
\renewcommand{\arraystretch}{1}
\caption{Modal analysis results}
\label{tab:modal}
\centering
\begin{tabular}{c|c|c}
\hline
Mode & Housing alone & Housing mounted inside \\
 & (Hz) & the spacecraft (Hz) \\
\hline
1st & 2596 & 227 \\ \hline
2nd & 2598 & 391 \\ \hline
3rd & 5920 & 497 \\ \hline
4th & 6112 & 628 \\ \hline
5th & 9519 & 695 \\ \hline
6th & 9525 & 719 \\
\hline
\end{tabular}
\end{table}

\begin{center}
\begin{figure}[ht!]
\caption{Von Mises stress and the deformation (enhanced) of the housing in caged static status.}
\label{fig:housingmises}
\includegraphics[width=\columnwidth]{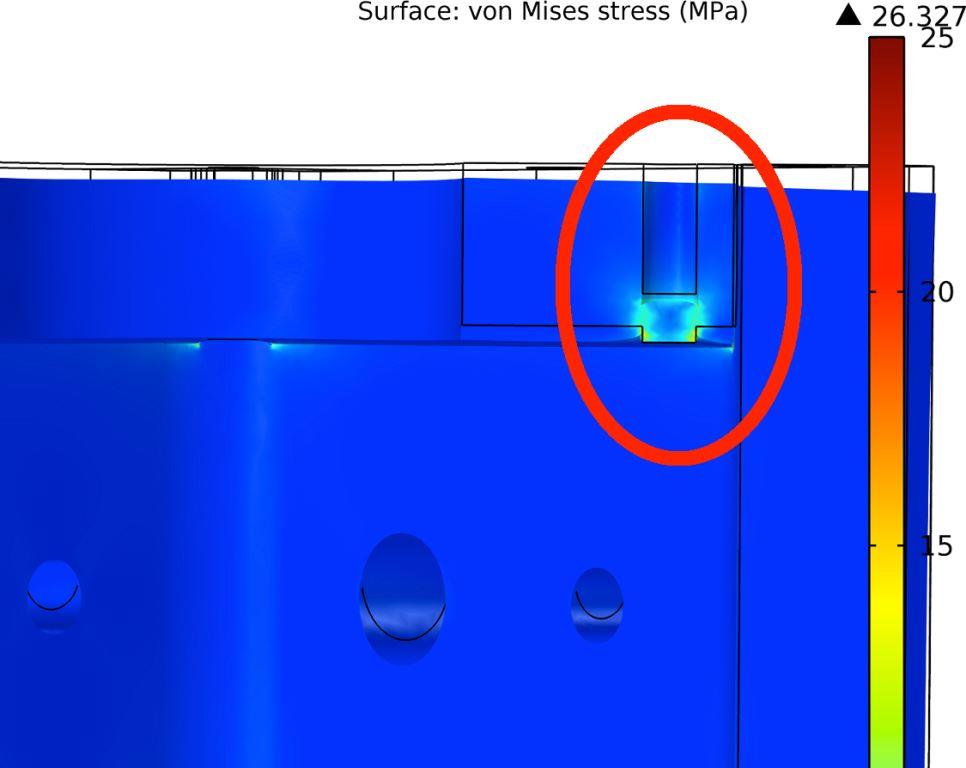}
\end{figure}
\end{center}

The static analysis of the housing with a caged TM (200 N of load) provides a maximum Von Mises stress of 27.5 MPa (safety factor: 11 and 16.5 for Al6061 and Al7075 respectively). This value does not take into account dynamic effects. On the other hand, this analysis provides precious information on the most critical features of the housing (Fig.~\ref{fig:housingmises}). The highest stress is located where the housing is fastened with the bulkhead (and the caging).
The mandatory shaking test will provide the final word on the design, on its capability to survive the launch environment and on potential damages on the TM and housing surface. The housing for the shaking test is slightly different from the proposed flight geometry. As a matter of facts, a load cell will be mounted on the +z cap to check the loads before, during and after the test.
 
The preferable spacecraft orbit is a sun-synchronous LEO. This means the satellite (as it happens for every CubeSat) will be close to Earth and its environment. One major effect is then the magnetic field that could induce forces on the TM. Also some systems inside the spacecraft generate a magnetic field. The housing has a major role in reducing this effect (reduction factor between the field inside and outside the housing: 0.01). In order to reach this goal a thin cover of a specific material has to be applied on the housing aluminum box. A FE analysis is performed to compare two materials (mu-metal and cobalt based 2714A, \cite{2714A}) and find the thickness required. 

\begin{center}
\begin{figure}[ht!]
\caption{Reduction factor as a function of the thickness of the shielding material (worst case angle between spacecraft and field, i.e. field parallel to z).}
\label{fig:housingmagnetic}
\includegraphics[width=\columnwidth]{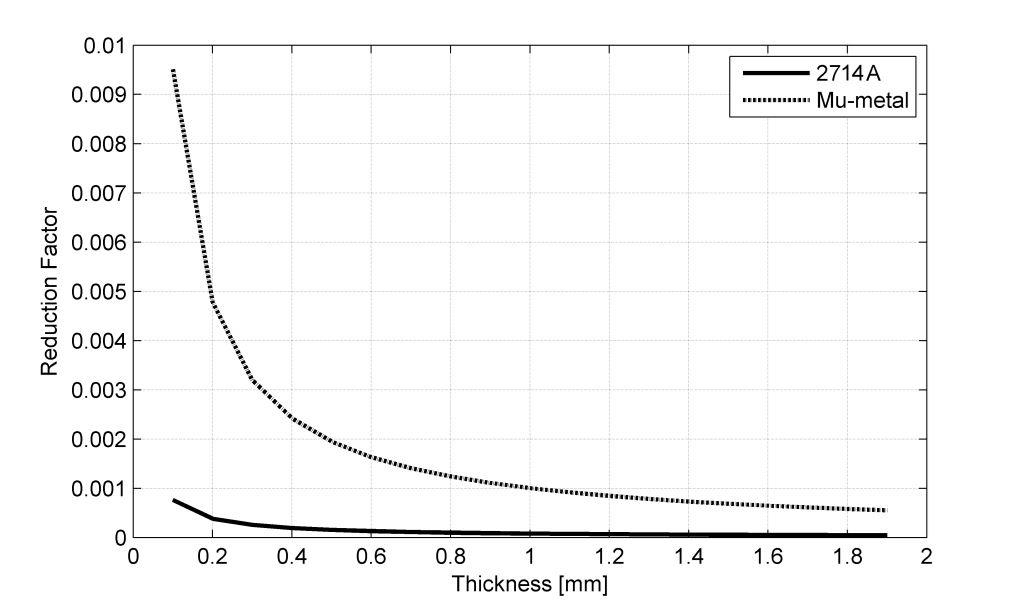}
\end{figure}
\end{center}

The geometry for the analysis is a simplified one (no small features, no caging). Besides, the shielding is modeled as a uniform cover around the housing. No real design on how this material is held has been defined. The excitation is a magnetic field with the highest value of the Earth magnetic field. Several analyses changing the angle between the field and the spacecraft are performed. The reduction factor is defined as the ratio between the average field inside the housing and the boundary excitation.
The results show a 10X better behavior of 2714A compared to mu-metal (Fig.~\ref{fig:housingmagnetic}). The reason for this behavior is the extremely high permeability of 2714A. In any case, the shielding material will be thinner than 0.1 mm. This means a foil, instead of a rigid plate, is enough to comply with the requirement.

\begin{center}
\begin{figure}[ht!]
\caption{Contributions to the spacecraft thermal equilibrium.}
\label{fig:thermal}
\includegraphics[width=0.9\columnwidth]{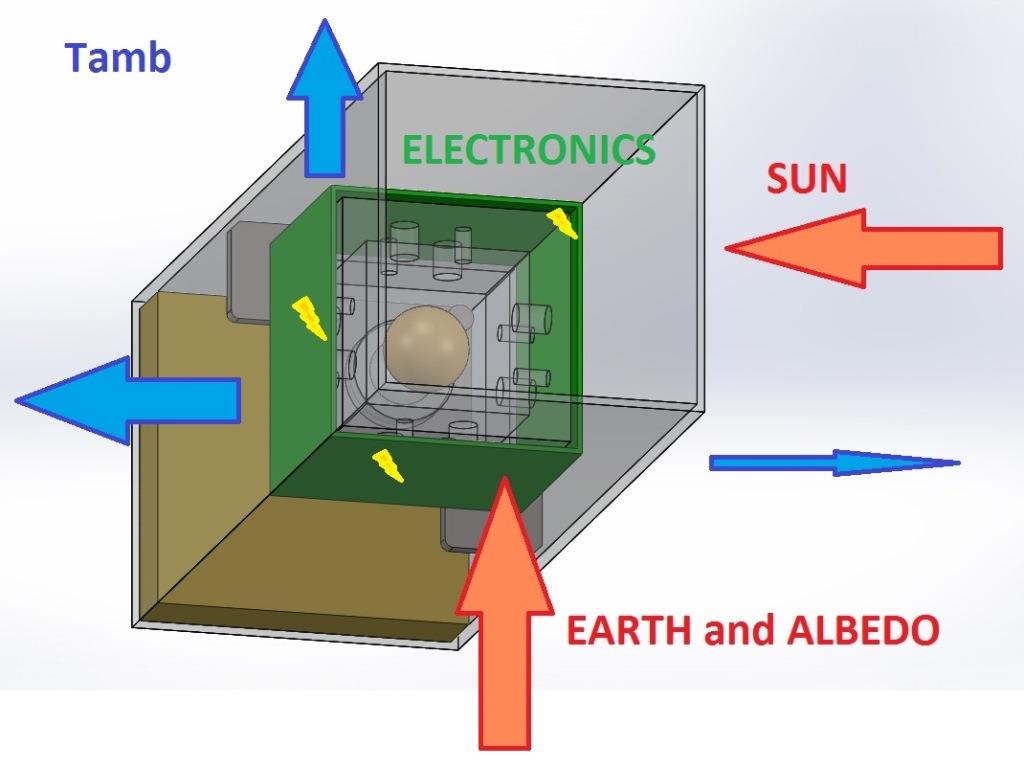}
\end{figure}
\end{center}

Thermal environment is one of the most critical aspects for every satellite. The drag-free performance requirement enhances this aspect. A temperature difference between opposed internal faces of the housing acts on the residual gas around the TM and results in a net force on the TM. In the current requirement, the maximum temperature difference is set to $10^{-3}$ $10^{-3}$ $\left(\frac{1 mHz}{f}\right)^{1/3}$ $K$ $Hz^{-1/2}$.
For the numerical analysis of this thermal problem accuracy is critical. The geometry of the model is therefore comprehensive of most of the mechanical features and power generating electronics.
These are the main aspects of the simulation (Fig.~\ref{fig:thermal}):
\begin{itemize}
	\item FE model includes spacecraft shell, thrusters, bulkhead, caging, housing, TM and DOSS electronics;
	\item entering heat flux: Sun (1000 $W/m^2$), Earth (400 $W/m^2$);
	\item outgoing heat flux: radiation with Tamb = 4 K;
	\item internal heat generation: 1 W (electronics);
	\item emissivity external surfaces: 0.7 (averaged between solar panels and spacecraft aluminum);
	\item internal radiation and conduction allowed.
\end{itemize}
The values and assumptions for the simulations are derived from \cite{gilmorethermal}.

In the first analysis, the model is subjected to step excitations (heat flux, power generation...) from the nominal conditions (293 K). The result provides an estimation of the performance in case of eclipse. This analysis has also the purpose of validating the model thanks to a comparison with a 1 degree of freedom analytical model. The steady-state temperature on the external sunny (hot) side is different by less than 3 K despite the assumptions for the analytical model (no 3D geometry, the spacecraft is a unique object).
The result of the simulation is then analyzed in the frequency domain and compared with the requirement. The temperature difference inside the housing along z is clearly above the requirement limits.
 
However, a more realistic analysis is that one simulating the effect of a heat flux with harmonic terms. More specifically, the Earth heat flux contributes with time-dependent components like $A \sin(2 \pi f t)$, where A is considered with two values: 80 $W/m^2$ and 40 $W/m^2$ and f is the frequency. However, as first approximation, in order to keep the FE analysis on a simple level, the maximum frequency is $2\times10^{-3}$ Hz. Otherwise, the period of the harmonic term and the time-constant of the spacecraft would be too different, requiring lengthy simulations. A certain amount of simulated time is required before the system reaches a steady state harmonic response. If the frequency is high the number of needed steps becomes higher.
 
\begin{center}
\begin{figure}[ht!]
\caption{PSD of the behavior of the temperature difference inside the housing (legend from top to bottom).}
\label{fig:housingPSD}
\includegraphics[width=\columnwidth]{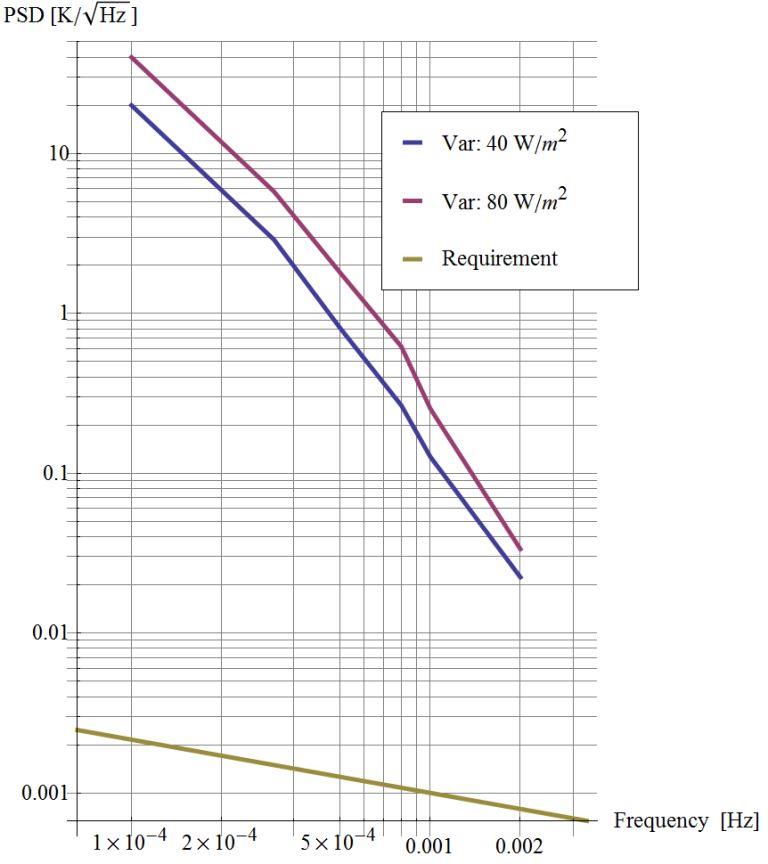}
\end{figure}
\end{center}

The thermal performance is clearly above the requirement limits. Several reasons can be identified for this behavior:
\begin{itemize}
	\item the geometry of the housing is not symmetrical as on one side there is the caging which results in a different thermal path and inertia;
	\item the requirement alone is particularly demanding as requires the uniformity to the level of mK;
	\item the FE simulation requires a large number of elements and nodes on a complex geometry. This hinders the accuracy of the model.
\end{itemize}
This problem has to be further understood. The first step would be an improvement in the analysis with software more specific for thermal and space problems. Particular care has to be taken in the definition of the mesh. The requirement has then to be analyzed and possibly relaxed. Finally, with a better understanding, a more symmetrical and better geometry has to be suggested. One of the reasons for the non-compliant behavior is the presence of the Caging System. Its presence alone is a strong violation of symmetry. A more complex solution has then to be found. Once a promising geometry and configuration is found, experimental measurements should be performed to validate the thermal design.



\chapter{Conclusion}\label{ch:conclusion} 

The use of drag-free technology in space is considered very promising. In facts, it can reduce the cost of LEO missions and is required in several missions in aeronomy and fundamental physics.

The state-of-the-art of this technology is represented by LISA-Pathfinder, an ESA mission whose launch is planned at the end of September 2015. The aim of this mission is further increasing the technology readiness level towards a Gravitational Wave Telescope, eLISA, as well as testing the model of several non-gravitational forces in interplanetary space.

Indeed, LISA-Pathfinder is a challenge and presents several critical points. One of them is the injection into geodesic trajectory of its cubic reference objects (or test masses, TM). The velocity of each of these objects, once all the constraints are removed, is limited to 5~$\mu$m/s to allow the capture and centering on behalf of the capacitive actuation. The Grabbing Positioning and Release Mechanism (GPRM) is the device in charge of such operation and of the compliance of the residual velocity with the specification. In order to assess this critical point, any phenomenon capable of transferring momentum from the GPRM to the test mass has to be analyzed. The main drivers of momentum are identified in the adhesion between the release tips and the relevant test mass and in the non-symmetrical GPRM action on the 2 sides of the test mass.

The effect of adhesion is estimated by means of several test campaigns performed on ground with a TM mock-up of reduced mass. The overall release velocity is then extrapolated by means of a large set of simulations, that takes into account also all the possible lacks of symmetry. The estimated worst-case velocity is equal to 1.50~$\mu$m/s. Thus, the TM velocity after the action of the release mechanism appears compliant with the requirement with a reasonable margin. The GPRM is one of the most critical aspects of LISA-Pathfinder and went through several years of both design and qualification. Such a result is very promising for the mission and provides important guidelines for the future eLISA space mission. In facts, some of the design parameters could be optimized in order to further mitigate the transferred momentum. 

Another key factor in the success of a drag-free space mission is the limitation of disturbances acting on the test mass that is free-floating inside the spacecraft. One of the systems involved in the disturbance reduction is the housing that also hosts the TM, supports the sensors and participate in the caging and release. In the scope of this research activity, the design of the housing for a low-cost drag-free mission is proposed. This design is described along with the mechanical, magnetic and thermal analyses. The mechanical and magnetic performance of this system appears easily compliant with the plans. On the other hand, the thermal performance shows some criticalities. This is another proof of drag-free projects still being a huge technological challenge.

The development of the drag-free concept is an important milestone for space industry and space science. The activities here described, as small part of a huge effort carried on all around the world, provide a small step towards a mature drag-free technology as a novel way for exploring the universe.


\appendix
\cleardoublepage
\chapter{Mitigation of Non Linear Motion of the Ground Release Tip}

The RTmu is moved by a PI ultrasonic piezo. In order to reproduce the absence of gravity, shear motion of the tip against the pendulum must be avoided. Unfortunately, the high speed camera shows that it is present a non negligible motion of the RTmu in the plane of the TMmu surface. This observation raises a concern over the goodness of a few tests, that -once the issue has been solved- have to be repeated.

The high speed camera is able to record grayscale images with a sampling frequency as high as 31466 Hz (for 60 $\times$ 80 pixels frames). The camera sees the shadow of the RTmu determined by a strong and constant light source. The voltage input to such a light source can be set, allowing repeatable lightning conditions.

\begin{figure}
\begin{center}
\includegraphics[width=0.7\columnwidth]{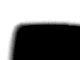}
\end{center}
\caption{Example of frame captured by the high speed camera.}
\label{fig:shadow}
\end{figure}

\begin{figure}
\begin{center}
\includegraphics[width=\columnwidth]{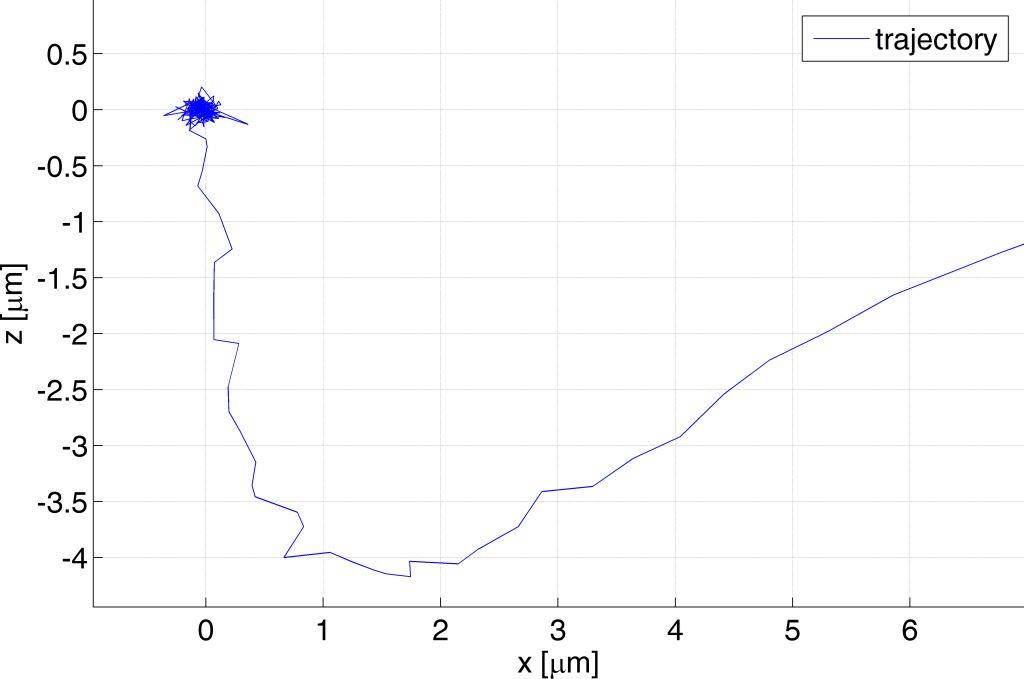}
\end{center}
\caption{Trajectory of the RTmu in the very first instants.}
\label{fig:trajbad}
\end{figure}

Fig.~\ref{fig:shadow} shows an example of frame captured by the high speed camera. The coordinates of the RTmu are estimated in this way:
\begin{enumerate}
	\item a threshold level of gray is chosen and a loop that analyzes all the frames is started;
	\item the bottom horizontal line of the frame is selected and a piecewise function of the graylevels is defined;
	\item the position of the threshold, $x_{RT}$, is found on this function;
	\item a vertical line is selected such that its position is $x_{RT} + \Delta$. $\Delta$ is equal for all the frames and is chosen in such a way that the vertical line cuts the frame where the RTmu boundary is qualitatively horizontal. A piecewise function of the graylevels is defined on this line.
	\item the position of the threshold, $z_{RT}$, is found on this function;
	\item the loop is repeated a few times with different thresholds.
\end{enumerate}
The trajectory obtained with this procedure, before the shear motion is mitigated, is depicted in Fig.~\ref{fig:trajbad}. The MATLAB script that executes the analysis is reported at the end of this Appendix.

Several effects allow to highlight the most-likely causes of this shear motion that:
\begin{itemize}
	\item depends on the commanded retraction. If the retraction is negative (i.e. an advancement), the z motion has also an opposed sign;
	\item depends on the RTmu initial position, but not on the tilting of the actuator;
	\item is visible not only on the RTmu, but also on the platform fized on the ultrasonic piezo. The scale of the motion measured on the platform is smaller by a factor 2 or 3;
	\item does not depend on the misalignement between the barycenter and the linear guide of the actuator. This has been the main hypothesis on the causes, but was refuted by the fact that moving the barycenter did not improved the behaviour significantly.	
\end{itemize}
These observations suggest that the motion is a dynamic effect and is due to the actuator itself, that due to aging or lack of performance presents also a vertical motion once it is actuated. The main hypothesis is that the force applied on the platform and the linear guide are not perfectly aligned. Thus, the non parallel component of the force pushes up or down the platform.

\begin{figure}
\begin{center}
\includegraphics[width=\columnwidth]{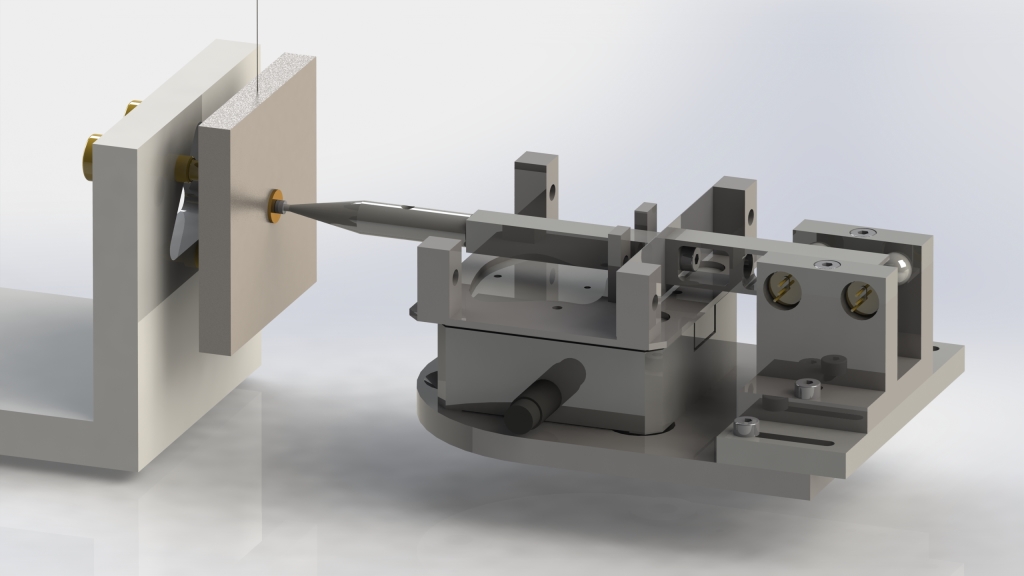}
\end{center}
\caption{Rendered view of the core of the experiment as improved for mitigation of shear motion.}
\label{fig:1bladeapp}
\end{figure}

\begin{figure}
\begin{center}
\includegraphics[width=\columnwidth]{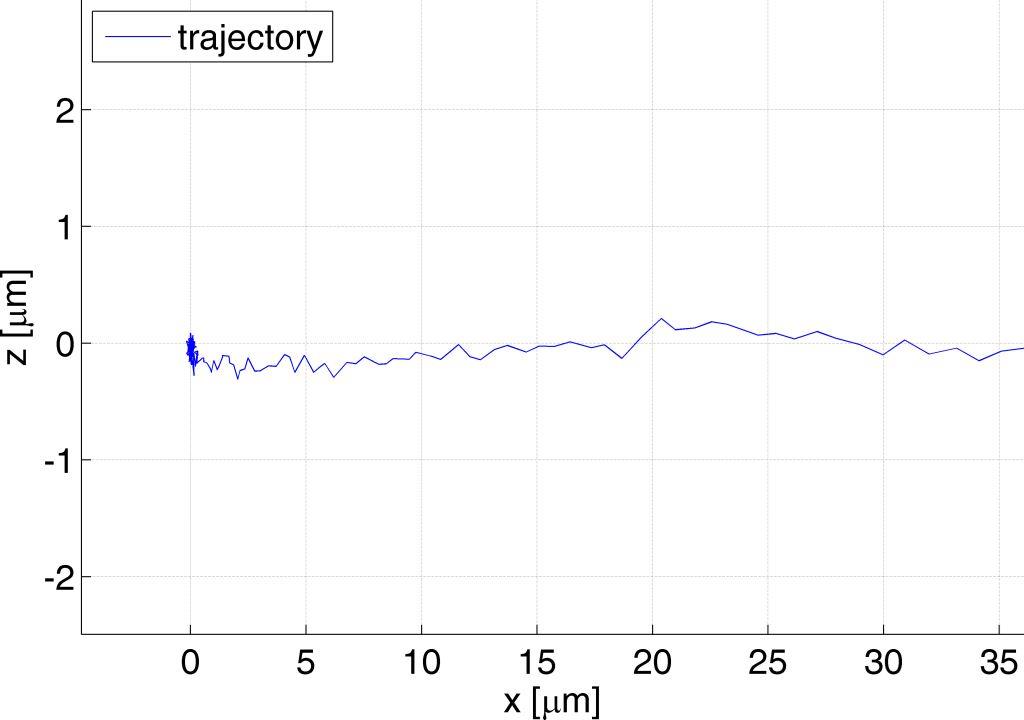}
\end{center}
\caption{Trajectory of the RTmu in the very first instants.}
\label{fig:trajgood}
\end{figure}

The shear motion can then be mitigated by reducing the stiffness of the blades against rotations and vertical motions. This is obtained by simply removing one of the two blades, Fig.~\ref{fig:1blade}. As a consequence the RTmu filters the non ideal motion and performs a much better trajectory, Fig.~\ref{fig:trajgood}.

\section{MATLAB script}

\begin{footnotesize}
\begin{verbatim}

close all
clear all

cd <directoryname>
flist = dir;
nframes = length(flist)-4;
no = str2double(flist(end).name(11:15))-nframes;
dx = 15;

disp('Start')

for j = 1:6
    
    glevel = 50+25*j;
    
for i = 1:nframes
   frameind = i + no;
   input = double(imread(['retraction' num2str(frameind) '.bmp']))+1;
   [nc,nr] = size(input);
   nc = nc - 1;
   Hline = input(nc-1,:);
   Hind = find(Hline < glevel, 1, 'first');
   posxa = (glevel - Hline(Hind-1))/(Hline(Hind)-Hline(Hind-1))+Hind-1;
   Hline = input(nc,:);
   Hind = find(Hline < glevel, 1, 'first');
   posx(i) = ((glevel-Hline(Hind-1))/(Hline(Hind)-Hline(Hind-1))+Hind ...
				- 1 + posxa)/2;
    
   Vlinesx = input(:,Hind - 1 + dx);
   Vlinedx = input(:,Hind + dx);
   Vindsx = find(Vlinesx < glevel, 1, 'first');
   posysx = (glevel-Vlinesx(Vindsx-1))/(Vlinesx(Vindsx)-Vlinesx(Vindsx-1))...
					+ Vindsx;
   Vinddx = find(Vlinedx < glevel, 1, 'first');
   posydx = (glevel-Vlinedx(Vinddx-1))/(Vlinedx(Vinddx)-Vlinedx(Vinddx-1))...
					+ Vinddx;
   posy(i) = (posx(i) - Hind + 1)*(posydx - posysx) + posysx;

end

posxj(j,:) = posx-mean(posx(1:10));
posyj(j,:) = posy-mean(posy(1:10));

end

%%

posx = mean(posxj);
posy = mean(posyj);

disp('Cicle done...now plotting')

px0 = mean(posx(1:10));
py0 = mean(posy(1:10));
pxf = mean(posx(end-100:end)-px0);

posx = 200*(posx - px0)/pxf;
posy = -200*(posy - py0)/pxf;

%%
close all

t = 0:1:length(posx)-1;
t = t/31466;

figure
hold on
plot(t,posx)
plot(t,posy, 'Color', [0.4 0.9 0.4])
legend('X','Z')
xlabel('t [s]')
ylabel('[\mu m]')
grid on

startind = find(abs(posx) > 0.4, 1, 'first') - 20;

figure
axes('fontsize', 16)
hold on
plot(posx,posy)
legend('trajectory','linewidth', 2)
xlabel('x [\mum]')
ylabel('z [\mum]')
grid on

traj = [posx;posy];

\end{verbatim}
\end{footnotesize}

\chapter{Post-Acquisition Removal of the Interferometer Noise}

The interferometer used in the on-ground experimental facility, a SIOS SPS 2000, demonstrates a lack of performance at the nm range. 

It can be observed that, the spurious vibration magnitude is a function of the strength of the interferometer signal and in reality depends on the TM position. This dependence is a consequence of the behaviour of the electronics at certain laser wavelength submultiples (1/8 among all). Fig.~\ref{fig:TMdet} shows the motion of TMmu without the linear trend, that has been subtracted. The distortion of the signal is clear in the saw-tooth shape. Such a shape is fitted with a polynomial and a sum of trigonometric functions:
\begin{equation}
	x_{TMmu,fit} = \sum\limits_{m=1}^{30} a_m t^m + \sum\limits_{n=1}^9 b_n cos(2\pi n x_{TM}/\lambda) + \sum\limits_{n=1}^9 c_n sin(2\pi n x_{TM}/\lambda) 
\end{equation}
where $a_m$, $b_n$, $c_n$ and $\lambda$ are fit parameters. $\lambda$ is the wavelength and the fit result is consistent with the nominal value.

\begin{figure}
\begin{center}
\includegraphics[width=1\columnwidth]{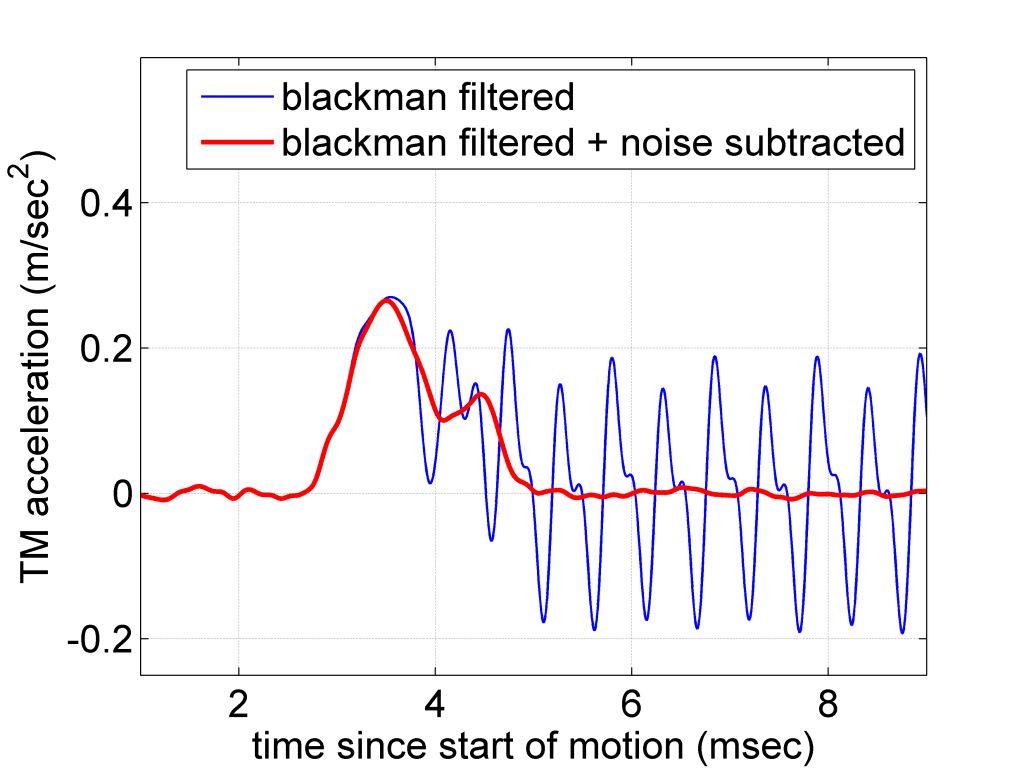}
\end{center}
\caption{Example of TM acceleration as measured and after correction in the high preload test campaign.}
\label{fig:acch2}
\end{figure}

\begin{figure}
\begin{center}
\includegraphics[width=1.1\columnwidth]{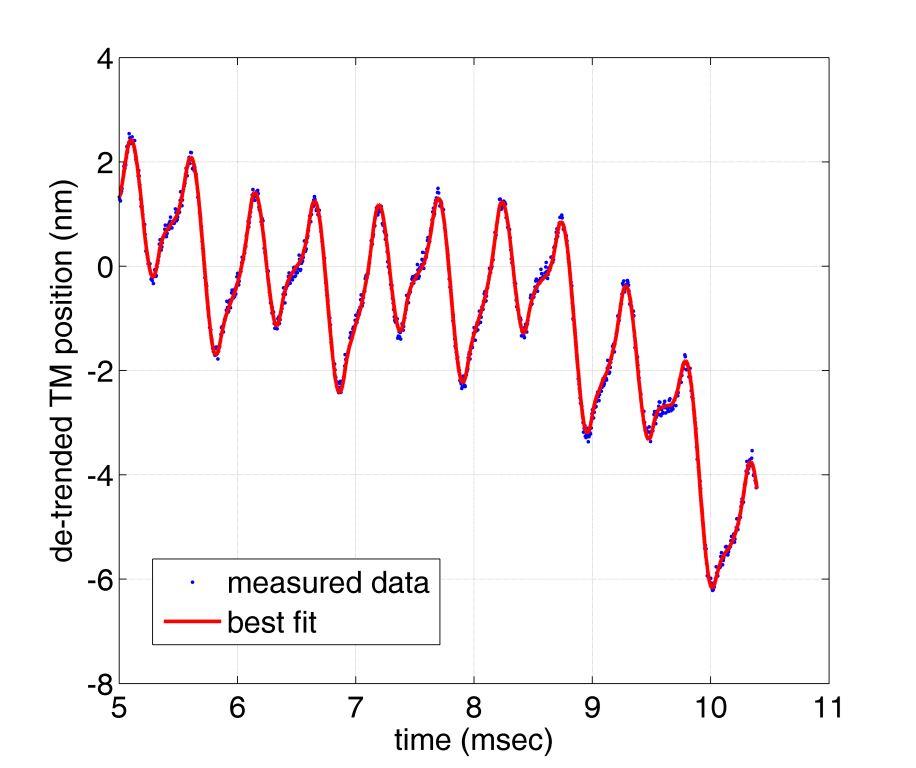}
\end{center}
\caption{TM motion without the linear trend. The distortion of the signal is represented by the saw-tooth behaviour.}
\label{fig:TMdet}
\end{figure}

For each test, the fit is performed on the part of the acquisition in which the TMmu motion is nominally linear. The fit parameters of the trigonometric functions are then used to subtract the periodic motion from all the data-set. Fig.~\ref{fig:acch2} shows the acceleration of the TMmu before and after the disturbance removal. Indeed, the acceleration is always filtered (3 kHz cut-frequency with a Blackman filter).

Being position-dependent, this disturbance is very critical when the impulse applied on the TMmu is long and the displacement in the same timeframe is not negligible. The test campaign with high residual load is an example of this situation. In the other campaigns, with low initial load, the disturbance determines a deformation of the measured motion that is mostly relevant after the impulse.

\chapter{Surfaces matching}

The matching of two surfaces (before and after the contact) is performed according to the following procedure:
\begin{enumerate}
	\item definition of a threshold that selects only the points on the bottom (or valleys).
	\item the user picks two points on each image (i.e. surface) that coincide with similar features. Their coordinates will be used to estimate a first guess.
	\item the root mean square between the valleys is minimized. The coordinates that result in a NaN are excluded from the computation.
\end{enumerate}

\section{MATLAB scripts}
\begin{footnotesize}
\begin{verbatim}
close all
clear all

%% surf PRE
filenamepre = <filename>;
data = importsurf(filenamepre);
[Xmbe,Ymbe,Zmbe] = cleansurf(data);

res = 1.10365e-007;

medZmbe = medfilt2(Zmbe,[30 30]);
Vbe = Zmbe;
Vbe(Zmbe > 0.2) = NaN;

figure
hold on
surf(Xmbe(1:5:end,1:5:end),Ymbe(1:5:end,1:5:end),Vbe(1:5:end,1:5:end))
zlim([-5 1])
caxis([-5 2])
shading interp
grid on
 for i = 1:2
    [xbe(i),ybe(i)] = ginput(1);
    plot(xbe(i),ybe(i),'ro','linewidth', 2)
    text(xbe(i),ybe(i),1,num2str(i),'Color',[0 0 0],'FontSize',20)
 end

Vbe(Zmbe > -0.7) = NaN;
 
%% surf POST

[Xmaf,Ymaf,Zmaf] = cleansurf(data);
filenamepost = <filename>;
data = importsurf(filenamepost);

medZmf = medfilt2(Zmaf,[30 30]);
Vaf = Zmaf;
Vaf(Zmaf > 0.35) = NaN;

figure
hold on
surf(Xmaf(1:5:end,1:5:end),...
		Ymaf(1:5:end,1:5:end),Vaf(1:5:end,1:5:end))
zlim([-5 1])
caxis([-5 2])
shading interp
grid on
for i = 1:2
    [xaf(i),yaf(i)] = ginput(1);
     plot(xaf(i),yaf(i),'ro','linewidth', 2)
     text(xaf(i),yaf(i),1,num2str(i),'Color',[0 0 0],...
					'FontSize',20)
end

Vaf(Zmaf > -0.4) = NaN;

%% compute from selection

XGtot = (Xmaf(1,end)+Xmaf(1,1))/2;
YGtot = (Ymaf(1,1)+Ymaf(end,1))/2;

dth0 = - atan((yaf(2)-yaf(1))/(xaf(2) - xaf(1)))....
		+ atan((ybe(2)-ybe(1))/(xbe(2) - xbe(1)));
xm = (xaf(2) + xaf(1))/2 - XGtot;
ym = (yaf(2) + yaf(1))/2 - YGtot;
xma = xm*cos(dth0) - ym*sin(dth0) + XGtot;
yma = ym*sin(dth0) + ym*cos(dth0) + YGtot;
dx0 = - xma + (xbe(2) + xbe(1))/2;
dy0 = - yma + (ybe(2) + ybe(1))/2;

Nminr = 10;
Nmaxr = 1590;

Nminc = 10;
Nmaxc = 1190;

Xp = Xmaf(Nminr:Nmaxr,Nminc:Nmaxc);
Yp = Ymaf(Nminr:Nmaxr,Nminc:Nmaxc);
Vp = Vaf(Nminr:Nmaxr,Nminc:Nmaxc);
XG = (Xp(1,end)+Xp(1,1))/2;
YG = (Yp(1,1)+Yp(end,1))/2;
Xp = Xp - XG;
Yp = Yp - YG;

% Vpf = medfilt2(Vp,[20 20]);
% Vbef = medfilt2(Vbe,[20 20]);

Vpf = Vp;
Vbef = Vbe;

%% Match
close all

clearvars conf fval

fvalopt = 100;

for i = 1:10
%    close all
    dxr = dx0 + heaviside(i-2)*10*randn;
    dyr = dy0 + heaviside(i-2)*10*randn;
    dthr = dth0 + heaviside(i-2)*0.02*randn;
    options = optimset('MaxIter',15,'MaxFunEvals',1000,...
					'TolFun',1e-2,'TolX',1e-3);  
    [confr,fval] = fminsearch(@(conf)RMS(conf, Xp, Yp, Vpf,...
					Xmbe, Ymbe, Vbef),[XG+dxr YG+dyr 0 dthr 0 0], options);
    if fvalopt > fval
        fvalopt = fval;
        dxopt = dxr;
        dyopt = dyr;
        dthopt = dthr;
    end
end

options = optimset('MaxIter',1000,'MaxFunEvals',1000,...
	'TolFun',1e-2,'TolX',1e-3,'PlotFcns', {@optimplotfval @optimplotx});  
[conf,fval] = fminsearch(@(conf)RMS(conf, Xp, Yp, Vpf,...
			Xmbe, Ymbe, Vbef),[XG+dxopt YG+dyopt 0 dthopt 0 0], options);

%% compute difference

x = conf(1);
y = conf(2);
z = conf(3);
th = conf(4);
alpha = conf(5);
beta = conf(6);

Xmaff = x + (Xmaf-XG)*cos(th) - (Ymaf-YG)*sin(th);
Ymaff = y + (Xmaf-XG)*sin(th) + (Ymaf-YG)*cos(th);
Zmaff = Zmaf + (Xmaf-XG)*alpha + (Ymaf-YG)*beta + z;
Zmbef = Zmbe;
diff =  Zmaff - interp2(Xmbe, Ymbe, Zmbef, Xmaff, Ymaff);
nandiff = isnan(diff);
diff(nandiff) = 0;

function D = RMS(conf, Xp, Yp, Zp, Xm, Ym, Zm)
if length(conf)==3
    x = conf(1);
    y = conf(2);
    th = conf(3);
    Xpr = x + Xp*cos(th) - Yp*sin(th);
    Ypr = y + Xp*sin(th) + Yp*cos(th);

    diff = (Zp - interp2(Xm, Ym, Zm, Xpr, Ypr)).^2;
    nandiff = isnan(diff);
    diff(nandiff) = 0;

    D = sqrt(sum(sum(diff))/...
					(numel(diff)-length(diff(nandiff))).^2);
    
elseif length(conf)==6
    x = conf(1);
    y = conf(2);
    z = conf(3);
    th = conf(4);
    alpha = conf(5);
    beta = conf(6);
    
    Xpr = x + Xp*cos(th) - Yp*sin(th);
    Ypr = y + Xp*sin(th) + Yp*cos(th);
    Zpr = Zp + Xp*alpha + Yp*beta + z;
    
    diff = (Zpr - interp2(Xm, Ym, Zm, Xpr, Ypr)).^2;
    nandiff = isnan(diff);
    diff(nandiff) = 0;

    D = sqrt(sum(sum(diff))/...
					(numel(diff)-length(diff(nandiff))).^2);

end
end

\end{verbatim}
\end{footnotesize}

\chapter{Mathematica Notebooks}

Most of the other scripts (simulation of the release, mechanism identification...) have been written in Wolfram Mathematica and reporting them here would be space-consuming and useless. They can be provided on-request by the author.

\cleardoublepage
\manualmark
\markboth{\spacedlowsmallcaps{\bibname}}{\spacedlowsmallcaps{\bibname}} 
\phantomsection 
\refstepcounter{dummy}
\addtocontents{toc}{\protect\vspace{\beforebibskip}} 
\addcontentsline{toc}{chapter}{\tocEntry{\bibname}}
\bibliographystyle{plainnat}
\label{app:bibliography} 
\bibliography{Journals,Technical}


%
\cleardoublepage
\pdfbookmark[1]{Acknowledgments}{acknowledgments}


\bigskip

\begingroup
\let\clearpage\relax
\let\cleardoublepage\relax
\let\cleardoublepage\relax
\chapter*{Acknowledgments}

\rmfamily
\vspace{2.5cm}

{\footnotesize
First and foremost I want to thank prof. Daniele Bortoluzzi. It has been a luck being his Ph.D. student. Among all, he's a patient and very careful person. Besides, his scientific guide has always been exceptional in all instances. He's able to focus on both tiny details and the big plan, inspire people in carrying on tenaciously even difficult tasks as well as aim at realistic goals.

At the same time, I am grateful to prof. Stefano Vitale for giving me the chance of contributing to such a huge enterprise as LISA-Pathfinder. He is not only a great scientist, but also a precise and capable manager and a professional example. All Europe will benefit from his work and abnegation. 

Last, but not least, I want to thank John W. Conklin for his help and guide for some parts of my studies. He's an extremely gifted researcher and engineer. His contribution has been a key factor in some of the results of this thesis and, at the same time, his support has been invaluable in building a successful and useful plan for my research.

Together with these persons, I am extremely grateful to Matteo Benedetti and prof. Luca Zaccarian and all the other professors, researchers in the Dept. of Industrial Engineering as well as the LISA group of the Dept. of Physics at UniTN. Together with them, I thank Luca Gambini, Andrea Zambotti and Blondo Seutchat for continuing and expanding my work with their contribution. A special mention goes also to Maximiliano, Antonio and the other brave Ph.D. students.

The design of the Drag-Free CubeSat was performed during a stay as a Visiting Researcher at the Hansen Experimental Physics Lab at Stanford University. I am in debt to Dan DeBra, Sasha Buchman and Andrew Kalman for their hospitality, help and useful lessons. I also want to thank Andreas Zoellner and Grant Cutler for their support in both making my stay proficuous and with everyday life's matters.

On top of this, I am convinced that the opportunity of working in Airbus Space, before the beginning of my Ph.D., has been of key importance in learning a positive work-ethic and an optimal balance between life and work. This wouldn't have been easy without the friendly, but rigorous attitude of Alex Schleicher, Nico Brandt and Tobias Ziegler. I'm very proud that, together with them and prof. Bortoluzzi, we were able to give the same chance I had to other master students, after me.

One last mention goes to CERN, that decided to believe in me after the Ph.D. and is offering everyday a fascinating and rewarding work opportunity. All the research institutes and international organizations have much to learn from CERN's commitment to peace, inclusiveness, equality, creativity and open-access.

\vspace{1cm}
\selectlanguage{italian}

Un ringraziamento speciale va alla mia famiglia che ha supportato le mie scelte e le mie necessità per il dottorato sia moralmente che economicamente. Sono convinto che la passione, l'interesse, l'attenzione siano caratteristiche che non si improvvisano, ma devono essere coltivate con tempo, pazienza e fatica. Spero qualcosa sia maturato, ma quel poco che lo ha fatto per gran parte non è stato merito mio.

In ordine sparso, un grazie sentito va a tutti i miei amici, che in maggioranza più che assoluta non hanno idea di come risolvere un'equazione di secondo grado (ma non mi perdo d'animo). 

Non voglio infine dimenticare i miei lupetti, esploratori, bambini e ragazzi con cui ho avuto la fortuna di fare volontariato, servizio e campeggi. Grazie per gli autentici momenti di felicità e anche di divertimento (passati e \textit{futuri}).

\vspace{1cm}
\selectlanguage{english}

Thanks also to all the other people that crossed my life, from Greece to Latvia, from UK to California, from China to Australia. Meeting people all around the world is a huge fortune. The Earth is a small spaceship, we are all equally passengers and every fight is a non-sense.

Thank you all for your contribution so far and keep up the good work!

\vspace{2cm}
\selectlanguage{italian}

\textit{''Lassù la montagna è silenziosa e deserta. La neve che in questi giorni è caduta abbondante ha cancellato i sentieri dei pastori, le aie dei carbonai, le trincee della Grande guerra, le avventure dei cacciatori. \\ E sotto quella neve vivono i miei ricordi.''}
\begin{flushright}
--- Mario Rigoni Stern, Sentieri sotto la neve
\end{flushright}



}
\bigskip

\noindent 

\endgroup


\end{document}